\documentclass[11pt]{amsart}

\usepackage{geometry}
\usepackage[utf8]{inputenc}
\usepackage[T1]{fontenc}
\usepackage{amsfonts}
\makeatletter
\let\old@tocline\@tocline
\let\section@tocline\@tocline

\makeatletter
\renewcommand{\tocsubsection}[3]{
  \indentlabel{\@ifnotempty{#2}{\ignorespaces#1 #2\quad}}#3}
\renewcommand{\tocsubsection}[3]{
  \indentlabel{\@ifnotempty{#3}{\ignorespaces#1 #2\quad}}#3}
\newcommand\@dotsep{4.5}
\def\@tocline#1#2#3#4#5#6#7{\relax
  \ifnum #1>\c@tocdepth 
  \else
    \par \addpenalty\@secpenalty\addvspace{#2}
    \begingroup \hyphenpenalty\@M
    \@ifempty{#4}{
      \@tempdima\csname r@tocindent\number#1\endcsname\relax
    }{
      \@tempdima#4\relax
    }
    \parindent\z@ \leftskip#3\relax \advance\leftskip\@tempdima\relax
    \rightskip\@pnumwidth plus1em \parfillskip-\@pnumwidth
    #5\leavevmode\hskip-\@tempdima{#6}\nobreak
    \leaders\hbox{$\m@th\mkern \@dotsep mu\hbox{.}\mkern \@dotsep mu$}\hfill
    \nobreak
    \hbox to\@pnumwidth{\@tocpagenum{\ifnum#1=1\bfseries\fi#7}}\par
    \nobreak
    \endgroup
  \fi}
\AtBeginDocument{
\expandafter\renewcommand\csname r@tocindent0\endcsname{0pt}
}
\def\l@subsection{\@tocline{2}{0pt}{2.5pc}{5pc}{}}
\makeatother

\DeclareRobustCommand{\SkipTocEntry}[4]{}

\usepackage{graphicx,stmaryrd}
\usepackage{amsmath}
\usepackage{amssymb,color}
\usepackage{mathrsfs}
\usepackage{subcaption}  

\usepackage{mathtools}

\def\br{\begin{remark}\rm\small}
\def\er{\end{remark}}
\newcommand{\dd}{\mathrm{d}}
\newcommand{\beq}{\begin{equation}}
\newcommand{\eeq}{\end{equation}}
\newcommand{\bea}{\begin{eqnarray}}
\newcommand{\eea}{\end{eqnarray}}
\newcommand{\Res}{\mathop{\,\rm Res\,}}

\newtheorem{theorem}{Theorem}
\newtheorem{example}[theorem]{Example}
\newtheorem{algorithm}[theorem]{Algorithm}
\newtheorem{proposition}[theorem]{Proposition}
\newtheorem{lemma}[theorem]{Lemma}
\newtheorem{corollary}[theorem]{Corollary}
\newtheorem{conjecture}[theorem]{Conjecture}
\theoremstyle{definition}
\newtheorem{definition}[theorem]{Definition}
\theoremstyle{remark}
\newtheorem{remark}[theorem]{Remark}

\numberwithin{equation}{section}
\numberwithin{theorem}{section}

\let\oldtocsubsection=\tocsubsection
\renewcommand{\tocsubsection}[2]{\hspace{1em}\oldtocsubsection{#1}{#2}}

\calclayout

\begin{document}

\title[Simple maps, Hurwitz numbers, and topological recursion]{Simple maps, Hurwitz numbers, and topological recursion}
\date{\today}
\author{Ga\"etan Borot}
\address{Max Planck Institut f\"ur Mathematik, Vivatsgasse 7, 53111 Bonn, Germany}
\email{gborot@mpim-bonn.mpg.de}
\author{Elba Garcia-Failde}
\address{Max Planck Institut f\"ur Mathematik, Vivatsgasse 7, 53111 Bonn, Germany}
\email{elba@mpim-bonn.mpg.de}
\thanks{We thank Roland Speicher for free discussions, Di Yang for explaining us the result of \cite{DiYang}, Dominique Poulhalon and Guillaume Chapuy for bringing to our attention the literature on non-separable maps, Sergey Shadrin and Danilo Lewa\'{n}ski for useful discussions on references and clearer presentations of some results, and Max Karev for discussing references. This work benefited from the support of the Max-Planck Gesellschaft.}

\begin{abstract}
We introduce the notion of fully simple maps, which are maps with non self-intersecting disjoint boundaries. In contrast, maps where such a restriction is not imposed are called ordinary.  We study in detail the combinatorics of fully simple maps with topology of a disk or a cylinder. We show that the generating series of simple disks is given by the functional inversion of the generating series of ordinary disks. We also obtain an elegant formula for cylinders. These relations reproduce the relation between moments and (higher order) free cumulants established by Collins et al \cite{Secondorderfreeness}, and implement the symplectic transformation $x \leftrightarrow y$ on the spectral curve in the context of topological recursion.  We conjecture that the generating series of fully simple maps are computed by the topological recursion after exchange of $x$ and $y$. We propose an argument to prove this statement conditionally to a mild version of the symplectic invariance for the $1$-hermitian matrix model, which is believed to be true but has not been proved yet. Our conjecture can be considered as a combinatorial interpretation of the property of symplectic invariance of the topological recursion.

Our argument relies on an (unconditional) matrix model interpretation of fully simple maps, via the formal hermitian matrix model with external field. We also deduce a universal relation between generating series of fully simple maps and of ordinary maps, which involves double monotone Hurwitz numbers. In particular, (ordinary) maps without internal faces -- which are generated by the Gaussian Unitary Ensemble -- and with boundary perimeters $(\lambda_1,\ldots,\lambda_n)$ are strictly monotone double Hurwitz numbers with ramifications $\lambda$ above $\infty$ and $(2,\ldots,2)$ above~$0$. Combining with a recent result of Dubrovin et al.~\cite{DiYang}, this implies an ELSV-like formula for these Hurwitz numbers.
\end{abstract}

\maketitle

\tableofcontents

\section{Introduction}

Maps are surfaces obtained from gluing polygons, and their enumeration by combinatorial methods has been intensively studied since the pioneering works of Tutte \cite{Tutte2}. In physics, summing over maps is a well-defined discrete replacement for the non-obviously defined path integral over all possible metrics on a given surface which underpin two-dimensional quantum gravity. The observation by t'Hooft \cite{tHooft} that maps are Feynman diagrams for the large rank expansion of gauge theories led Br\'ezin--Itzykson--Parisi--Zuber \cite{BIPZ} to the discovery that hermitian matrix integrals are generating series of maps. The rich mathematical structure of matrix models -- integrability, representation theory of $U(\infty)$, Schwinger--Dyson equations, etc. -- led to further insights into the enumeration of maps, \textit{e.g.} \cite{ACM,dFZJ}. It also inspired further developments, putting the problem of counting maps into the more general context of enumerative geometry of surfaces, together with geometry on the moduli space of curves \cite{HarerZagier,Witten,Kontsevich}, volumes of the moduli space \cite{Mirzakhani}, Gromov-Witten theory \cite{Witten,BKMP}, Hurwitz theory \cite{EMS,EynardHarnad}, etc. and unveiling a common structure of ``topological recursion'' \cite{EORev,EynardBook}.

We say that a face $f$ of a map is \emph{simple} when at most two edges of $f$ are incident to every vertex in $f$. In the definition of maps, polygons may be glued along edges without restrictions, in particular faces may not be simple. This leads to singular situations, somehow at odds with the intuition of what a neat discretization of a surface should look like. This definition is the one naturally prescribed by the Feynman diagram expansion of hermitian matrix models. It is also one for which powerful combinatorial (generalized Tutte's recursion, Schaeffer bijection, etc.) and algebraic/geometric (matrix models, integrability, topological recursion, etc.) methods can be applied. Within such methods, it is possible to count maps with restrictions of a global nature (topology, number of vertices, number of polygons of degree $k$) and on the number of marked faces (also called boundaries) and their perimeters. Combined with probabilistic techniques, they helped in the development of a large corpus of knowledge about the geometric properties of random maps.

Tutte introduced in \cite{Tutte4} the notion of planar non-separable map, in which faces must all be simple. Some of the combinatorial methods aforementioned have been extended to handle non-separable maps -- see \textit{e.g.} \cite{Nonsep1,Nonsep2} --, but the analytic methods have not been explored and the probabilistic aspects not as much.

Brown, a student of Tutte, studied non-separable maps of arbitrary genus \cite{Nonsep}, which were refined later in \cite{Nonsep0}, distinguishing between the notions of graph-separability and map-separability, which only coincide for planar maps. Maps with only simple faces are still non-separable for arbitrary genus. However, our notion of simplicity is much stronger for non-planar maps than both notions of non-separability. 

\begin{figure}[h!]
 \begin{center}
        \def\svgwidth{0.7\columnwidth}
        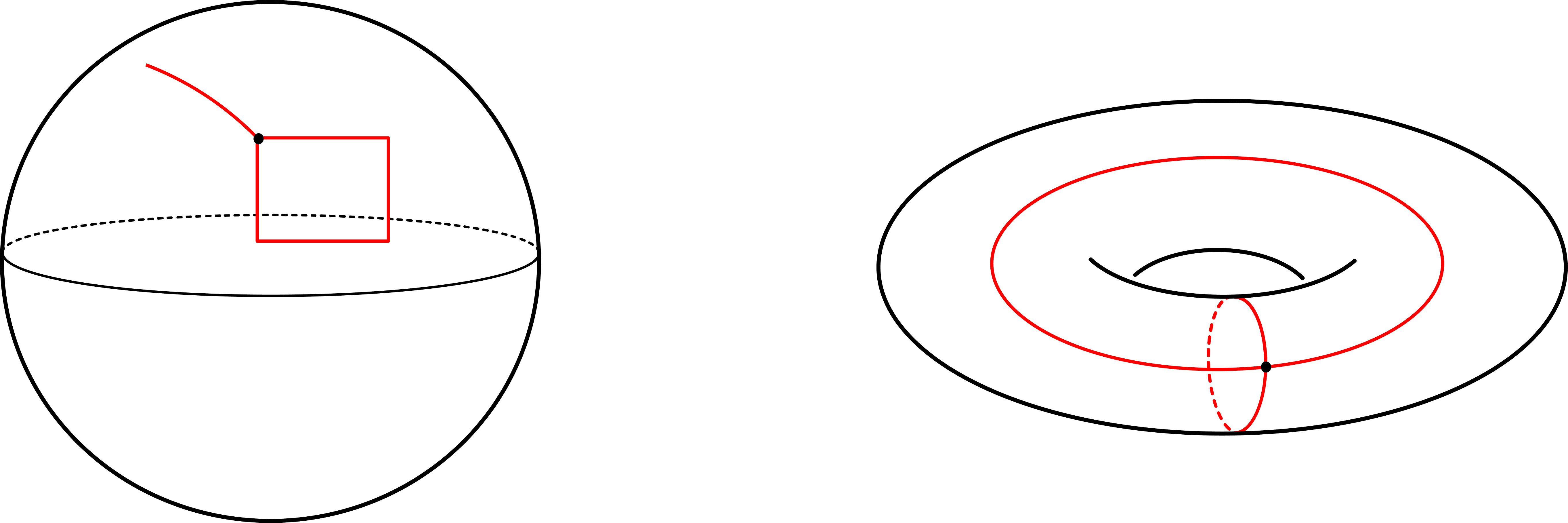
        	\caption{The map on the left is separable and the faces are not simple. For genus $0$, a map is non-separable if and only if all faces are simple. This is not true for higher genus. For example, the map on the right is non-separable, but its face is not simple. This type of non-simplicity in which removing the problematic vertex does not disconnect the map appears only for higher genus and is more complicated to deal with than the one in separable maps, which is the only type of non-simplicity present for planar maps.}
\end{center}
\end{figure}

In the present work, regarding the planar case, we consider an intermediate problem, \textit{i.e.} the enumeration of maps where only the boundaries are imposed to be simple. This more refined problem for genus $0$, and much more refined for higher genus, is interesting by itself. Actually, when there are several boundaries, we are led to distinguish maps in which each boundary is simple from an even more restrictive type of maps in which every vertex of a boundary $b$ belongs to at most two edges of the boundaries (whether $b$ or another one). The latter are called \emph{fully simple maps}. We call \emph{simple} the maps in which each boundary is simple, and \emph{ordinary} the maps in which no restriction is imposed on vertices on the boundary. Moreover, we find remarkable combinatorial and algebraic properties that also justify the relevance of this problem \textit{a posteriori}.

We discovered later\footnote{We thank Timothy Budd for bringing these references to our attention.} that Krikun \cite{Krikun} enumerated planar fully simple triangulations (which he called ``maps with holes''), using a combinatorial identity due to Tutte. Bernardi and Fusy recently recovered Krikun's formula and also provided an expression for the number of planar fully simple quadrangulations (which they call simply ``maps with boundaries'') with even boundary lengths, via a bijective procedure \cite{BernardiFusy}. The general enumeration problem of fully simple maps that we consider in this article regards maps of any genus with internal faces of arbitrary bounded degrees.

Some of our arguments will apply to more general models of maps, where faces are not necessarily homeomorphic to disks. For instance, maps carrying an $O(n)$ loop model \cite{GK}, an Ising model \cite{Kazakov86} or a Potts model \cite{BBG3}, can be described as maps having faces which are either disks or cylinders \cite{BEO}. We refer to \cite{Eformal} for more examples. The vocabulary we adopt is summarized in the following tables. Here, $(b_i)_{i = 1}^n$ are the boundary faces.

\begin{center}
\begin{figure}[h!]
\begin{tabular}{|c|c|}
\hline
\textit{boundary type} & \textit{description} \\
\hline
ordinary & no restriction \\
\hline simple & $b_i$ non self-intersecting \\
\hline
fully simple & $b_i$ disjoint from $\bigcup_{j\neq i} b_j$ \\
\hline
\end{tabular}
\caption{\label{Fig1}}
\end{figure}

\vspace{0.2cm}
\begin{figure}[h!]
\begin{tabular}{|c|c|c|}
\hline
\textit{type of maps} & \textit{topology of inner faces} & \textit{matrix model \eqref{mesmap}-\eqref{mesmapgeneral}} \\
\hline
usual & disks & $t_{d}$ \\
\hline
with loops \cite{BEO} & disks and cylinders & $t_{0;d_1}$ and $t_{0;d_1,d_2}$ \\
\hline
stuffed \cite{Bstuff} & arbitrary & all $t_{h;d_1,\ldots,d_k}$ \\
\hline
\end{tabular}
\caption{\label{Fig2}}
\end{figure}
\end{center}

\subsection{Disks and cylinders via combinatorics}

For planar maps with one boundary (disks) or two boundaries (cylinders), we give in Section~\ref{Section3}-\ref{Section4} a bijective algorithm which reconstructs ordinary maps from fully simple maps. This algorithm is not sensitive to the assumption -- included in the definition of usual maps -- that faces must be homeomorphic to disks. Therefore, it applies to all types of maps described in Figure~\ref{Fig2}.

We deduce two remarkable formulas for the corresponding generating series. Let $F_{\ell}$ (resp. $H_{\ell}$) be the generating series of ordinary (resp. fully simple) disks with perimeter $\ell$, and
$$
W(x) = \frac{1}{x} + \sum_{\ell \geq 1} \frac{F_{\ell}}{x^{\ell + 1}},\qquad X(w) = \frac{1}{w} + \sum_{\ell \geq 1} H_{\ell}w^{\ell - 1}\,.
$$
\begin{proposition}
\label{P10} For all types of maps in Figure~\ref{Fig2}, $X(W(x)) = x$.
\end{proposition}

Let $F_{\ell_1,\ell_2}$ (resp. $H_{\ell_1,\ell_2})$ be the generating series of ordinary (resp. fully simple) cylinders with perimeters $(\ell_1,\ell_2)$, and
$$
W_{2}^{[0]}(x_1,x_2) = \sum_{\ell_1,\ell_2 \geq 1} \frac{F_{\ell_1,\ell_2}}{x_1^{\ell_1 + 1}x_2^{\ell_2 + 1}},\qquad X_{2}^{[0]}(w_1,w_2) = \sum_{\ell_1,\ell_2 \geq 1} H_{\ell_1,\ell_2}\,w_{1}^{\ell_1 - 1}w_{2}^{\ell_2 - 1}\,.
$$
\begin{proposition}
\label{P20}
For all types of maps in Figure~\ref{Fig2}, if one sets $x_i = X(w_i)$ or equivalently $w_i = W(x_i)$, 
$$
\bigg(W_{2}^{[0]}(x_1,x_2) + \frac{1}{(x_1 - x_2)^2}\bigg)\dd x_1\dd x_2 = \bigg(X_{2}^{[0]}(w_1,w_2) + \frac{1}{(w_1 - w_2)^2}\bigg)\dd w_1\dd w_2.
$$
\end{proposition}
The identities of Propositions~\ref{P10}-\ref{P20} are equalities of formal series in $x_i \rightarrow \infty$ and $w_i \rightarrow 0$.

\subsection{Matrix model interpretation and consequences}

It is well-known that the generating series of ordinary maps with prescribed boundary perimeters $(\ell_i)_{i = 1}^n$ are computed as the moments $\langle {\rm Tr}\,M^{\ell_1} \cdots {\rm Tr}\,M^{\ell_n} \rangle$ in the formal hermitian matrix model
\beq
\label{mesmap} \dd\mu(M) = \dd M\,\exp[-N\,{\rm Tr}\,V(M)],\qquad V(x) = \frac{x^2}{2} - \sum_{d \geq 1} \frac{t_dx^d}{d}\,,
\eeq
where $t_d$ is the weight per $d$-gon, and the weight of a map of Euler characteristic $\chi$ is proportional to $N^{\chi}$.  Restricting to connected maps amounts to considering the cumulant expectation values $\kappa_n({\rm Tr}\,M^{\ell_1},\ldots,{\rm Tr}\,M^{\ell_n})$ instead of the moments. More generally, the measure
\beq
\label{mesmapgeneral} \dd\mu(M) = \dd M\,\exp\bigg(- N\,{\rm Tr}\,\frac{M^2}{2} + \sum_{h \geq 0} \sum_{k \geq 1} \sum_{d_1,\ldots,d_k \geq 1} \frac{N^{2 - 2h - k}}{k!}\,\frac{t_{h;d_1,\ldots,d_k}}{d_1\cdots d_k} {\rm Tr}\,M^{d_1} \cdots {\rm Tr}\,M^{d_k}\bigg)
\eeq
generates maps with loops or stuffed maps.

We show in Section~\ref{Matcomb} that the generating series of fully simple maps with prescribed boundary perimeters $(\ell_i)_{i = 1}^n$ in these models are computed as $\langle \mathcal{P}_{\gamma_1}(M) \cdots \mathcal{P}_{\gamma_n}(M)\rangle$, where $\gamma$ is a permutation of $\{1,\ldots,L\}$ with $n$ disjoint cycles $(\gamma_i)_{i = 1}^n$ of respective lengths $(\ell_i)_{i = 1}^n$, $L = \sum_{i = 1}^n \ell_i$, and $\mathcal{P}_{\gamma_i}(M) = \prod_{j} M_{j,\gamma_i(j)}$. This quantity does not depend on the permutation $\gamma$, but only the lengths $(\ell_i)_{i = 1}^n$, which are encoded into a partition $\lambda$, and we write $\langle \mathcal{P}^{(\ell_1)}(M) \cdots \mathcal{P}^{(\ell_n)}(M)\rangle=\langle\mathcal{P}_{\lambda}(M)\rangle$.  Again, the cumulants
$$
\kappa_{n}\big(\mathcal{P}_{\gamma_1}(M),\ldots,\mathcal{P}_{\gamma_n}(M)\big) = \kappa_{n}\big(\mathcal{P}^{(\ell_1)}(M), \ldots, \mathcal{P}^{(\ell_n)}(M)\big)
$$
generate only connected maps.

\subsubsection{Relation between ordinary and fully simple via Hurwitz theory}

The expression $\prod_{i} \mathcal{P}_{\gamma_i}(M)$ is a function of $M$ which is not invariant under $U_N$-conjugation. Yet, as the measure $\mu$ is unitary invariant, its expectation value must be expressible in terms of $U_N$-invariant observables, \textit{i.e.} as a linear combination of $\langle \prod_{i} {\rm Tr}\,M^{m_i} \rangle$. In other words, we can express the fully simple generating series in terms of the ordinary generating series. The precise formula is derived via Weingarten calculus.
\begin{theorem}\label{EHH}
If $\mu$ is a unitarily invariant measure on $\mathcal{H}_N$, in particular for the measures \eqref{mesmap}-\eqref{mesmapgeneral} generating any type of map in Figure~\ref{Fig2},
\bea
\frac{\langle \mathcal{P}_{\lambda}(M) \rangle}{|{\rm Aut}\,\lambda|} & = & \sum_{\mu \vdash |\lambda|} N^{-|\mu|} \bigg(\sum_{k \geq 0} (-N)^{-k} [H_{k}]_{\lambda,\mu}\bigg) \Big\langle \prod_{i = 1}^{\ell(\mu)} {\rm Tr}\,M^{\mu_i} \Big\rangle\,, \\
\label{subfree}\frac{\Big\langle \prod_{i = 1}^{\ell(\mu)} {\rm Tr}\,M^{\mu_i} \Big\rangle}{|{\rm Aut}\,\mu|} & = & \sum_{\lambda \vdash |\mu|} N^{|\lambda|} \bigg( \sum_{k \geq 0} N^{-k}\,[E_{k}]_{\mu,\lambda}\bigg)\, \langle \mathcal{P}_{\lambda}(M) \rangle\,.
\eea
\end{theorem}
The notation $\mu \vdash L$ means that $\mu$ is a partition of $L\in\mathbb{Z}_{\geq 0}$, and $|{\rm Aut}\,\lambda| = L!/|C_{\lambda}|$, where $|C_{\lambda}|$ is the number of permutations in the conjugacy class which corresponds to the partition $\lambda$, that is the number of permutations of cycle type $\lambda$. The transition kernels $[H_{k}]_{\lambda,\mu}$ and $[E_{k}]_{\lambda,\mu}$ are universal numbers expressed via character theory of the symmetric group. Invoking the general relations \cite{HarnadGuay,NovakJ} between the enumeration of branched covers of $\mathbb{P}^1$, paths in the Cayley graph of the symmetric groups, and representation theory, we identify
\begin{itemize}
\item[$\bullet$] $[H_{k}]_{\lambda,\mu}$ with the double, weakly monotone Hurwitz numbers;
\item[$\bullet$] $[E_k]_{\lambda,\mu}$ with the double, strictly monotone Hurwitz numbers.
\end{itemize}
In terms of branched covers of $\mathbb{P}^1$, $\lambda$ and $\mu$ encode the ramification profiles over $0$ and $\infty$, and $k$ is the number of simple ramifications. Formula \eqref{subfree} is appealing as it is subtraction-free, and suggests the existence of a bijection describing ordinary maps as gluing of a fully simple map ``along'' a strictly monotone branched cover. We postpone such a bijective proof of \eqref{subfree} to a future work.

\subsubsection{Combinatorial interpretation of the matrix model with external field}

As a by-product, we show that the partition function of the formal hermitian matrix model with external field $A \in \mathcal{H}_{N}$
$$
\check{Z}(A) = \int_{\mathcal{H}_{N}} \dd\mu(M)\,\exp[N{\rm Tr}(MA)]
$$
is a generating series of fully simple maps in the following sense
\begin{proposition} If $\mu$ is a unitary invariant measure on $\mathcal{H}_{N}$ -- in particular for all types of maps in Figure~\ref{Fig2},
$$
\frac{\check{Z}(A)}{\check{Z}(0)} = \sum_{\lambda} \frac{|C_{\lambda}|}{|\lambda|!}\,N^{|\lambda|}\langle \mathcal{P}_{\lambda}(M) \rangle \prod_{i = 1}^{\ell(\lambda)} {\rm Tr}\,A^{\lambda_i}\,.
$$ 
\end{proposition}

\subsubsection{Application: an ELSV-like formula}

If we have a model in which the generating series of fully simple maps are completely known, \eqref{subfree} can be used to compute a certain set of monotone Hurwitz numbers in terms of generating series of maps. This is the case for the Gaussian Unitary Ensemble, \textit{i.e.} $t_d = 0$ for all $d$ in \eqref{mesmap}. As the matrix entries are independent
$$
\langle \mathcal{P}_{\lambda}(M) \rangle_{{\rm GUE}} = \prod_{i = 1}^{\ell(\lambda)} \frac{\delta_{\lambda_i,2}}{N}\,.
$$ 
Combinatorially, this formula is also straightforward: as the maps generated by the GUE have no internal faces, the only connected fully simple map is the disk of perimeter 2. Dubrovin et al. \cite{DiYang} recently proved a formula relating the GUE moments with all $\ell_i$ even, to cubic Hodge integrals. Combining their result with our \eqref{subfree} specialized to the GUE, we deduce in Section~\ref{Section11} an ELSV-like formula for what could be called $2$-orbifold strictly monotone Hurwitz numbers.

\begin{proposition}
Let $[E^{\circ}_{g}]_{\lambda,\mu}$ be the number of connected, strictly monotone branched covers from a curve of genus $g$ to $\mathbb{P}^1$ with ramifications $\lambda$ and $\mu$ above $0$ and $\infty$, and simple ramifications otherwise. If $m_1,\ldots,m_n \geq 0$, set $|m| = \sum_{i} m_i$. We have
\bea
&& |{\rm Aut}\,(2m_1,\ldots,2m_n)|\,[E^{\circ}_{g}]_{(2m_1,\ldots,2m_n),(2,\ldots,2)} \nonumber \\
& = & 2^{g} \int_{\overline{\mathcal{M}}_{g,n}} [\Delta] \cap \Lambda(-1)\Lambda(-1)\Lambda(\tfrac{1}{2})\exp\Big(-\sum_{j \geq 1} \frac{\kappa_{j}}{j}\Big) \prod_{i = 1}^n \frac{m_i\,{2m_i \choose m_i}}{1 - m_i\psi_i}\,, \nonumber
\eea
where 
$$
[\Delta] = \sum_{h = 0}^{g} \frac{[\Delta_h]}{2^{3h}(2h)!},
$$
and $[\Delta_h]$ is the class of $\overline{\mathcal{M}}_{g - h,n + 2h}$ included in $\overline{\mathcal{M}}_{g,n}$ by identifying pairwise the $2h$ last punctures.
\end{proposition}

\subsection{Topological recursion interpretation}

\subsubsection{Review}

It was proved in \cite{E1MM,C05,EynardBook} for maps, and \cite{BEOn,BEO} for maps with loops, that the generating series of ordinary maps $W_{n}^{[g]}$ satisfies the topological recursion (hereafter, TR) formalized by Eynard and Orantin \cite{EORev}. This is a universal recursion in minus the Euler characteristic $2g - 2 + n$ (of a surface of genus $g$ with $n$ boundary components), which takes as input data
\begin{itemize}
\item[$\bullet$] a Riemann surface $\mathcal{C}$ realized as a branched cover $p\,:\,\mathcal{C} \rightarrow \mathbb{C}$;
\item[$\bullet$] an analytic function $\lambda$ on $\mathcal{C}$;
\item[$\bullet$] a bidifferential $B$ with a double pole on the diagonal in $\mathcal{C}^2$.
\end{itemize}
The outcome of TR is a sequence of multidifferentials $(\omega_{n}^{[g]})_{g,n}$ on $\mathcal{C}^n$, including for $n = 0$ a sequence of scalars $\omega_{0}^{[g]} = \mathcal{F}_{g}$. We call them ``TR amplitudes''. In the context of maps, $\mathcal{C}$ is the curve on which the generating series of disks can be maximally analytically continued with respect to its parameter $x$ coupled to the boundary perimeter, and $\mathcal{C}$ has a distinguished point $[\infty]$ corresponding to $x \rightarrow \infty$.
\begin{theorem}\label{TRRRR} The TR amplitudes for the initial data
\beq
\label{inidataordmap} \left\{\begin{array}{lll} p = x \\ \lambda = w = W_{1}^{[0]}(x) \\ B(z_1,z_2) = \Big(W_{2}^{[0]}(x(z_1),x(z_2)) + \frac{1}{(x(z_1) - x(z_2))^2}\Big)\dd x(z_1)\dd x(z_2) \end{array}\right.
\eeq
compute the generating series of usual maps or maps with loops, through
\beq
\label{WngTR}W_{n}^{[g]}(x(z_1),\ldots,x(z_n)) = \frac{\omega_{g,n}(z_1,\ldots,z_n)}{\dd x(z_1)\cdots \dd x(z_n)},\qquad 2g - 2 + n > 0.
\eeq
Here $z_i$ is a generic name for points in $\mathcal{C}$, and \eqref{WngTR} means the equality of Laurent expansions near $z_i \rightarrow [\infty]$.
\end{theorem}

\subsubsection{Symplectic invariance}

The most remarkable, and still mysterious property of the topological recursion is its symplectic invariance:
\begin{conjecture}
\label{thxy} Assume $\mathcal{C}$ is compact and $\lambda$ and $p$ are meromorphic. Let $\check{\omega}_{n}^{[g]}$ be the TR amplitudes for the initial data in which the role of $\lambda$ and $p$ is exchanged and $B$ remains the same. We have
$$
\forall g \geq 2,\qquad \check{\mathcal{F}}^{[g]} = \mathcal{F}^{[g]}
$$
and this formula also holds up to an explicit corrective term for $g = 0,1$. The TR amplitudes with $n \geq 1$ for the two initial data differ.
\end{conjecture}
This statement first appeared in \cite{EO2MM} with a tentative proof but some additive constants in the definition of $\mathcal{F}^{[g]}$ were overlooked. A corrigendum was proposed in \cite{EOxy} but the completeness of the proof and validity of the description of these corrective terms is still under scrutiny. This property is called symplectic invariance because the change $(p,\lambda) \rightarrow (-\lambda,p)$ preserves the symplectic form $\dd\lambda \wedge \dd p$ on $\mathbb{C}^2$. The invariance is also believed to hold with weaker assumptions. Symplectic invariance has been checked for many examples of initial data. In topological strings on toric Calabi-Yau threefolds, symplectic invariance is expected on physical grounds as it corresponds to the framing independence of the closed sector, albeit involving curves given by a polynomial relation between $e^{p}$ and $e^{\lambda}$. However, this is one of the few instances where the reason behind symplectic invariance is understood.

Propositions~\ref{P10}-\ref{P20} tell us that swapping $\lambda$ and $p$ in the initial data \eqref{inidataordmap} amounts to replacing the generating series of ordinary disks and cylinders with their fully simple version. We see it as the planar tip of an iceberg.

\begin{conjecture}
\label{MainConj} For usual maps or maps with loops, let $\check{\omega}_{n}^{[g]}$ be the TR amplitudes for the initial data \eqref{inidataordmap} after the exchange of $x$ and $w$. We have
\beq
\label{XngTR}X_{n}^{[g]}(w(z_1),\ldots,w(z_n)) = \frac{\check{\omega}_{n}^{[g]}(z_1,\ldots,z_n)}{\dd w(z_1)\cdots \dd w(z_n)},\qquad 2g - 2 + n > 0.
\eeq
This is an equality of formal Laurent series when $z_i \rightarrow [\infty]$.
\end{conjecture}
The validity of this conjecture would give a combinatorial interpretation to the symplectic invariance. Gaining understanding of this deep feature constituted an important motivation for us to study fully simple maps. Observe that if $n=0$, that is we consider maps without boundaries, we have in a very natural way that $X_{0}^{[g]} = W_{0}^{[g]}$, as the condition for maps to be fully simple affects only the boundaries.  

Using non-combinatorial techniques, we give in Section~\ref{Section9} the path to a possible proof of this conjecture for usual maps. We manage to reduce the problem to a technical condition regarding a milder version of the symplectic invariance for the family of spectral curves of the so-called $1$-hermitian matrix model with external field.

Our argument is however not combinatorial, and relies on the study of the formal $1$-hermitian matrix model with external field. It is still desirable to prove Conjecture~\ref{MainConj} in a combinatorial way (and thus unconditionally), as it would give an independent proof of symplectic invariance for the initial data related to maps -- \textit{i.e.} a large class of curves of genus $0$ --, and it may be naturally generalizable for all types of maps in the Figure~\ref{Fig2}.

Apart from the general proof for disks and cylinders, and the ideas towards the full proof for usual maps, we gathered some combinatorial evidence supporting our conjecture in Section~\ref{Quadrangulations}. In fact, there is no \textit{a priori} reason for the coefficients of expansion of $\check{\omega}_{n}^{[g]}$ to be positive integers. Besides, for the same given perimeters, there should be less fully simple maps than ordinary maps. For the initial data corresponding to quadrangulations, we have checked in Sections~\ref{ToriSection} and \ref{PairPantsSection} that, for the topologies $(1,1)$ and $(0,3)$, for the topologies $(1,1)$ and $(0,3)$, positivity and the expected inequalities hold for the coefficients of~$\check{\omega}_{1,1}$ and $\check{\omega}_{0,3}$ obtained after fixing the number of internal quadrangles, for boundaries up to length~$14$.

For the pair of pants case (topology $(0,3)$) the evidence is much stronger. In 2017, O.~Bernardi and \'{E}.~Fusy gave a formula for the number of fully simple planar quadrangulations with boundaries of prescribed even lengths in \cite{BernardiFusy}. We computed the outcome of their formula and they perfectly match our conjectural numbers for the cases of even lengths. Their formula also agrees with our enumeration for disks and cylinders. From our results for cylinders and our conjectural numbers for pairs of pants, one can observe that an analogous formula seems to be true also in presence of some odd boundaries.

\begin{conjecture}
Let $Q$ be the number of internal quadrangles and $k_1,\ldots,k_n$ positive integers with $L=\sum_{i=1}^n k_i$ the total boundary length. If $v=2Q-L-n+2 \geq 0$, in which case it counts the number of internal vertices, we have that the number of planar fully simple quadrangulations is given by
\beq
\alpha(Q,L,n)\prod_{i=1}^n \varepsilon(k_i),
\eeq
where $\alpha(Q,L,n)\coloneqq\frac{3^{Q-\frac{L}{2}}(e-1)!}{v!(L+Q)!}$, with $e=\frac{L}{2}+2Q$ the total number of edges, and
$$
\varepsilon(k)\coloneqq \begin{cases}
\frac{(3l)!}{l!(2l-1)!}, & \text{ if } k=2l, \\
\sqrt{3}\frac{(3l+1)!}{l!(2l)!}, & \text{ if } k=2l+1. \\
\end{cases}
$$
\end{conjecture}
This formula reproduces the theorem of O.~Bernardi and \'{E}.~Fusy for even boundary lengths and generalizes it to include the presence of odd boundary lengths.

If Conjecture~\ref{MainConj} is true, it would solve theoretically the problem of enumeration of fully simple maps in full generality. Moreover, the algorithm of TR allows to solve explicitly the first cases of the iteration. So our conjecture would produce another proof of the formula of \cite{BernardiFusy} for cylinders and pairs of pants with even lengths, would allow to prove the cases in presence of odd lengths and would produce the first explicit formulas for numbers of fully simple maps of positive genus. We give explicit formulas for ordinary and conjecturally fully simple maps of genus $1$ with~$1$ boundary in Section~\ref{ToriSection}.

Furthermore, the fact that the conjecture produces the right numbers for pairs of pants seems to indicate that our technical condition should hold in general; otherwise, we believe the presence of non-zero correction terms in our technical condition should be manifested from the beginning of the recursion. Finally, even if the conjecture was not true and there were some non-zero correction terms modifying our technical condition, the data suggests there is a combinatorial problem behind since we obtain positive integers, so the correction terms may give rise to a simpler combinatorial problem complementing the number of fully simple maps.

\subsection{Application to free probability} 

The results of Theorem~\ref{P10} for simple disks and Theorem~\ref{P20} for fully simple cylinders coincide with the formulas found for generating series of the first and second order free cumulants in \cite{Secondorderfreeness}. We proved these formulas via combinatorics of maps, independently of \cite{Secondorderfreeness}, and also explained that they are natural in light of the topological recursion. The restriction of Conjecture~\ref{MainConj} to genus $0$ would give a recursive algorithm to compute the higher order free cumulants of the matrix $M$ sampled from the large $N$ limit of the measure \eqref{mesmap}. This is interesting as the relation at the level of generating series between $n$-th order free cumulants and $n$-th correlation moments, called $R$-transform machinery, is not otherwise known for $n \geq 3$ as of writing, thus imposing to work with their involved combinatorial definition via so-called partitioned permutations.

We explain in Section~\ref{SFree} that a possible generalization of Conjecture~\ref{MainConj} to stuffed maps -- for which generating series of ordinary maps are governed by a generalization of TR -- would shed light on computation of generating series of higher order cumulants in the full generality of \cite{Secondorderfreeness}. Given the universality of the TR structure, one may also wonder if a universal theory of approximate higher order free cumulants can be formulated taking into account the higher genus amplitudes.

\part{Combinatorics}

\section{Objects of study}\label{objects}

\subsection{Maps}

Maps can intuitively be thought as graphs drawn on surfaces or discrete surfaces obtained from gluing polygons, and they receive different names in the literature: ribbon graphs, discrete surfaces, fat graphs...

\begin{definition}\label{usual_map}
An \emph{embedded graph} of genus $g$ is a connected graph $\Gamma$ embedded into a connected orientable surface $X$ of genus $g$ such that $X \setminus \Gamma$ is a disjoint union of connected components, called ${\it faces}$, each of them homeomorphic to an open disk. 

Every edge belongs to two faces (which may be the same) and we call {\it length} of a face the number of edges belonging to it.

We say that an embedded graph has $n$ {\it boundaries}, when it has $n$ marked faces, labeled $1,\ldots,n$, which we require to contain a marked edge, called {\it root}, represented by an arrow following the convention that the marked face sits on the left side of the root. A face which is not marked receives the name of {\it inner face}.

Two embedded graphs $\Gamma_i\subset X_i$, $i=1,2$, are {\it isomorphic} if there exists an orientation preserving homeomorphism $\varphi: X_1 \rightarrow X_2$ such that  $\left.\varphi\right|_{\Gamma_1}$ is a graph isomorphism between $\Gamma_1$ and $\Gamma_2$, and the restriction of $\varphi$ to the marked edges is the identity.

A \emph{map} is an isomorphism class of embedded graphs.
\end{definition}

Note that we include the connectedness condition in the definition of map. We will also work with disjoint unions of maps and we will specify we are dealing with a non-connected map whenever it is necessary.
Observe that $\sum_{f\in F} {\rm length}(f) = 2|E|$, where the sum is taken over the set of faces $F$ of the map and $E$ denotes the set of edges.

We call {\it planar} a map of genus $0$. We call {\it map of topology} $(g,n)$ a map of genus $g$ with $n$ boundaries. For the cases $(0,1)$ and $(0,2)$ we use the special names: {\it disks} and {\it cylinders}, respectively.

\subsubsection{Permutational model}

The embedding of the graph into an oriented surface provides the extra information of a cyclic order of the edges incident to a vertex. More precisely, we consider half-edges, each of them incident to exactly one vertex. Let $H$ be the set of half-edges and observe that $|H|=2|E|$. We label the half-edges by $1,\ldots, 2|E|$ in an arbitrary way.

\begin{figure}[h!]
 \begin{center}
        \def\svgwidth{\columnwidth}
        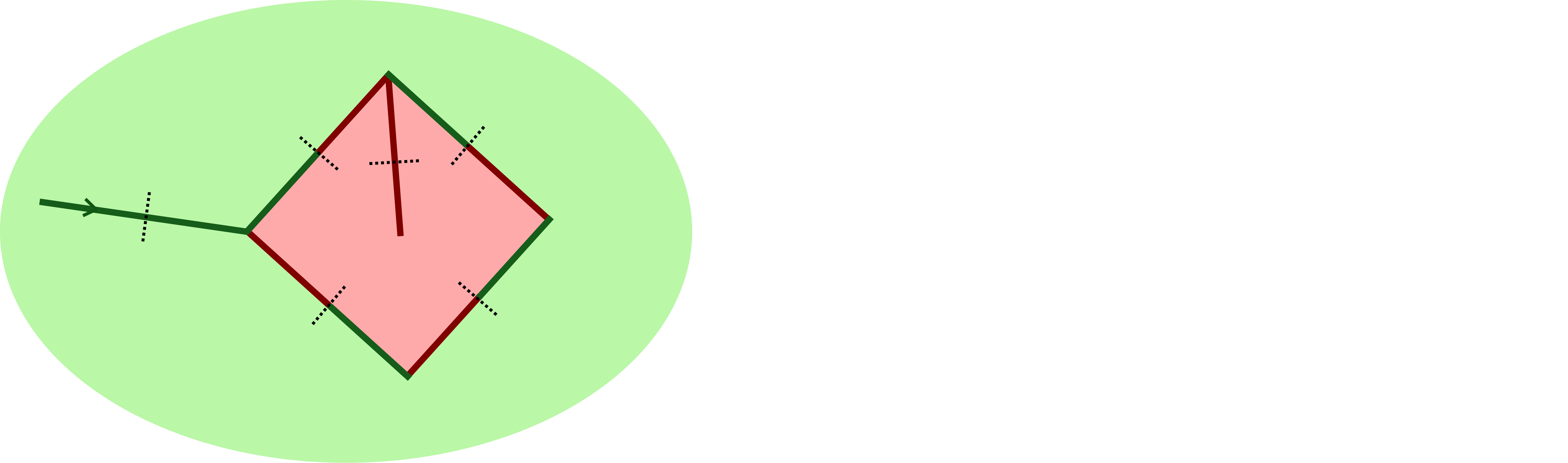
        	\caption{Two different ways of representing half-edges: in both cases by numbered segments, but on the left the two half-edges forming an edge are drawn consecutively, while on the right two consecutive half-edges belong to the same face and two half-edges forming an edge are drawn parallel. When maps are depicted as on the right, they are usually called ribbon graphs. We will sometimes use this representation because it can be a bit clearer, but will usually use the simpler representation on the left. On the left, the half-edges incident to a vertex are clearly the ones touching the vertex in the drawing; on the right, with our conventions, the half-edges incident to a vertex are the ones on the left viewed from the vertex in question.}\label{permutations}
\end{center}
\end{figure}

Every map with labeled half-edges can be encoded by a so-called {\it combinatorial map}, which consists of a pair of permutations $(\sigma,\alpha)$ acting on $H$ such that all cycles of $\alpha$ have length $2$. Given a half-edge $h\in H$, let $\sigma(h)$ be the the half-edge after $h$ when turning around its vertex according to the orientation fixed for the underlying surface (by convention counterclockwise). On the other hand, let $\alpha(h)$ be the other half-edge of the edge to which $h$ belongs. The information that $\alpha$ provides is encoded in the graph structure of the map, while $\sigma$ characterizes the additional data of a map given by the embedding of the graph in the surface.
\begin{itemize}
\item[$\bullet$] A cycle of $\sigma$ corresponds to a vertex in the map.
\item[$\bullet$] Every cycle of $\alpha$ corresponds to an edge of the map.
\item[$\bullet$] The faces may be represented by cycles of a permutation, called $\varphi$, of $H$.
\end{itemize}
Observe that with the convention that the face orientation is also counterclockwise, we obtain
\beq
\label{abc1}\sigma\circ\alpha\circ\varphi = {\rm id},
\eeq
and hence $\varphi$ can be determined by $\sigma^{-1}\alpha^{-1}$.

Rooting an edge in a face amounts to marking the associated label in the corresponding cycle of $\varphi$. Such cycles containing a root will be ordered and correspond to boundaries of the map.

\begin{example}{\rm
In Figure \ref{permutations}, we have
$$
\begin{array}{rcl}
\sigma & = & (5\ 11)(4\ 12)(3\  9\  7)(2\  6\  10), \\
\alpha & = & (1\  6)(2\  9)(7\  8)(3\  12)(4\  11)(5\  10), \\
\varphi & = & \sigma^{-1}\circ\alpha^{-1} = (1\  2\  3\  4\  5\  6)(7\  8\  9\  10\  11\  12), 
\end{array}
$$
where the root is the half-edge labeled $1$.
}
\end{example}

The lengths of the cycles of $\sigma$ and $\varphi$ correspond to the degrees of vertices and faces, respectively. The Euler characteristic is given by
$$
\chi(\sigma,\alpha)= |\mathcal{C}(\sigma)| - |\mathcal{C}(\alpha)| + |\mathcal{C}(\varphi)| - n,
$$
where $\mathcal{C}(\cdot)$ denotes the set of cycles of a permutation and $n$ is the number of cycles of $\varphi$ containing a root.

The group $G = \langle \sigma,\alpha \rangle$ is called the {\it cartographic group}. Its orbits on the set of half-edges determine the connected components of the map. If the action of $G$ on $H$ is transitive, the map is connected, and its genus $g$ is given by the formula:
$$
2 - 2g - n= \chi(\sigma,\alpha),
$$
where $n$ is the number of boundaries. If all orbits contain a root, the map is called {\it $\partial$-connected}.

\subsubsection{Automorphisms}\label{aut}
Let us consider the decomposition $H= H^{u}\sqcup H^{\partial}$, where $H^{u}$ is the set of half-edges belonging to unmarked faces and $H^{\partial}$ is the set of half-edges belonging to boundaries.

Observe that from a combinatorial map one can recover all the information of the original map. Therefore, there is a one-to-one correspondence between maps with labeled half-edges and combinatorial maps. There is a canonical way of labeling half-edges in boundaries: assigning the first label to the root, continuing by cyclic order of the boundary and taking into account that boundaries are ordered. However, we can label half-edges of unmarked faces in many different ways. To obtain a bijective correspondence with unlabeled maps, we have to identify configurations which differ by a relabeling of $H^u$, \textit{i.e.}~$(\sigma,\alpha)\sim (\gamma\sigma\gamma^{-1},\gamma\alpha\gamma^{-1})$ with $\gamma$ any permutation acting on $H$ such that $\left.\gamma\right|_{H^{\partial}}={\rm Id}_{H^{\partial}}$. We call such an equivalence class {\it unlabeled combinatorial map} and we denote it by $[(\sigma,\alpha)]$. Note that unlabeled combinatorial maps are in bijection with the unlabeled maps we defined at the beginning of this section.

\begin{definition}
Given a combinatorial map $(\sigma, \alpha)$ acting on $H$, we call $\gamma$ an {\it automorphism} if it is a permutation acting on $H$ such that $\left.\gamma\right|_{H^{\partial}}={\rm Id}_{H^{\partial}}$ and
$$
\sigma = \gamma\sigma\gamma^{-1}, \ \ \ \ \ \ \alpha=\gamma\alpha\gamma^{-1}.
$$
\end{definition}
Observe that for connected maps with $n\geq 1$ boundaries, the only automorphism is the identity. Note also that these special relabelings that commute with $\sigma$ and $\alpha$, and we call automorphisms, exist because of a symmetry of the (unlabeled) map. The symmetry factor $\left|{\rm Aut}(\sigma,\alpha)\right|$ of a map is its number of automorphisms. 

We denote $\text{Gl}(\sigma,\alpha)$ the number of elements in the class $[(\sigma,\alpha)]$ and $\text{Rel}(\sigma,\alpha)$ the total number of relabelings of $H^u$, which is $|H^u|!$, if we consider completely arbitrary labels for the half-edges. By the orbit-stabilizer theorem, we have $\text{Gl}(\sigma,\alpha)=\frac{\text{Rel}(\sigma,\alpha)}{\left|{\rm Aut}(\sigma,\alpha)\right|}$.

We refer the interested reader to the book \cite{LandoZvonkin} for further details on the topic of ordinary maps, although the conventions, notations and even concepts differ a little bit.

\subsection{Simple and fully simple maps}

\begin{definition}
We call a boundary $B$ ${\it simple}$ if no more than two edges belonging to $B$ are incident to a vertex. We say a map is {\it simple} if all boundaries are simple.
\end{definition}

\begin{figure}[h!]
 \begin{center}
        \def\svgwidth{\columnwidth}
        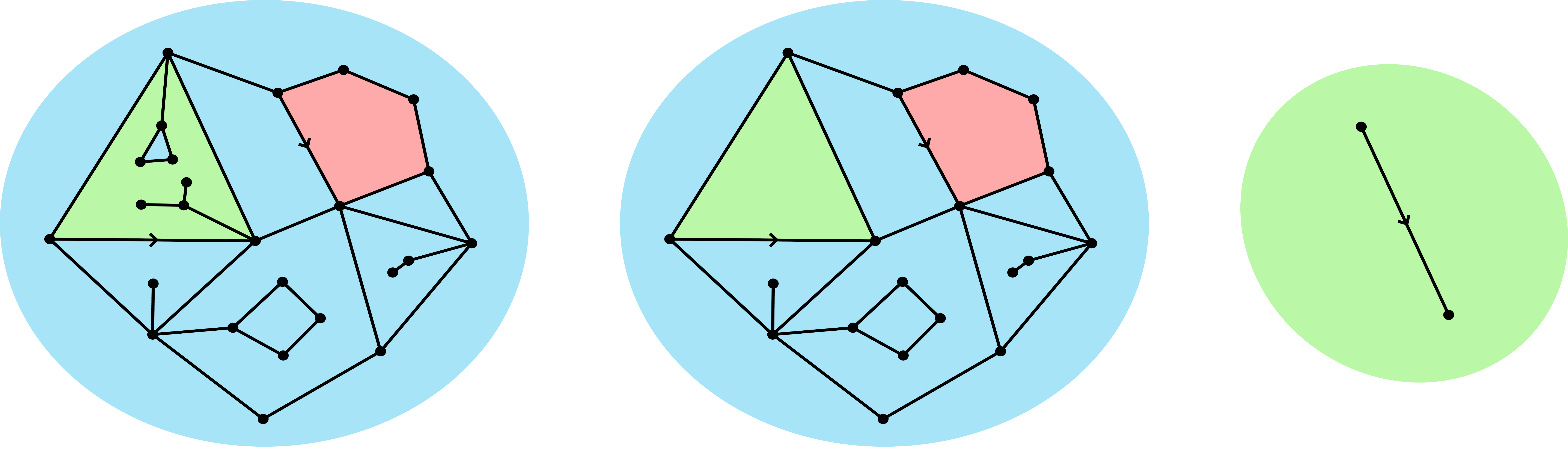
        	\caption{(a) is an ordinary cylinder where the green boundary is non-simple and (b) is a simple cylinder. (c) is the only simple map where two edges in the boundary are identified.}\label{degenerate simple map}
\end{center}
\end{figure}

To acquire an intuition about what this concept means, observe that the condition for a boundary to be simple is equivalent to not allowing edges of polygons corresponding to the boundary to be identified, except for the degenerate case of a boundary with only two edges which are identified, which is indeed considered to be a simple map (see Figure \ref{degenerate simple map}.(c)).

We will call {\it ordinary} the maps introduced in the previous section to emphasize that they are not necessarily simple.

\begin{definition}
We say a boundary $B$ is {\it fully simple} if no more than two edges belonging to any boundary are incident to a vertex of $B$.
We say a map is {\it fully simple} if all boundaries are fully simple.
\end{definition}

One can visualize the concept of a fully simple boundary as a simple boundary which moreover does not share any vertex with any other boundary.

\begin{figure}[h!]
 \begin{center}
        \def\svgwidth{\columnwidth}
        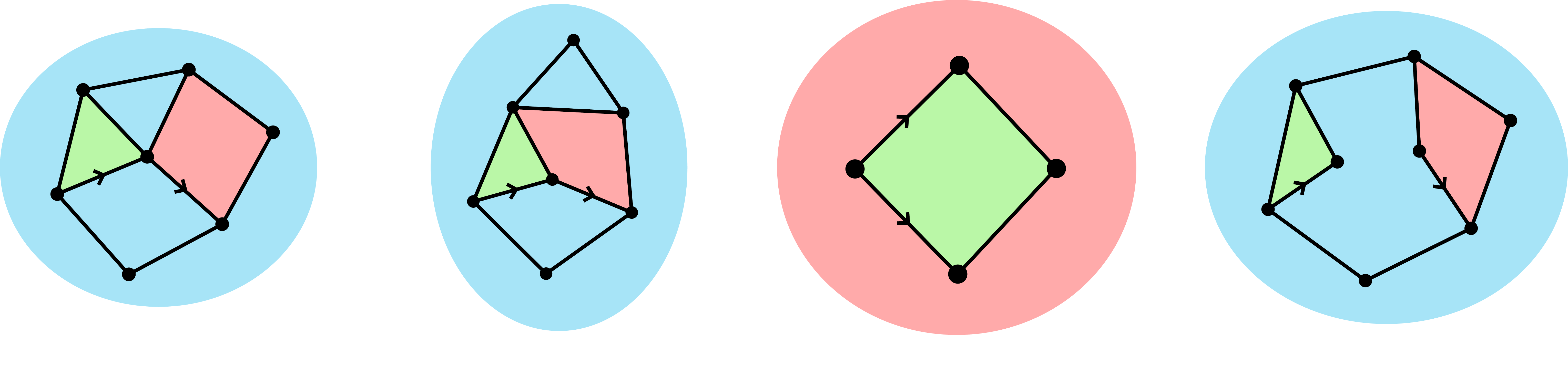
        	\caption{Four simple maps: (a), (b) and (c) are non-fully simple, and (d) is fully-simple. In (a) the two boundaries share a vertex, in (b) an edge, and in (c) they are completely glued to each other.}\label{totally glued}
\end{center}
\end{figure}

\subsection{Generating series}

We introduce now the notations and conventions for the generating series of our objects of study.

Let $\mathbb{M}_n^{[g]}$ be the set of maps of genus $g$ with $n$ boundaries and we denote $\mathbb{M}_n^{[g]}(v)$ the subset where we also fix the number of vertices to $v$.

Then we define the generating series of maps of genus $g$ and $n$ boundaries of respective lengths $l_1,\ldots, l_n$:
\beq
F_{l_1,\ldots,l_n}^{[g]}\coloneqq \sum_{\mathcal{M}\in\mathbb{M}_n^{[g]}}\ \  \frac{\prod_{j\geq 1}t_j^{n_j(\mathcal{M})}}{|{\rm Aut}\, \mathcal{M}|}\prod_{i=1}^n \delta_{l_i,\ell_i(\mathcal{M})}, \nonumber
\eeq
where $n_j(\mathcal{M})$ denotes the number of unmarked faces of length $j$, and $\ell_i(\mathcal{M})$ denotes the length of the $i$-th boundary of $\mathcal{M}$. For $n=0$, we denote $F^{[g]}$ the generating series of closed maps of genus $g$.

We have that 
$$
F^{[g]}, F_{l_1,\ldots,l_n}^{[g]} \in \mathbb{Q}\left[[t_1, t_2,\ldots\right]],
$$
that is the number of maps is finite after fixing the topology $(g,n)$ and the number of internal faces $n_j$ of every possible length $j\geq 1$.

\begin{remark}{\rm
Unmarked faces are often required to have length $\geq 3$ and $\leq d <\infty$ in the literature. In this typical setting, it can be easily checked that the set $\mathbb{M}_n^{[g]}(v)$ is always finite, without having to fix the numbers $n_j$ of internal faces. As a consequence, if we consider the generating series with an extra sum over the number of vertices $v\geq 1$ and a weight $u^v$, then it would belong to $\mathbb{Q}\left[t_3,\ldots, t_d\right][[u]]$. We do not need this further restriction, so from now on we consider the more general setting we introduced.
}
\end{remark}

We take the convention that $\mathbb{M}_1^{[0]}(1)$ contains only one map which consists of a single vertex and no edges; it is the map of genus $0$ with $1$ boundary of length $0$, that is $F_0^{[0]}=1$. Apart from this degenerate case, we always consider that boundaries have length $\geq 1$.

Summing over all possible lengths, we define the generating series of maps of genus $g$ and $n$ boundaries as follows:
$$
W_n^{[g]}(x_1,\ldots, x_n)\coloneqq \sum_{l_1,\ldots,l_n \geq 0} \frac{F^{[g]}_{l_1,\ldots,l_n}}{x_1^{1+l_1}\cdots x_n^{1+l_n}}.
$$
We have that $W_n^{[g]}(x_1,\ldots, x_n)\in \mathbb{Q}\left[\frac{1}{x_1},\ldots, \frac{1}{x_n}\right][[t_1, t_2,\ldots]]$ and observe that
$$
F_{l_1,\ldots,l_n}^{[g]} = (-1)^n \underset{x_1\rightarrow\infty}{{\rm Res}}\cdots \underset{x_n\rightarrow\infty}{{\rm Res}} x_1^{l_1}\cdots x_n^{l_n} W_n^{[g]}(x_1,\ldots,x_n) \dd x_1\cdots \dd x_n.
$$

We denote $H^{[g]}_{k_1,\ldots,k_n}$ the analogous generating series for fully simple maps of genus $g$ and $n$ boundaries of fixed lengths $k_1,\ldots,k_n$ and we introduce the following more convenient generating series for fully simple maps with boundaries of all possible lengths:
$$
X_{n}^{[g]}(w_1,\ldots,w_n) \coloneqq \sum_{k_1,\ldots,k_n \geq 0} H^{[g]}_{k_1,\ldots,k_n} w_1^{k_1 - 1} \ldots w_n^{k_n - 1}.
$$

Finally, we denote $G^{[g]}_{k_1,\ldots,k_m | l_1, \ldots, l_n}$ the generating series of maps with $m$ simple boundaries of lengths $k_1,\ldots, k_m$ and $n$ ordinary boundaries of lengths $l_1, \ldots, l_n$. We write
$$
Y_{m|n}^{[g]}(w_1,\ldots,w_m\mid x_1,\ldots,x_n) = \sum_{(\mathbf{k},\mathbf{l}) \in \mathbb{N}^{m}\times\mathbb{N}^n} \frac{w_1^{k_1 - 1}\cdots w_m^{k_m - 1}}{x_1^{l_1 + 1}\cdots x_n^{l_n + 1}}\,G_{\mathbf{k}|\mathbf{l}}^{[g]}.
$$

We use the following simplification for maps with only simple boundaries: $G^{[g]}_{k_1,\ldots,k_m}$ for $G^{[g]}_{k_1,\ldots,k_m |}$, and $Y_{m}^{[g]}$ for $Y_{m|}^{[g]}$.

Observe that for maps with only one boundary the concepts of simple and fully simple coincide. Therefore $G^{[g]}_{k}=H^{[g]}_{k}$ and $Y_{1}^{[g]}=X_{1}^{[g]}$.

For all the generating series introduced we allow to omit the information about the genus in the case of $g=0$. We also use the simplification of removing the information about the number of boundaries if $n=1$. In this way, $W$ and $X$ stand for $W_1^{[0]}$ and $X_1^{[0]}$.

\section{Simple disks from ordinary disks}\label{disks}
\label{Section3}

We can decompose an ordinary disk $\mathcal{M}$ with boundary of length $\ell >0$ into a simple disk $\mathcal{M}^s$ with boundary of length $1\leq \ell ' \leq \ell$ and ordinary disks of lengths $\ell_i < \ell$, using the following procedure:
\begin{algorithm}[from ordinary to simple]\label{alg1}{\rm
Set $\mathcal{M}^s:=\mathcal{M}$. We run over all edges of $\mathcal{M}$, starting at the root edge $e_1$ and following the cyclic order around the boundary. When we arrive at a vertex $v_i$ from an edge $e_i$, we create two vertices out of it: the first remains on the connected component containing $e_i$, while the second one glues together the remaining connected components, giving a map $\mathcal{M}_i$. We then update $\mathcal{M}^s$ to be the first connected component and proceed to the next edge on it. Every $\mathcal{M}_i$, for $i=1, \ldots, \ell '$, is an ordinary map consisting of
\begin{itemize}
\item a single vertex whenever $v_i$ was simple, or
\item a map with a boundary of positive length with the marked edge being the edge in $\mathcal{M}_i$ following $e_i$ in $\mathcal{M}$.
\end{itemize}}
\end{algorithm}

\begin{example} {\rm Consider a non-simple map with a boundary of length $11$ (non-simple vertices are circled). Applying the algorithm we obtain the simple map $\mathcal{M}^s$ of length $3$, and $3$ ordinary maps $\mathcal{M}_1$, $\mathcal{M}_2$, $\mathcal{M}_3$.
    \begin{center}
        \def\svgwidth{\columnwidth}
\begingroup%
  \makeatletter%
  \providecommand\color[2][]{%
    \errmessage{(Inkscape) Color is used for the text in Inkscape, but the package 'color.sty' is not loaded}%
    \renewcommand\color[2][]{}%
  }%
  \providecommand\transparent[1]{%
    \errmessage{(Inkscape) Transparency is used (non-zero) for the text in Inkscape, but the package 'transparent.sty' is not loaded}%
    \renewcommand\transparent[1]{}%
  }%
  \providecommand\rotatebox[2]{#2}%
  \ifx\svgwidth\undefined%
    \setlength{\unitlength}{4040.3394043bp}%
    \ifx\svgscale\undefined%
      \relax%
    \else%
      \setlength{\unitlength}{\unitlength * \real{\svgscale}}%
    \fi%
  \else%
    \setlength{\unitlength}{\svgwidth}%
  \fi%
  \global\let\svgwidth\undefined%
  \global\let\svgscale\undefined%
  \makeatother%
  \begin{picture}(1,0.19554978)%
    \put(0,0){\includegraphics[width=\unitlength,page=1]{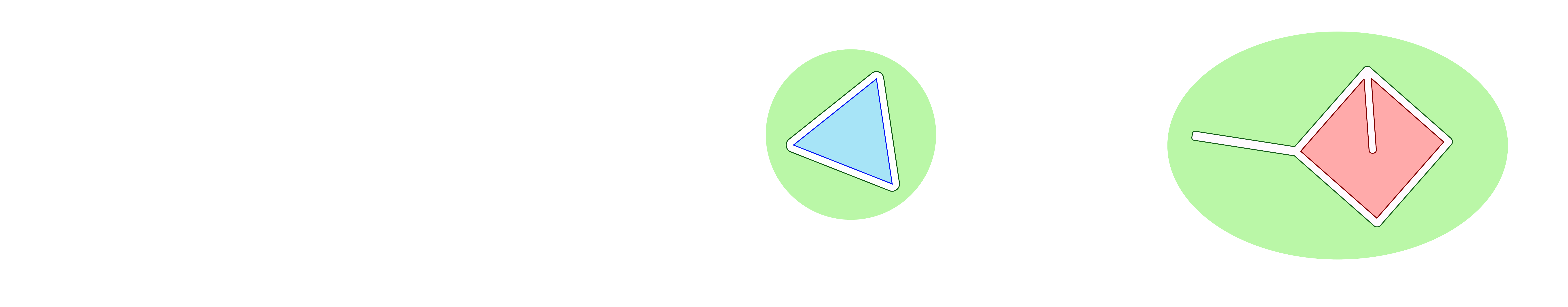}}%
    \put(0.52434125,0.19502656){\color[rgb]{0,0,0}\makebox(0,0)[lt]{\begin{minipage}{0.12640847\unitlength}\raggedright \scalebox{0.75}{$\ell '\,=\,3$} \end{minipage}}}%
    \put(0.76713341,0.19742672){\color[rgb]{0,0,0}\makebox(0,0)[lt]{\begin{minipage}{0.24087363\unitlength}\raggedright \scalebox{0.75}{$\ell_1+\ell_2+\ell_3 \,= \,2+0+6\,=\,8$}\end{minipage}}}%
    \put(0.74619734,0.02264315){\color[rgb]{0,0,0}\makebox(0,0)[lb]{\smash{}}}%
    \put(0,0){\includegraphics[width=\unitlength,page=2]{AlgDisks.pdf}}%
    \put(0.53122614,0.00905833){\color[rgb]{0,0,0}\makebox(0,0)[lt]{\begin{minipage}{0.07600509\unitlength}\raggedright $\mathcal{M}^s$\end{minipage}}}%
    \put(0.77322024,0.00976241){\color[rgb]{0,0,0}\makebox(0,0)[lt]{\begin{minipage}{0.18801261\unitlength}\raggedright $\bigsqcup_{i=1}^{\ell '} \mathcal{M}_i$ \end{minipage}}}%
    \put(0,0){\includegraphics[width=\unitlength,page=3]{AlgDisks.pdf}}%
    \put(0.56697728,0.16744481){\color[rgb]{0,0,0}\makebox(0,0)[lt]{\begin{minipage}{0.09959554\unitlength}\raggedright $\scalebox{0.65}{$2$}$\end{minipage}}}%
    \put(0.57859012,0.06983829){\color[rgb]{0,0,0}\makebox(0,0)[lt]{\begin{minipage}{0.11136998\unitlength}\raggedright $\scalebox{0.65}{$3$}$\end{minipage}}}%
    \put(0,0){\includegraphics[width=\unitlength,page=4]{AlgDisks.pdf}}%
    \put(0.68090853,0.13561759){\color[rgb]{0,0,0}\makebox(0,0)[lt]{\begin{minipage}{0.10475285\unitlength}\raggedright $\scalebox{0.65}{$1$}$\end{minipage}}}%
    \put(0.76379851,0.13223442){\color[rgb]{0,0,0}\makebox(0,0)[lt]{\begin{minipage}{0.09600645\unitlength}\raggedright $\scalebox{0.65}{$3$}$\end{minipage}}}%
    \put(0.68779342,0.0738466){\color[rgb]{0,0,0}\makebox(0,0)[lt]{\begin{minipage}{0.09280621\unitlength}\raggedright \end{minipage}}}%
    \put(0,0){\includegraphics[width=\unitlength,page=5]{AlgDisks.pdf}}%
    \put(0.4841838,0.09944829){\color[rgb]{0,0,0}\makebox(0,0)[lt]{\begin{minipage}{0.09035188\unitlength}\raggedright $\scalebox{0.65}{$1$}$\end{minipage}}}%
    \put(0,0){\includegraphics[width=\unitlength,page=6]{AlgDisks.pdf}}%
    \put(0.72436269,0.19113486){\color[rgb]{0,0,0}\makebox(0,0)[lt]{\begin{minipage}{0.12566904\unitlength}\raggedright $\scalebox{0.65}{$2$}$\end{minipage}}}%
    \put(0.77339915,-0.09016442){\color[rgb]{0,0,0}\makebox(0,0)[lb]{\smash{}}}%
    \put(0.41503318,0.12199346){\color[rgb]{0,0,0}\makebox(0,0)[lt]{\begin{minipage}{0.05374372\unitlength}\raggedright \end{minipage}}}%
    \put(0.42465047,0.11294188){\color[rgb]{0,0,0}\makebox(0,0)[lt]{\begin{minipage}{0.04469215\unitlength}\raggedright $\leadsto$\end{minipage}}}%
    \put(0,0){\includegraphics[width=\unitlength,page=7]{AlgDisks.pdf}}%
  \end{picture}%
\endgroup%

    \end{center}
    \vspace{0.3cm}
The maps should be regarded as drawn on the sphere and the outer face is in all cases the boundary.}
\end{example}

Using the decomposition given by the algorithm, we find that $X$ and $W$ are reciprocal functions:
\begin{proposition}\label{0,1}
\beq
\label{relmaster1} x = X(W(x)).
\eeq
\end{proposition}
\noindent \textbf{Proof.}
Since the algorithm establishes a bijection, we find that
\beq
\forall \ell \geq 1,\qquad F_{\ell} = \sum_{\ell' = 1}^{\ell} H_{\ell'} \sum_{\substack{\ell_1,\ldots,\ell_{\ell'} \geq 0 \\ \sum_i (\ell_i + 1) = \ell}} \prod_{i = 1}^{\ell'} F_{\ell_i},
\eeq
which implies, at the level of resolvents:
\begin{align*}
W(x) & = \sum_{\ell \geq 0} \frac{F_{\ell}}{x^{\ell+1}} = \frac{F_0}{x} + \sum_{\ell\geq 1} \frac{F_\ell}{x^{\ell+1}} = \frac{H_0}{x} + \sum_{\ell \geq 1} \frac{1}{x^{\ell+1}} \sum_{\ell' = 1}^{\ell} H_{\ell'} \sum_{\substack{\ell_1,\ldots,\ell_{\ell'} \geq 0 \\ \sum_i (\ell_i + 1) = \ell}} \prod_{i = 1}^{\ell'} F_{\ell_i}  \\
 & = \frac{1}{x}\sum_{\ell' \geq 0} H_{\ell'}(W(x))^{\ell'} = \frac{W(x)}{x} X(W(x)). 
\end{align*}
\hfill $\Box$

\section{Cylinders}\label{cyl}
\label{Section4}
\subsection{Replacing an ordinary boundary by a simple boundary}
Let us consider a planar map $\mathcal{M}$ with one ordinary boundary of length $\ell$, and one simple boundary of length $k$. We apply the procedure described in Algorithm \ref{alg1} to the ordinary boundary. We have to distinguish two cases depending on the nature of $\mathcal{M}^s$:
\begin{itemize}
\item either $\mathcal{M}^s$ is a planar map with one simple boundary of some length $\ell'$, and another simple boundary of length $k$ (which we did not touch). Then, the rest of the pieces $\mathcal{M}_1, \ldots, \mathcal{M}_{\ell'}$ are planar maps with one ordinary boundary of lengths $\ell_1,\ldots,\ell_{\ell'}$.
\item or $\mathcal{M}^s$ is a planar map with one simple boundary of some length $\ell'$. And the rest consists of a disjoint union of:
\begin{itemize}
\item a planar map with the simple boundary of length $k$ which bordered $\mathcal{M}$ initially, and another ordinary boundary with some length $\ell_1$, 
\item and  $\ell' - 1$ planar maps with one ordinary boundary of lengths $\ell_2,\ldots,\ell_{\ell'}$.
\end{itemize}
\end{itemize}
This decomposition is again a bijection, and hence
\beq
G_{k|\ell} = \sum_{\ell' = 1}^{\ell}  \sum_{\substack{\ell_1,\ldots,\ell_{\ell'} \geq 0 \\ \sum_i (\ell_i + 1) = \ell}} \left(G_{k,\ell'}\prod_{i = 1}^{\ell'} F_{\ell_i}+ \ell'\,G_{\ell'} G_{k|\ell_1}\prod_{i = 2}^{\ell'} F_{\ell_i}\right).
\eeq
We deduce, at the level of resolvents, that
\beq
Y_{1|1}(w\mid x) = \frac{W(x)}{x} Y_2(w,W(x))+\frac{Y_{1|1}(w\mid x)}{x} (\partial_w(w X(w)))_{w=W(x)}.
\eeq
Isolating $Y_2$, we obtain:
\beq
\label{2simple} Y_{2}(w,W(x)) = -Y_{1|1}(w\mid x) (\partial_{w}X(w))_{w=W(x)} .
\eeq

\subsection{From ordinary cylinders to simple cylinders}

We consider the following operator
\beq
\frac{\partial}{\partial V(x)} = \sum_{k \geq 0} \frac{k}{x^{k + 1}} \frac{\partial}{\partial t_{k}}\,,
\eeq
which creates an ordinary boundary of length $k$ weighted by $x^{-(k + 1)}$. Therefore, we have
\begin{align*}
W_{n}^{[g]}(x_1,\ldots,x_n) & = \frac{\partial}{\partial V(x_2)}\cdots\frac{\partial}{\partial V(x_n)}\,W_{1}^{[g]}(x_1),  \\
Y_{1|n}^{[g]}(w \mid x_1,\ldots,x_n) & = \frac{\partial}{\partial V(x_1)}\cdots \frac{\partial}{\partial V(x_n)}\,Y_{1}^{[g]}(w). 
\end{align*}
Applying $\frac{\partial}{\partial V(x_1)}$ to equation (\ref{relmaster1}), we obtain
\beq
0 =  (\partial_w X(w))_{w=W(x_1)} \frac{\partial}{\partial V(x_1)}W(x_1) + Y_{1|1}(W(x_1)\mid x_2) \nonumber
\eeq
and hence
\beq
\label{ge1}Y_{1 | 1}(W(x_1) \mid x_2) = - W_{2}(x_1,x_2)\,(\partial_w X(w))_{w = W(x_1)}.
\eeq
Finally, combining equations (\ref{2simple}) and (\ref{ge1}), we obtain the following relation between ordinary and simple cylinders:
\beq\label{simpleOrdCyl}
Y_2(W(x_1),W(x_2))= W_2(x_1,x_2) (\partial_w X(w))_{w = W(x_1)} (\partial_w X(w))_{w = W(x_2)}.
\eeq

\subsection{From simple cylinders to fully simple cylinders}

We describe an algorithm which expresses a planar map $\mathcal{M}$ with two simple boundaries in terms of planar fully simple maps. The idea is to merge simple boundaries that touch each other. By definition, a simple boundary which is not fully simple shares at least one vertex with another boundary. By convention, whenever we refer to cyclic order in the process, we mean cyclic order of the first boundary. Let $B_1$ and $B_2$ denote the first and the second boundaries respectively.
\begin{definition}
A {\it pre-shared piece of length $m >0$} is a sequence of $m$ consecutive edges in $B_1$ which are shared with $B_2$. We define a {\it pre-shared piece of length $m =0$} to be a vertex which both boundaries $B_1$ and $B_2$ have in common. 

The first vertex $sv_1$ of the first edge and the second vertex $sv_2$ of the last edge in a pre-shared piece are called the {\it endpoints}.
If $m=0$, the endpoints coincide by convention with the only vertex of the pre-shared piece: $sv_1=sv_2$.

We say that a  pre-shared piece of length $m \geq 0$ is a {\it shared piece of length $m \geq 0$} if the edge in $B_1$ that arrives to $sv_1$ and the edge in $B_1$ outgoing from $sv_2$ are not shared with $B_2$.

We define the {\it interior} of a shared piece to be the shared piece minus the two endpoints.
The interior of a shared vertex is empty.
\end{definition}

Before describing the decomposition algorithm, we describe a special case which corresponds to maps as in Figure \ref{totally glued}.(c), which we will exclude. Consider a map whose two only faces are the two simple boundaries. The only possibility is that they have the same length and are completely glued to each other. We will count this kind of maps apart.
\begin{algorithm}\label{AlgCyls}{\rm (From simple cylinders to fully simple disks or cylinders)
\begin{enumerate}
\item Save the position of the marked edge on each boundary.

\item Denote $r$ the number of shared pieces. If $r=0$, we already have a fully simple cylinder and we stop the algorithm. Otherwise, denote the shared pieces by $S_0, \ldots, S_{r-1}$. Save their lengths $m_0, \ldots, m_{r-1}$, labeled in cyclic order, and shrink their interiors so that only shared vertices remain.

Since we have removed all common edges and boundaries are simple,  every shared vertex has two non-identified incident edges from $B_1$ and two from $B_2$.

\item Create two vertices $v_1, v_2$ out of each shared vertex $v$ in such a way that each $v_j$ has exactly one incident edge from $B_1$ and one from $B_2$, which were consecutive edges for the cyclic order at $v$ in the initial map.

In this way, we got rid of all shared pieces, and we obtain a graph drawn on the sphere formed by $r$ connected components which are homeomorphic to a disk. We consider each connected component separately, and we glue to their boundary a face homeomorphic to a disk.

\item For $i = 0,\ldots , r-1$, call $\mathcal{M}_i$ the connected component which was sharing a vertex with $S_i$ and $S_{i + 1\,({\rm mod}\, r)}$. Mark the edge in $\mathcal{M}_i$ which belonged to the first boundary and was outgoing from $sv_2^i$ in $\mathcal{M}_i$. Then, $\mathcal{M}_i$ becomes a simple disk. Denote by $\ell_i'$ (resp. $\ell_i''$) the number of edges of the boundary of $\mathcal{M}_i$ previously belonging to $B_1$ (resp. $B_2$). Then, the boundary of $\mathcal{M}_i$  has perimeter $\ell_i'+\ell_i''$.
\end{enumerate}}
\end{algorithm}

\begin{figure}[h!]
 \begin{center}
        \def\svgwidth{\columnwidth}
        \input{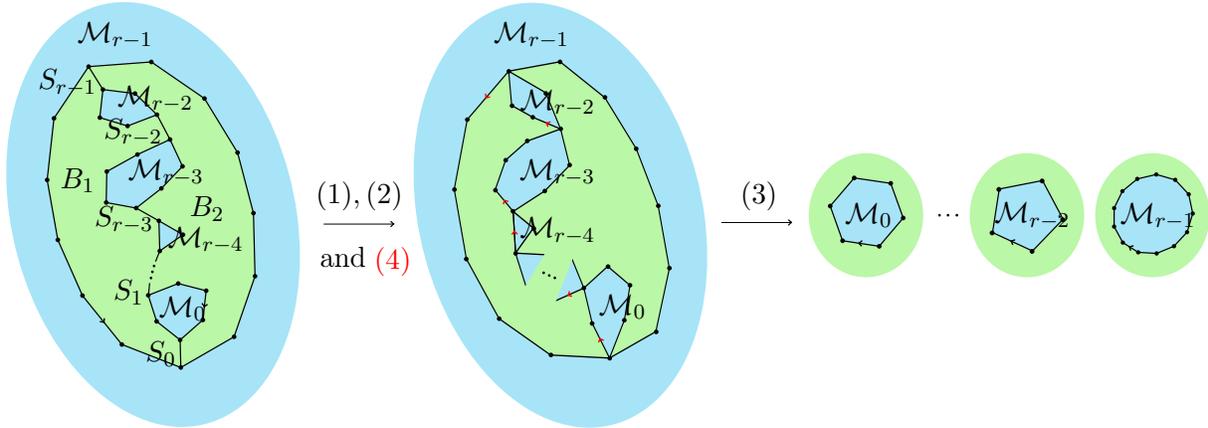}
        	\caption{Schematic representation of Algorithm \ref{AlgCyls}. The two green faces are the two simple, but non-fully simple, boundaries and the blue part represents the inner faces. The shared pieces $S_0,\ldots,S_{r-1}$ are drawn schematically as shared pieces of length $1$. $\mathcal{M}_{r-1}$ is drawn as the outer face, but it does not play a special role.}
\end{center}
\end{figure}

Observe that by construction, $\ell_i', \ell_i'' \geq 1$ and $m_i\geq 0$. Moreover, note that the only map of length $\ell_i'+\ell_i''\geq 2$ and considered simple which is not allowed as $\mathcal{M}_i$ is the map with one boundary of length $2$ where the two edges are identified as in Figure \ref{degenerate simple map}.(c), since this would correspond to a shared piece of length $1$ and it would have been previously removed.

This decomposition is a bijection, since we can recover the original map from all the saved information and the obtained fully simple maps. To show this, we describe the inverse algorithm:

\begin{algorithm}{\rm (From fully simple disks or cylinders to simple cylinders)
\begin{enumerate}
\item Let $r$ be the number of given (fully) simple discs. If $r=0$, we already had a fully simple cylinder and the algorithms become trivial. Otherwise, observe that every $\mathcal{M}_i$, for $i = 0,\ldots , r-1$, is a planar disk with two distinguished vertices $v_1^i$ and $v_2^i$. The first one $v_1^i$ is the starting vertex of the root edge $e_1^{i}$ and the second one $v_2^i$ is the ending vertex of the edge $e_{\ell_i'}^{i}$, where the edges are labeled according to the cyclic order of the boundary.

\item For $i = 0,\ldots , r-1$, consider shared pieces $S_i$ of lengths $m_i$.

\item Glue $sv_1^{i}$ of a shared piece $S_{i}$  to $v_2^{i-1\,({\rm mod}\, r)}$ in $\mathcal{M}_{i - 1\,({\rm mod}\, r)}$ and $sv_2^{i}$ of $S_{i}$ to $v_1^{i}$ in $\mathcal{M}_{i}$. 

All the marked edges in the $\mathcal{M}_i$'s should belong to the same simple face, which we call $B_1$. We call $B_2$ the other face, which is bordered by following the edges from $v_2^i$ to $v_1^i$ in every $\mathcal{M}_i$, and the shared piece $S_i$ from $sv_2^i$ to $sv_1^i$, for $i = 1,\ldots,r$.

\item Remove the $r$ markings in $B_1$ and recover the roots in $B_1$ and $B_2$, which are now part of our data.

We have glued the $r$ simple disks and shared pieces into a map with two simple (not fully simple) boundaries $B_1$ and $B_2$.
\end{enumerate}}
\end{algorithm}

This bijection translates into the following relation between generating series of simple and fully simple cylinders:
\begin{proposition}\label{0,2}
\beq
Y_2(w_1,w_2) =  X_2(w_1,w_2) + \partial_{w_1}\partial_{w_2}\ln\left(\frac{w_1 - w_2}{X(w_1) - X(w_2)}\right).
\eeq
\end{proposition}
\noindent \textbf{Proof.}
Let us introduce:
$$
\tilde{X}(w) = X(w) - w^{-1} - w = \sum_{\ell \geq 1} \tilde{H}_{\ell}\,w^{\ell - 1},
$$
the generating series of (fully) simple disks, excluding the disk with boundary of length $0$ which consists of a single vertex, and the simple disk with boundary of length $2$ in which the two edges of the boundary are identified, as in Figure \ref{degenerate simple map}.(c).

Then, using the bijection we established, we obtain that
\beq
G_{L_1,L_2} = H_{L_1,L_2} + \delta_{L_1,L_2}\,L_1  + \sum_{r \geq 1} \sum_{\substack{\ell_i',\,\,\ell_i'' > 0,\,\,m_i \geq 0 \\ \sum_{i=0}^{r-1} \ell_i' + \sum_{i} m_i = L_1 \\ \sum_{i=0}^{r-1}  \ell_i'' + \sum_{i} m_i = L_2}} \frac{L_1L_2}{r} \prod_{i = 1}^r \tilde{H}_{\ell_i' + \ell_i''}, \nonumber 
\eeq
where the first term of the right hand side counts the case $r=0$ in which the simple cylinders were already fully simple and the second term counts the degenerate case we excluded from the algorithm. We already observed that this degenerate case can only occur if $L_1=L_2$ and there are $L_1^2$ possibilities for the two roots, but we also divide by $L_1$ because of the cyclic symmetry of this type of cylinders.

Summing over lengths $L_1, L_2 \geq 1$ with a weight $w_1^{L_1 - 1}w_2^{L_2 - 1}$, we get:
\bea
Y_2(w_1,w_2) & = & X_2(w_1,w_2) - \partial_{w_1}\partial_{w_2}\ln(1 - w_1w_2) \nonumber \\
& & + \partial_{w_1}\partial_{w_2}\left(\sum_{r \geq 1} \frac{1}{r}\bigg(\sum_{m \geq 0} (w_1w_2)^m\bigg)^r\bigg(\sum_{\ell' ,\ell' > 0} \tilde{H}_{\ell' + \ell''} w_1^{\ell'}w_2^{\ell''}\bigg)^{r}\right). \nonumber
\eea
Let us remark that
$$
\sum_{\substack{\ell',\ell'' > 0 \\ \ell' + \ell'' = \ell  }} w_1^{\ell'}w_2^{\ell''} = \frac{w_{1}^{\ell+ 1} - w_2^{\ell + 1}}{w_1 - w_2} - w_1^{\ell} - w_2^{\ell}.
$$
Therefore,
\bea
\sum_{\ell',\ell'' \geq 1} w_1^{\ell'}w_2^{\ell''}\,\tilde{H}_{\ell' + \ell''} & = & \frac{w_1^2\tilde{X}(w_1) - w_2^2\tilde{X}(w_2)}{w_1 - w_2} - w_1\tilde{X}(w_1) - w_2\tilde{X}(w_2) \nonumber \\
& = & \frac{w_1w_2(\tilde{X}(w_1) - \tilde{X}(w_2))}{w_1 - w_2} \nonumber \\
& = & w_1w_2\,\frac{(X(w_1) - X(w_2))}{w_1 - w_2} - (1 - w_1w_2). \nonumber
\eea
And finally, 
\bea
Y_2(w_1,w_2) & = & X_2(w_1,w_2) - \partial_{w_1}\partial_{w_2}\ln(1 - w_1w_2) \nonumber \\
& & - \partial_{w_1}\partial_{w_2}\ln\left[1 - \frac{1}{1 - w_1w_2}\left(w_1w_2\,\frac{X(w_1) - X(w_2)}{w_1 - w_2} - (1 - w_1w_2)\right)\right] \nonumber \\
& = & X_2(w_1,w_2) - \partial_{w_1}\partial_{w_2}\ln\left[-w_1w_2\,\frac{X(w_1) - X(w_2)}{w_1 - w_2}\right] \nonumber \\
& = & X_2(w_1,w_2) + \partial_{w_1}\partial_{w_2}\ln\left(\frac{w_1 - w_2}{X(w_1) - X(w_2)}\right). \nonumber
\eea
\hfill $\Box$

\section{Combinatorial interpretation of symplectic invariance}\label{symplinv}
\label{Section5}
\subsection{General result for usual maps}

In this section we recall that the generating series of ordinary maps satisfy the topological recursion, we explain how the formulas obtained in the previous section fit naturally in the universal setting of topological recursion, state a precise conjecture on how this would generalize for higher topologies, and give some numerical evidence and illustration in the particular case of quadrangulations.

We consider the reparametrization
\beq
x(z)=\alpha + \gamma \left( z+\frac{1}{z}\right),\nonumber
\eeq
where $\alpha$ and $\gamma$ are parameters determined in terms of the weights $u, t_3, t_4, \ldots$, and which makes $w(z)\coloneqq W_1^{[0]}(x(z))$ a rational function of $z$. This computation of $\alpha$, $\gamma$ and $W_{1}^{[0]}(x(z))$ is summarized later in Lemma~\ref{Tutte000}.

We introduce the fundamental differential of the second kind:
\beq
B(z_1,z_2)\coloneqq \frac{\dd z_1 \dd z_2}{(z_1-z_2)^2}. \nonumber
\eeq

Up to a correction term, and written as a bidifferential form, the cylinder generating function is the fundamental differential of the second kind:
\begin{theorem}\cite{E1MM}\label{ordcyl}
\beq 
W_2^{[0]}(x(z_1), x(z_2)) \dd x(z_1) \dd x(z_2) + \frac{\dd x(z_1) \dd x(z_2)}{(x(z_1)-x(z_2))^2} = B(z_1,z_2). \nonumber
\eeq
\end{theorem}

In general, we rewrite the generating functions of maps in terms of the variables $z_i$ and as multi-differential forms in $\mathbb{CP}^1$:
\beq
\omega_{g,n}(z_1,\ldots,z_n)=W_n^{[g]}(x(z_1),\ldots, x(z_n)) \dd x(z_1)\cdots \dd x(z_n) + \delta_{g,0}\delta_{n,2}\frac{\dd x(z_1) \dd x(z_2)}{(x(z_1)-x(z_2))^2}. \nonumber
\eeq
The forms $\omega_{g,n}$ satisfy the so-called topological recursion with spectral curve given by $(x,w)$ and initial data 
$$
(\omega_{0,1}(z_1), \omega_{0,2}(z_1,z_2))=(w(z_1)\dd x(z_1), B(z_1,z_2)).
$$ 
More concretely:
\begin{theorem}\cite{E1MM}\label{ordmaps}
For all $g \geq 0$, $n\geq 1$ with $2g-2+n >0$,
\bea
\displaystyle
\omega_{g,n}(z_1,\ldots,z_n)= & \underset{z=\pm 1}{{\rm Res}} \, K(z_1,z)\Big( \omega_{g-1,n+1}(z,1/z,z_2,\ldots,z_n) \nonumber \\
& +\displaystyle \ \  \sum_{\mathclap{\substack{h =0,\ldots, g \\ I \sqcup J =\{2,\ldots,n\}}}} \limits \, \mathclap{{}^{'}} \ \ \  \omega_{h, |I|+1}(z,z_I)\, \omega_{g-h,|J|+1}(1/z,z_J)\Big), \nonumber
\eea
with $\sum{}^{'}$ meaning that we omit $(h,I)=(0,\emptyset)$ and $(h,J)=(g,\emptyset)$, and the following recursion kernel
\beq 
K(z_1,z)\coloneqq\frac{\frac{1}{z_1-z}-\frac{1}{z_1-1/z}}{2(w(z)-w(1/z))}\frac{\dd z_1}{\gamma (1-z^{-2})\dd z}. \nonumber
\eeq
\end{theorem}

For further details on these results for the generating series of ordinary maps, see the book \cite{EynardBook}.

The main feature of the $\mathcal{F}_g=\omega_{g,0}$ produced by the topological recursion is the so-called symplectic invariance. If two spectral curves $\mathcal{S}$ and $\check{\mathcal{S}}$ are symplectically equivalent, that is $|\dd x\wedge \dd w |=|\dd \check{x}\wedge \dd \check{w}|$, then a relation between $\mathcal{F}_g[\mathcal{S}]$ and $\mathcal{F}_g[\check{\mathcal{S}}]$ is expected. More concretely, they are proved to be equal for any symplectic transformation not generated using the transformation which exchanges $x$ and $w$ (see \cite{EOFg}). For this reason, the $\mathcal{F}_g$'s are called \emph{symplectic invariants}. The exchanging transformation $(x,w)\mapsto(w,x)$ is then considered to be the most mysterious and interesting one. These invariants are also expected to be equal after exchanging $x$ and $w$ up to some correction terms whose exact form is still under scrutiny (see \cite{EO2MM,EOxy} for some progress towards the relation for algebraic compact curves). One of the motivations to study fully simple maps is to gain some understanding of this property in this complicated case.

Let us consider now the the spectral curve given by $(\check{x},\check{w})=(w,x)$, with initial data 
$$
(\check{\omega}_{0,1}(z_1), \check{\omega}_{0,2}(z_1,z_2)) \coloneqq  (x(z) \dd w(z), B(z_1,z_2)).
$$
For $2g-2+n >0$, we call $\check{\omega}_{g,n}$ the TR amplitudes for this spectral curve.

It is natural to wonder whether the $\check{\omega}_{g,n}$ also solve some enumerative problem and, in that case, which kind of objects they are counting. We propose an answer which also offers a combinatorial interpretation of the important property of symplectic invariance:
\begin{conjecture}\label{conj}
The invariants $\check{\omega}_{g,n}$ enumerate fully simple maps of genus $g$ and $n$ boundaries in the following sense:
\beq
\check{\omega}_{g,n}(z_1,\ldots,z_n)=X_n^{[g]}(w(z_1),\ldots, w(z_n)) \dd w(z_1)\cdots \dd w(z_n) + \delta_{g,0}\delta_{n,2}\frac{\dd w(z_1) \dd w(z_2)}{(w(z_1)-w(z_2))^2}. \nonumber
\eeq
\end{conjecture}

Using non-combinatorial techniques, we give in Section~\ref{PRof} the path to a possible proof of this conjecture for usual maps. We reduce the problem to a technical condition regarding a milder version of symplectic invariance of the so-called 1-hermitian matrix model with external field, which will be given in Section~\ref{PRof}.

Observe that if $n=0$, that is we consider maps without boundaries, we have in a very natural way that $W_0^{[g]}=X_0^{[g]}$. It would be interesting to investigate the relation between the TR $n=0$ invariants $\mathcal{F}_g[\mathcal{S}]$ and $\mathcal{F}_g[\mathcal{\check{S}}]$. We believe there should be a combinatorial justification for their relation which could be explored with a combinatorial proof of our conjecture, but this is beyond the scope of this article.

We dedicate the rest of this section to prove some first cases of this conjecture in a combinatorial way, give some evidence for the conjecture in general and  comment on some possible generalizations.

Using our formulas relating the generating series of fully simple disks and cylinders with the ordinary ones, we obtain a combinatorial proof of the first two base cases of the conjecture:
\begin{theorem}
\label{simplTh} Conjecture~\ref{conj} is true for the two base cases $(g,n)= (0,1)$ and $(0,2)$.
\end{theorem}
\noindent \textbf{Proof.}
By Proposition \ref{0,1}, we obtain
$$
\check{\omega}_{0,1}(z) \coloneqq x(z) \dd w(z) = X(W(x(z))) \dd w(z),
$$
which is equal to $X(w(z)) \dd w(z)$ by definition. This proves the theorem for $(g,n)= (0,1)$.

For cylinders, we have, by definition: $\check{\omega}_{0,2}(z_1,z_2)=B(z_1,z_2)$. Substituting the expression for the generating series of ordinary cylinders in terms of the one for simple cylinders given by formula (\ref{simpleOrdCyl}) in the equation for the fundamental differential of the second kind from Theorem \ref{ordcyl}, we obtain
\bea
\omega_{0,2}(z_1,z_2) = B(z_1,z_2)  = & W_2(x(z_1), x(z_2)) \dd x(z_1) \dd x(z_2) + \frac{\dd x(z_1) \dd x(z_2)}{(x(z_1)-x(z_2))^2} \nonumber \\
 = & Y_2(w(z_1), w(z_2)) \dd w(z_1) \dd w(z_2) + \frac{\dd x(z_1) \dd x(z_2)}{(x(z_1)-x(z_2))^2}. \nonumber
\eea
Finally, using Proposition \ref{0,2}, we get the theorem for $(g,n)= (0,2)$:
\beq
\check{\omega}_{0,2}(z_1,z_2)=X_2(w(z_1), w(z_2)) \dd w(z_1) \dd w(z_2) + \frac{\dd w(z_1) \dd w(z_2)}{(w(z_1)-w(z_2))^2}. \nonumber
\eeq
\hfill $\Box$

\subsection{Supporting data for quadrangulations}\label{Quadrangulations}

In this section we compare the number of fully simple, simple and ordinary disks and cylinders in the case in which all the internal faces are quadrangulations, which allows us to make computations explicitly using our results for the base topologies: $(0,1)$ and $(0,2)$. We also compare the conjectural number of fully simple quadrangulations to the number of ordinary ones for topologies $(1,1)$ and $(0,3)$, whose outcomes are given by the first iteration of the algorithm of topological recursion. The reasonable outcomes support our conjecture that, after the exchange transformation, TR counts some more restrictive kind of maps. 

The $(1,1)$ topology is especially interesting since it is the first case with genus $g>0$. For topology $(1,1)$, we provide explicit general formulas for the number of ordinary maps and for the conjectural number of fully simple maps, which we extract from TR. We also give a combinatorial argument that indeed shows that the conjecture provides the right numbers for the first possible length: $\ell = 2$. 

The $(0,3)$ topology is also particularly relevant since in that case we find substantial evidence for our conjecture, using the formulas proved in \cite{BernardiFusy} for fully simple planar quadrangulations with even boundary lengths. Moreover, this is one of the most relevant cases for one of our motivations coming from free probability, since TR for fully simple maps of topology $(0,3)$ would provide new interesting formulas relating the third order free cumulants to the third order correlation moments (see Section~\ref{Highfree}).

We consider maps whose internal faces are all quadrangles \cite{Tutte2}, that is $t_{j} = t \delta_{j,4}$ where we denote here $t$ the weight per internal quadrangle. The spectral curve is given by
$$
x(z) = c\Big(z + \frac{1}{z}\Big),\qquad w(z) = \frac{1}{cz} - \frac{tc^3}{z^3},\qquad 
$$
with
\beq\label{cSeries}
c = \sqrt{\frac{1 - \sqrt{1 - 12t}}{6t}} = 1 + \frac{3t}{2} + \frac{63}{8}t^2 + \frac{891}{16}t^3 + \frac{57915}{128}t^4 + O(t^4).
\eeq
The zeroes of $\dd x$ are located at $z = \pm 1$, and the deck transformation is $\iota(z) = \frac{1}{z}$.  The zeroes of $\dd w$ are located at $z = \pm c^2 \sqrt{3t}$, and the deck transformation is
$$
\check{\iota}(z) = \frac{c^2z(c^2t + \sqrt{4tz^2 - 3c^4t^2})}{2(z^2 - tc^4)}.
$$
Consider the multidifferentials $\omega_{g,n}$ and $\check{\omega}_{g,n}$ on $\mathbb{P}^1$ as at the beginning of the section but with initial data specialized for quadrangulations.
We define
\bea
F_{\ell_1,\ldots,\ell_n}^{[g]} & \!\!\!\! = \!\!\!\! & (-1)^n \Res_{z_1 \rightarrow \infty} (x(z_1))^{\ell_1} \cdots \Res_{z_n \rightarrow \infty} (x(z_n))^{\ell_n} \bigg(\omega_{g,n}(z_1,\ldots,z_n) - \delta_{g,0}\delta_{n,2} \frac{\dd x(z_1)\dd x(z_2)}{(x(z_1) - x(z_2))^2}\bigg), \nonumber \\
\check{F}_{k_1,\ldots,k_n}^{[g]} & \!\!\!\! = \!\!\!\! & \Res_{z_1 \rightarrow \infty} (w(z_1))^{-k_1} \cdots \Res_{z_n \rightarrow \infty} (w(z_n))^{-k_n}\bigg(\check{\omega}_{g,n}(z_1,\ldots,z_n) - \delta_{g,0}\delta_{n,2}\frac{\dd w(z_1)\dd w(z_2)}{(w(z_1) - w(z_2))^2}\bigg). \nonumber
\eea
We know that $\check{F}_{k_1}=H_{k_1}$ and $\check{F}_{k_1,k_2}=H_{k_1,k_2}$, and we conjecture $\check{F}_{k_1,\ldots,k_n}^{[g]}=H_{k_1,\ldots,k_n}^{[g]}$ in general.

\subsubsection{Disks}
We explore two tables to compare the coefficients $[t^{Q}]\,F_{\ell}$ and $[t^{Q}]\,\check{F}_{k}$. It is remarkable that all the $[t^{Q}]\,\check{F}_{k}$ are nonnegative integers, which already suggested \textit{a priori} that they may be counting some objects. Theorem \ref{simplTh} identifies $[t^{Q}]\,\check{F}_{k}$ with the number of (fully) simple disks $[t^{Q}]\,H_{k}$.

If the length of the boundary is odd, the number of disks is obviously $0$.

Observe that if we consider a boundary of length $\ell=2$, the number of ordinary disks is equal to the number of (fully) simple disks because the only two possible boundaries of length $2$ are simple in genus $0$. If the two vertices get identified in the non-degenerate case, either the genus is increased or an internal face of length $1$ appears, which is not possible because we are counting quadrangulations.

\begin{center}
\begin{figure}[h!]
\begin{center}
{\scriptsize \begin{tabular}{|c||c|c|c|c|c|c|c|c|c|}
\hline
$\ell$ & $Q = 0$ & $1$ & $2$ & $3$ & $4$ & $5$ & $6$ & $7$ & $8$ \\
\hline\hline
$\mathbf{2}$ & $1$ & $2$ & $9$ & $54$ & $378$ & $2916$ & $24057$ & $208494$ & $1876446$ \\
\hline
$\mathbf{4}$ & $2$ & $9$ & $54$ & $378$ & $2916$ & $24057$ & $208494$ & $1876446$ & $17399772$ \\
\hline
$\mathbf{6}$ & $5$ & $36$ & $270$ & $2160$ & $18225$ & $160380$ & $1459458$ & $13646880$ & $130489290$ \\
\hline
$\mathbf{8}$ & $14$ & $140$ & $1260$ & $11340$ & $103950$ & $972972$ & $9287460$ & $90221040$ & $890065260$ \\
\hline
\end{tabular}}
\caption{Number of ordinary disks with boundary of length $\ell$ and $Q$ quadrangles.}
\end{center}
\end{figure}
\end{center}

We also remark that the $[t^{Q}]\,H_{k}$ in the second table are (much) smaller than the corresponding $[t^{Q}]\,F_{k}$, and for small number of quadrangles some of them are $0$, which makes sense due to the strong geometric constraints to form maps with simple boundaries and a small number of internal faces.

\begin{center}
\begin{figure}[h!]
\begin{center}
{\scriptsize \begin{tabular}{|c||c|c|c|c|c|c|c|c|c|}
\hline
$k$ & $Q = 0$ & $1$ & $2$ & $3$ & $4$ & $5$ & $6$ & $7$ & $8$ \\
\hline\hline
$\mathbf{2}$ & $1$ & $2$ & $9$ & $54$ & $378$ & $2916$ & $24057$ & $208494$ & $1876446$ \\
\hline
$\mathbf{4}$ & $0$ & $1$ & $10$ & $90$ & $810$ & $7425$ & $69498$ & $663390$ & $6444360$ \\
\hline
$\mathbf{6}$ & $0$ & $0$ & $3$ & $56$ & $756$ & $9072$ & $103194$ & $1143072$ & $12492144$ \\
\hline
$\mathbf{8}$ & $0$ & $0$ & $0$ & $12$ & $330$ & $5940$ & $89100$ & $1211760$ & $15540822$ \\
\hline
\end{tabular}}
\caption{\label{FSdisks} Number of simple disks with boundary of length $k$ and $Q$ quadrangles.}
\end{center}
\end{figure}
\end{center}

\subsubsection{Cylinders}

We explore now the number of cylinders imposing different constraints to the boundaries: $[t^{Q}]\,F_{\ell_1,\ell_2}$ (ordinary) and $[t^{Q}]\,H_{k_1,k_2}$ (fully simple), and also $[t^{Q}]\,G_{k_1\mid\ell_1}$ (one simple boundary, one ordinary boundary) and $[t^{Q}]\,G_{k_1,k_2\mid}$ (simple) given by the formulas (\ref{ge1}) and (\ref{simpleOrdCyl}) respectively.

Since we know how to convert an unmarked quadrangle into an ordinary boundary of length $4$, we can relate the outcomes for cylinders with at least one of the boundaries being ordinary of length $4$ to the previous results for disks as follows:
\bea
\bullet & 4\frac{\partial}{\partial t} F_{\ell_1} = F_{\ell_1,4} & \Rightarrow \ \    4Q[t^Q]\,F_{\ell_1} = [t^{Q-1}]\, F_{\ell_1,4}, \nonumber \\
\bullet & 4\frac{\partial}{\partial t} G_{k_1} = G_{k_1\mid 4} & \Rightarrow \ \   4Q [t^Q]\,G_{k_1}= [t^{Q-1}]\, G_{k_1\mid 4}. \nonumber
\eea
If the sum of the lengths of the two boundaries is odd, the number of quadrangulations is obviously $0$.

Observe that the results also satisfy the following inequalities:
$$
[t^{Q}]\,F_{l_1,l_2} \geq [t^{Q}]\,G_{l_1\mid l_2}\geq [t^{Q}]\,G_{l_1,l_2} \geq [t^{Q}]\,H_{l_1,l_2},
$$
which are compatible with the combinatorial interpretation that Theorem \ref{simplTh} offers, since we are imposing further constraints whenever we force a boundary to be simple or, even more, fully simple.

We also obtain more and more zeroes for small number of quadrangles as we impose stronger conditions on the boundaries.
 
\begin{center}
\begin{figure}[h!]
\begin{center}
{\scriptsize 
\scalebox{0.95}{\begin{tabular}{|c||c|c|c|c|c|c|c|c|c|}
\hline
$(\ell_1,\ell_2)$ & $Q = 0$ & $1$ & $2$ & $3$ & $4$ & $5$ & $6$ & $7$ & $8$ \\
\hline\hline
$\mathbf{(1,1)}$ & $1$ & $3$ & $18$ & $135$ & $1134$ & $10206$ & $96228$ & $938223$ & $9382230$ \\
\hline
$\mathbf{(3,1)}$ & $3$ & $18$ & $135$ & $1134$ & $10206$ & $96228$ & $938223$ & $9382230$ & $95698746$\\
\hline
$\mathbf{(5,1)}$ & $10$ & $90$ & $810$ & $7560$ & $72900$ & $721710$ & $7297290$ & $75057840$ & $782989740$ \\
\hline
$\mathbf{(7,1)}$ & $35$ & $420$ & $4410$ & $45360$ & $467775$ & $4864860$ & $51081030$ & $541326240$ & $5785424190$ \\
\hline
$\mathbf{(9,1)}$ & $126$ & $1890$ & $22680$ & $255150$ & $2806650$ & $30648618$ & $334348560$ & $3653952120$ & $40052936700$ \\
\hline\hline
$\mathbf{(2,2)}$ & $2$ & $12$ & $90$ & $756$ & $6804$ & $64152$ & $625482$ & $6254820$ & $63799164$ \\
\hline
$\mathbf{(4,2)}$ & $8$ & $72$ & $648$ & $6048$ & $58320$ & $577368$ & $5837832$ & $60046272$ & $626391792$ \\
\hline
$\mathbf{(6,2)}$ & $30$ & $360$ & $3780$ & $38880$ & $400950$ & $4169880$ & $43783740$ & $463993920$ & $4958935020$ \\
\hline
$\mathbf{(8,2)}$ & $112$ & $1680$ & $20160$ & $226800$ & $2494800$ & $27243216$ & $297198720$ & $3247957440$ & $335602610400$ \\
\hline\hline
$\mathbf{(3,3)}$ & $12$ & $108$ & $972$ & $9072$ & $87480$ & $866052$ & $8756748$ & $90069408$ & $939587688$ \\
\hline
$\mathbf{(5,3)}$ & $45$ & $540$ & $5670$ & $58320$ & $601425$ & $6254820$ & $65675610$ & $695990880$ & $7438402530$ \\
\hline
$\mathbf{(7,3)}$ & $168$ & $2520$ & $30240$ & $340200$ & $3742200$ & $40864824$ & $445798080$ & $4871936160$ & $53403915600$ \\
\hline
$\mathbf{(9,3)}$ & $630$ & $11340$ & $153090$ & $1871100$ & $21891870$ & $250761420$ & $2841962760$ & $32042349360$ & $360476430300$ \\
\hline\hline
$\mathbf{(4,4)}$ & $36$ & $432$ & $4536$ & $46656$ & $481140$ & $5003856$ & $52540488$ & $556792704$ & $5950722024$ \\
\hline
$\mathbf{(6,4)}$ & $144$ & $2160$ & $25920$ & $291600$ & $3207600$ & $35026992$ & $382112640$ & $4175945280$ & $45774784800$ \\
\hline
$\mathbf{(8,4)}$ & $560$ & $10080$ & $136080$ & $1663200$ & $19459440$ & $222899040$ & $2526189120$ & $28482088320$ & $320423493600$ \\
\hline
\end{tabular}}}
\caption{Number of ordinary cylinders with boundaries of lengths $(\ell_1,\ell_2)$ and $Q$ quadrangles: $[t^{Q}]\,F_{\ell_1,\ell_2}$.}
\end{center}
\end{figure}
\end{center}

\begin{center}
\begin{figure}[h!]
\begin{center}
\scalebox{0.95}{{\scriptsize \begin{tabular}{|c||c|c|c|c|c|c|c|c|c|}
\hline
$(k_1,\ell_1)$ & $Q = 0$ & $1$ & $2$ & $3$ & $4$ & $5$ & $6$ & $7$ & $8$ \\
\hline\hline
$\mathbf{(1,1)}$ & $1$ & $3$ & $18$ & $135$ & $1134$ & $10206$ & $96228$ & $938223$ & $9382230$ \\
\hline
$\mathbf{(3,1)}$ & $0$ & $3$ & $36$ & $378$ & $3888$ & $40095$ & $416988$ & $4378374$ & $46399392$\\
\hline
$\mathbf{(5,1)}$ & $0$ & $0$ & $15$ & $315$ & $4725$ & $62370$ & $773955$ & $9287460$ & $109306260$ \\
\hline
$\mathbf{(7,1)}$ & $0$ & $0$ & $0$ & $84$ & $2520$ & $49140$ & $793800$ & $11566800$ & $158233824$ \\
\hline
$\mathbf{(9,1)}$ & $0$ & $0$ & $0$ & $0$ & $495$ & $19305$ & $463320$ & $8860995$ & $148551975$ \\
\hline\hline
$\mathbf{(2,2)}$ & $2$ & $12$ & $90$ & $756$ & $6804$ & $64152$ & $625482$ & $6254820$ & $63799164$ \\
\hline
$\mathbf{(4,2)}$ & $0$ & $8$ & $120$ & $1440$ & $16200$ & $178200$ & $1945944$ & $21228480$ & $231996960$ \\
\hline
$\mathbf{(6,2)}$ & $0$ & $0$ & $42$ & $1008$ & $16632$ & $235872$ & $3095820$ & $38864448$ & $474701472$ \\
\hline
$\mathbf{(8,2)}$ & $0$ & $0$ & $0$ & $240$ & $7920$ & $166320$ & $2851200$ & $43623360$ & $621632880$ \\
\hline\hline
$\mathbf{(1,3)}$ & $3$ & $18$ & $135$ & $1134$ & $10206$ & $96228$ & $938223$ & $9382230$ & $95698746$\\
\hline
$\mathbf{(3,3)}$ & $3$ & $36$ & $378$ & $3888$ & $40095$ & $416988$ & $4378374$ & $46399392$ & $495893502$\\
\hline
$\mathbf{(5,3)}$ & $0$ & $15$ & $315$ & $4725$ & $62370$ & $773955$ & $9287460$ & $109306260$ & $1271521800$ \\
\hline
$\mathbf{(7,3)}$ & $0$ & $0$ & $84$ & $2520$ & $49140$ & $793800$ & $11566800$ & $158233824$ & $2076818940$ \\
\hline
$\mathbf{(9,3)}$ & $0$ & $0$ & $0$ & $495$ & $19305$ & $463320$ & $8860995$ & $148551975$ & $2287700415$ \\
\hline\hline
$\mathbf{(2,4)}$ & $8$ & $72$ & $648$ & $6048$ & $58320$ & $577368$ & $5837832$ & $60046272$ & $626391792$ \\
\hline
$\mathbf{(4,4)}$ & $4$ & $80$ & $1080$ & $12960$ & $148500$ & $1667952$ & $18574920$ & $206219520$ & $2288739240$ \\
\hline
$\mathbf{(6,4)}$ & $0$ & $24$ & $672$ & $12096$ & $181440$ & $2476656$ & $32006016$ & $399748608$ & $4882643712$ \\
\hline
$\mathbf{(8,4)}$ & $0$ & $0$ & $144$ & $5280$ & $118800$ & $2138400$ & $33929280$ & $497306304$ & $6911094960$ \\
\hline
\end{tabular}}}
\caption{Number of cylinders with the first boundary simple of length $k_1$ and the second boundary ordinary of length $\ell_1$: $[t^{Q}]\,G_{k_1\mid\ell_1}$.}
\end{center}
\end{figure}
\end{center}

\begin{center}
\begin{figure}[h!]
\begin{center}
\scalebox{0.95}{{\scriptsize \begin{tabular}{|c||c|c|c|c|c|c|c|c|c|}
\hline
$(k_1,k_2)$ & $Q = 0$ & $1$ & $2$ & $3$ & $4$ & $5$ & $6$ & $7$ & $8$ \\
\hline\hline
$\mathbf{(1,1)}$ & $1$ & $3$ & $18$ & $135$ & $1134$ & $10206$ & $96228$ & $938223$ & $9382230$ \\
\hline
$\mathbf{(1,3)}$ & $0$ & $3$ & $36$ & $378$ & $3888$ & $40095$ & $416988$ & $4378374$ & $46399392$\\
\hline
$\mathbf{(1,5)}$ & $0$ & $0$ & $15$ & $315$ & $4725$ & $62370$ & $773955$ & $9287460$ & $109306260$ \\
\hline
$\mathbf{(1,7)}$ & $0$ & $0$ & $0$ & $84$ & $2520$ & $49140$ & $793800$ & $11566800$ & $158233824$ \\
\hline
$\mathbf{(1,9)}$ & $0$ & $0$ & $0$ & $0$ & $495$ & $19305$ & $463320$ & $8860995$ & $148551975$ \\
\hline\hline
$\mathbf{(2,2)}$ & $2$ & $12$ & $90$ & $756$ & $6804$ & $64152$ & $625482$ & $6254820$ & $63799164$ \\
\hline
$\mathbf{(2,4)}$ & $0$ & $8$ & $120$ & $1440$ & $16200$ & $178200$ & $1945944$ & $21228480$ & $231996960$ \\
\hline
$\mathbf{(2,6)}$ & $0$ & $0$ & $42$ & $1008$ & $16632$ & $235872$ & $3095820$ & $38864448$ & $474701472$ \\
\hline
$\mathbf{(2,8)}$ & $0$ & $0$ & $0$ & $240$ & $7920$ & $166320$ & $2851200$ & $43623360$ & $621632880$ \\
\hline\hline
$\mathbf{(3,3)}$ & $3$ & $27$ & $252$ & $2457$ & $24705$ & $253935$ & $2653560$ & $28089828$ & $300480678$\\
\hline
$\mathbf{(3,5)}$ & $0$ & $15$ & $270$ & $3690$ & $45900$ & $547560$ & $6395760$ & $73862280$ & $847681200$ \\
\hline
$\mathbf{(3,7)}$ & $0$ & $0$ & $84$ & $2268$ & $41076$ & $628992$ & $8808912$ & $116940348$ & $1499730876$ \\
\hline
$\mathbf{(3,9)}$ & $0$ & $0$ & $0$ & $495$ & $17820$ & $402435$ & $7341840$ & $118587645$ & $1772680140$ \\
\hline\hline
$\mathbf{(4,4)}$ & $4$ & $48$ & $536$ & $5952$ & $66132$ & $735696$ & $8196552$ & $91476864$ & $1022868648$ \\
\hline
$\mathbf{(4,6)}$ & $0$ & $24$ & $504$ & $7728$ & $105336$ & $1354752$ & $16855776$ & $205426368$ & $2469577896$ \\
\hline
$\mathbf{(4,8)}$ & $0$ & $0$ & $144$ & $4320$ & $85200$ & $1401120$ & $20856960$ & $291942144$ & $3922233840$ \\
\hline
\end{tabular}}}
\caption{Number of simple cylinders with boundaries of lengths $(k_1,k_2)$ and $Q$ quadrangles: $[t^{Q}]\,G_{k_1,k_2}$.}
\end{center}
\end{figure}
\end{center}

\begin{center}
\begin{figure}[h!]
\begin{center}
\scalebox{0.95}{{\scriptsize \begin{tabular}{|c||c|c|c|c|c|c|c|c|c|}
\hline
$(k_1,k_2)$ & $Q = 0$ & $1$ & $2$ & $3$ & $4$ & $5$ & $6$ & $7$ & $8$ \\
\hline\hline
$\mathbf{(1,1)}$ & $0$ & $1$ & $9$ & $81$ & $756$ & $7290$ & $72171$ & $729729$ & $7505784$ \\
\hline
$\mathbf{(1,3)}$ & $0$ & $0$ & $6$ & $108$ & $1458$ & $17820$ & $208494$ & $2388204$ & $27066312$\\
\hline
$\mathbf{(1,5)}$ & $0$ & $0$ & $0$ & $35$ & $945$ & $17010$ & $257985$ & $3572100$ & $46845540$ \\
\hline
$\mathbf{(1,7)}$ & $0$ & $0$ & $0$ & $0$ & $210$ & $7560$ & $170100$ & $3084480$ & $49448070$ \\
\hline
$\mathbf{(1,9)}$ & $0$ & $0$ & $0$ & $0$ & $0$ & $1287$ & $57915$ & $1563705$ & $33011550$ \\
\hline\hline
$\mathbf{(2,2)}$ & $0$ & $0$ & $6$ & $108$ & $1458$ & $17820$ & $208494$ & $2388204$ & $27066312$ \\
\hline
$\mathbf{(2,4)}$ & $0$ & $0$ & $0$ & $40$ & $1080$ & $19440$ & $294840$ & $4082400$ & $53537760$ \\
\hline
$\mathbf{(2,6)}$ & $0$ & $0$ & $0$ & $0$ & $252$ & $9072$ & $204120$ & $3701376$ & $59337684$ \\
\hline
$\mathbf{(2,8)}$ & $0$ & $0$ & $0$ & $0$ & $0$ & $1584$ & $71280$ & $1924560$ & $40629600$ \\
\hline\hline
$\mathbf{(3,3)}$ & $0$ & $0$ & $0$ & $48$ & $1296$ & $23328$ & $353808$ & $4898880$ & $64245312$\\
\hline
$\mathbf{(3,5)}$ & $0$ & $0$ & $0$ & $0$ & $315$ & $11340$ & $255150$ & $4626720$ & $74172105$ \\
\hline
$\mathbf{(3,7)}$ & $0$ & $0$ & $0$ & $0$ & $0$ & $2016$ & $90720$ & $2449440$ & $51710400$ \\
\hline
$\mathbf{(3,9)}$ & $0$ & $0$ & $0$ & $0$ & $0$ & $0$ & $12870$ & $694980$ & $21891870$ \\
\hline\hline
$\mathbf{(4,4)}$ & $0$ & $0$ & $0$ & $0$ & $300$ & $10800$ & $243000$ & $4406400$ & $70640100$ \\
\hline
$\mathbf{(4,6)}$ & $0$ & $0$ & $0$ & $0$ & $0$ & $2016$ & $90720$ & $2449440$ & $51710400$ \\
\hline
$\mathbf{(4,8)}$ & $0$ & $0$ & $0$ & $0$ & $0$ & $0$ & $13200$ & $712800$ & $22453200$ \\
\hline
\end{tabular}}}
\caption{\label{FScyl} Number of fully simple cylinders with boundaries of lengths $(k_1,k_2)$ and $Q$ quadrangles: $[t^{Q}]\,H_{k_1,k_2}$.}\label{fig:fscyl}
\end{center}
\end{figure}
\end{center}

\vspace{-0.5cm}

Observe that forcing a boundary of length $1$ or $2$ to be simple does not have any effect in the planar case and therefore the corresponding rows in the first three tables coincide. 
However, imposing that the cylinder is fully simple is much stronger, so in the last table (Figure \ref{fig:fscyl}) all the entries are (much) smaller.

\subsubsection{Tori with 1 boundary}\label{ToriSection}

We compute
\bea\label{tori1}
\omega_{1,1}(z) & = & \frac{z^3(tc^4z^4 + z^2(1 - 5tc^4) + tc^4)}{c(z^2 - 1)^5(1 - 3tc^4)^2}\,\dd z,  \\
\label{tori2}\check{\omega}_{1,1}(z) & = & \frac{3t^2c^9z^5[(3tc^4-2)z^4 + 3tc^4(9tc^4-1)z^2 - 27t^3c^{12}]}{(3tc^4-z^2)^5(1 - 3tc^4)^2}\,\dd z. 
\eea

We present in Figure~\ref{ordinaryTori} the number of tori with $1$ ordinary boundary of perimeter $\ell$ and $Q$ internal quadrangles, as given by Theorem~\ref{ordmaps}.
\begin{center}
\begin{figure}[h!]
\begin{center}
\scalebox{0.88}{{\scriptsize \begin{tabular}{|c||c|c|c|c|c|c|c|c|c|c|}
\hline
$\ell$ & $Q = 0$ & $1$ & $2$ & $3$ & $4$ & $5$ & $6$ & $7$ & $8$ \\
\hline\hline
$\mathbf{2}$ & $0$ & $1$ & $15$ & $198$ & $2511$ & $31266$ & $385398$ &  $4721004$ & $57590271$ \\
\hline
$\mathbf{4}$ & $1$ & $15$ & $198$ & $2511$ & $31266$ & $385398$ & $4721004$ & $57590271$ & $700465482$  \\
\hline
$\mathbf{6}$ & $10$ & $150$ & $1980$ & $25110$ & $312660$ & $3853980$ & $47210040$ & $575902710$ & $7004654820$  \\
\hline
$\mathbf{8}$ & $70$ & $1190$ & $16590$ & $216720$ & $2748060$ & $34286480$ & $423600030$ & $5199957000$ & $63549802260$  \\
\hline
$\mathbf{10}$ & $420$ & $8190$ & $122850$ & $1678320$ & $21925890$ & $279389250$ & $3505914090$ & $43551655560$ & $537235675200$  \\
\hline
$\mathbf{12}$ & $2310$ & $51282$ & $831600$ & $11962566$ & $162074682$ & $2121490602$ & $27174209832$ & $343061095608$ & $4287091638060$  \\ 
\hline
$\mathbf{14}$ & $12012$ & $300300$ & $5261256$ & $79891812$ & $1126377252$ & $15198795612$ & $199385314128$ & $2565902298960$ & $32572738238040$ \\
\hline
\end{tabular}}}
\caption{\label{ordinaryTori}$[t^{Q}]\,F_{\ell}^{[1]}$.}
\end{center}
\end{figure}
\end{center}

For comparison, we present the coefficients $[t^{Q}]\,\check{F}_{k}^{[1]}$ in the same range. Again, it is remarkable that they are all nonnegative integers; Conjecture~\ref{conj} proposes a combinatorial interpretation for them. We also remark that they are always (much) smaller than the corresponding $[t^{Q}]\,F_{\ell}^{[1]}$, and that some of them for small number of quadrangles are $0$, which indicates as before that they may be counting a subclass of ordinary maps.

Due to the strong geometric constraints to form maps with simple boundaries and few internal faces, our observations support that the $[t^{Q}]\,\check{F}_{k}^{[1]}$ may indeed be counting fully simple tori with $Q$ quadrangles.

Moreover, we can give the following simple combinatorial argument to prove the conjecture provides the right answer for $\ell=2$:

\begin{remark}\label{remarkl2}
$F_{2}^{[1]}=H_{1,1}+H_2^{[1]}$.
\end{remark}
\noindent \textit{Proof.}
Ordinary tori with a boundary of length $2$ can be of the following two types:
\begin{itemize}
\item The boundary is simple and the two edges are not identified. This type of ordinary tori are exactly the fully simple tori, counted by $H_2^{[1]}$.

\item The two edges of the boundary are not identified, but the two vertices are, hence the boundary forms a non-trivial cycle of the torus. This type of ordinary tori are obviously in bijection with fully simple cylinders with boundary lengths $(1,1)$, counted by $H_{1,1}$, since one can just glue the vertices of the two boundaries of the cylinder to recover the torus.
\hfill $\Box$
\end{itemize}

Observing our data, we find that for $Q=0,\ldots,8$, we have $[t^Q](\check{F}_2^{[1]}=F_2^{[1]}- H_{1,1})$. Thus, from the remark, we get $[t^Q]\check{F}_2^{[1]}=[t^Q]H_2^{[1]}$, up to at least $Q=8$ quadrangles. We are going to provide now explicit formulas for genus $1$, which will, in particular, help us prove this for all $Q\geq 0$.

Let $\phi_m=c^{2m}\frac{1+(m-1)\sqrt{1-12t}}{1-12t}$, where we recall from \eqref{cSeries} that $c^2= \frac{1 - \sqrt{1 - 12t}}{6t}.$
Then, 
\bea\label{formulasSeries1}
F_{2(m+1)}^{[1]} & = &\frac{(2m+1)!}{6\,n!^2}\,\phi_m, \ \text{ for } m\geq 0,\\
\label{formulasSeries2}\check{F}_{2m}^{[1]} & = &\frac{(3m)!\, t^{m+1}}{4\,m!(2m-1)!}\,\phi_{3m+1}, \ \text{ for } m\geq 1.
\eea
More explicitly, the number of ordinary (and of conjectural fully simple) tori with one boundary can be computed with the following expansion:
\beq\label{explicitphi}
\phi_m = \sum_{n\geq 0} \left(mr_{m,n} + (1-m)\sum_{i=0}^{n-1}r_{m,i}\frac{2\cdot 3^{n-i}}{n-i}\binom{2(n-i-1)}{n-i-1}\right) (3t)^n,
\eeq
where
$$
\frac{c^{2m}}{1-12t}=\sum_{i\geq 0}r_{m,i}(3t)^i,
$$
with 
\beq\label{explicitCoeff}
r_{m,i} = 2^{m+2i}-\frac{1}{2}\sum_{j=0}^{m/2}(-1)^j\binom{m-j-1}{j}\binom{2(m+i-j)}{m+i-j}.
\eeq
The formulas~\eqref{formulasSeries1}-\eqref{formulasSeries2} can be directly extracted from the expressions obtained from TR: \eqref{tori1}-\eqref{tori2}, and the explicit coefficients \eqref{explicitCoeff} can be computed using Lagrange inversion.

Remarkably, both~\eqref{formulasSeries1} and \eqref{formulasSeries2} are given in terms of the same $\phi_m$ with a shifted index, a shifted power of $t$ and different, but simple, combinatorial prefactors. This suggests that if our conjecture~\ref{conj} is true, there is an equivalent underlying combinatorial problem for ordinary and fully simple rooted tori.

We can now confirm our conjectural formula for fully simple rooted tori, for $\ell=2$:
\begin{remark}
$\check{F}_2^{[1]}=H_2^{[1]}.$
\end{remark}
\begin{proof}
From our result for cylinders~\ref{simplTh}, we can extract the simple closed formula:
$$
H_{1,1}=c^6t.
$$
Now, using our explicit expressions~\eqref{formulasSeries1} and \eqref{formulasSeries2}, it can be checked that $F_2^{[1]}=H_{1,1}+\check{F}_2^{[1]}$. Hence the claim follows from our previous Remark~\eqref{remarkl2}.
\end{proof}

\begin{center}
\begin{figure}[h!]
\begin{center}
{\scriptsize \begin{tabular}{|c||c|c|c|c|c|c|c|c|c|c|}
\hline  $k$ & $Q = 0$ & $1$ & $2$ & $3$ & $4$ & $5$ & $6$ & $7$ & $8$ \\
\hline\hline
 $\mathbf{2}$ & 0 & 0 & 6 & 117 & 1755 & 23976 & 313227 & 3991275 & 50084487 \\
\hline $\mathbf{4}$ & 0 & 0 & 0 & 105 & 2925 & 55215 & 885330 & 13009005 & 181316880 \\
\hline $\mathbf{6}$ & 0 & 0 & 0 & 0 & 1260 & 46116 & 1065960 & 19983348 & 332470656 \\
\hline $\mathbf{8}$ & 0 & 0 & 0 & 0 & 0 & 12870 & 585090 & 16073640 &346928670 \\
\hline $\mathbf{10}$ & 0 & 0 & 0 & 0 & 0 & 0 & 120120 & 6531525 & 208243035 \\
\hline $\mathbf{12}$ & 0 & 0 & 0 & 0 & 0 & 0 & 0 & 1058148 & 66997476 \\
\hline $\mathbf{14}$ & 0 & 0 & 0 & 0 & 0 & 0 & 0 & 0 & 8953560 \\
\hline 
\end{tabular}}
\caption{$[t^{Q}]\,H_{k}^{[1]}$.}
\end{center}
\end{figure}
\end{center}

If our conjecture is true, \eqref{explicitphi} would provide the first formula counting a class of fully simple maps for positive genus $g>0$.

\subsubsection{Pairs of pants: evidence for conjecture for even boundary lengths}\label{PairPantsSection}

Very recently, Bernardi and Fusy \cite{BernardiFusy} were able to count, via a bijective procedure, the number of planar fully simple quadrangulations with boundaries of prescribed even lengths. We write their formula here in terms of our notations:
\begin{theorem} 
Let $Q$ be the number of internal quadrangles and $k_1,\ldots,k_n$ positive even integers with $L=\sum_{i=1}^n k_i$ the total boundary length. If $v=2Q-L-n+2 \geq 0$, in which case it counts the number of internal vertices, we have that the number of planar fully simple quadrangulations is given by
\beq\label{evenLengths}
[t^Q]H_{k_1,\ldots,k_n} = \alpha(Q,L,n)\prod_{i=1}^n k_i\binom{\frac{3}{2}k_i}{k_i}.
\eeq
where $\alpha(Q,L,n)\coloneqq\frac{3^{Q-\frac{L}{2}}(e-1)!}{v!(L+Q)!}$, with $e=\frac{L}{2}+2Q$ the total number of edges.
\end{theorem}

This formula reproduces the number of fully simple disks $[t^Q]H_{k}$ and cylinders $[t^Q]H_{k_1,k_2}$ in our Figures~\ref{FSdisks} and~\ref{FScyl}, for even boundary lengths $k, k_1, k_2$. Therefore, it can be recovered from our Theorem~\ref{simplTh} for those base topologies.

More importantly, we checked that our conjectural numbers of fully simple pairs of pants give indeed the right numbers for even boundary lengths $1\leq k_1, k_2, k_3 \leq 8$:
$$
[t^Q] \check{F}_{k_1,k_2,k_3} = H_{k_1,k_2,k_3}, \ \ \ \text{ for } 0\leq Q \leq 8,
$$
thus providing solid evidence for our Conjecture~\ref{conj} in the case of quadrangulations of topology $(0,3)$.

From our data for cylinders and pairs of pants, one can propose a similar formula for fully simple quadrangulations with (any) prescribed boundary lengths:
\beq
[t^Q]H_{k_1,\ldots,k_n} = \alpha(Q,L,n)\prod_{i=1}^n \varepsilon(k_i),
\eeq
where
$$
\varepsilon(k)\coloneqq \begin{cases}
\frac{(3l)!}{l!(2l-1)!}, & \text{ if } k=2l, \\
\sqrt{3}\frac{(3l+1)!}{l!(2l)!}, & \text{ if } k=2l+1. \\
\end{cases}
$$
This formula is a conjectural generalization of \eqref{evenLengths} to include the presence of odd boundary lengths. Observe that since the total length $L$ has to be even, the number of odd lengths also has to be even, hence only factors of $3$ will appear for every extra pair of odd lengths.

Again, the case of $[t^Q]H_{k_1,k_2}$ with $k_1$ and $k_2$ odd can be proved from our Theorem~\ref{simplTh}.

\section{Fully simple pairs of pants}\label{RtransPairPants}
Since we have strong evidence for our Conjecture~\ref{conj} in the case of the topology $(0,3)$, we state here as a consequence a formula relating the generating series of ordinary and fully simple pairs of pants.

Observe that the differential of any meromorphic function $f$ is given by
$$
\dd f(p) = \Res_{z \rightarrow p} B(p,z)\,f(z).
$$
Let us denote collectively the zeroes of $\dd x$ by $\mathbf{a}$, and by $\mathbf{b}$ the zeroes of $\dd w$.
Using the formula in \cite[Theorem 4.1]{EOFg}, we get that
$$
\omega_{3}^{[0]}(z_1,z_2,z_3) = \Res_{z \rightarrow \mathbf{a}} -\frac{B(z,z_1)B(z,z_2)B(z,z_3)}{\dd x(z)\dd w(z)}
$$
and
$$
\check{\omega}_{3}^{[0]}(z_1,z_2,z_3) = \Res_{z \rightarrow \mathbf{b}} -\frac{B(z,z_1)B(z,z_2)B(z,z_3)}{\dd x(z)\dd w(z)}, 
$$
where the residue at a set means the sum over residues at every point in the set.

Using also that for any meromorphic $1$-form $\alpha$ on a compact algebraic curve, which is the case for the spectral curve for maps, we have that
$$
\sum_{p} \Res_{z \rightarrow p} \alpha(z) = 0,
$$
we obtain:
\begin{align}\label{03formula}
& \omega_{3}^{[0]}(z_1,z_2,z_3) + \check{\omega}_{3}^{[0]}(z_1,z_2,z_3) = \Res_{z \rightarrow z_1,z_2,z_3} \frac{B(z,z_1)B(z,z_2)B(z,z_3)}{\dd x(z)\dd y(z)}  \\
& =  \dd_{1}\Big[\frac{B(z_1,z_2)B(z_1,z_3)}{\dd x(z_1)\dd y(z_1)}\Big] + \dd_{2}\Big[\frac{B(z_2,z_1)B(z_2,z_3)}{\dd x(z_2)\dd y(z_2)}\Big] + \dd_3\Big[\frac{B(z_3,z_1)B(z_3,z_2)}{\dd x(z_3)\dd y(z_3)}\Big]. \nonumber
\end{align}

\section{Generalization to stuffed maps}
\label{Section6}

\subsection{Review of definitions}

We introduce stuffed maps as in \cite{Bstuff}, which encompass the previously studied maps since by substitution one may consider them as maps whose elementary cells are themselves maps. We will also see that all our results can be generalized to stuffed maps.
\begin{definition}
An {\it elementary 2-cell} of genus $h$ and $k$ boundaries of lengths $m_1,\ldots, m_k$ is a connected orientable surface of genus $h$ with boundaries $B_1, \ldots, B_k$ endowed with a set $V_i\subset B_i$ of $m_i\geq 1$ vertices. The connected components of $B_i\setminus V_i$ are called {\it edges}. We require that each boundary has a marked edge, called the root, and by following the cyclic order, the rooting induces a labeling of the edges of the boundaries. We say that such an elementary $2$-cell is {\it of topology} $(h,k)$.

A {\it stuffed map} of genus $g$ and $n$ boundaries of lengths $l_1,\ldots, l_n$ is the object obtained from gluing $n$ labeled elementary 2-cells of topology $(0,1)$ with boundaries of lengths $l_1,\ldots,l_n$, and a finite collection of unlabeled elementary 2-cells by identifying edges of opposite orientation and with the same label in such a way that the resulting surface has genus $g$. The labeled cells are considered as {\it boundaries} of the stuffed map, and the marked edges which do not belong to the boundary are forgotten after gluing. 
\end{definition}

A map in the usual sense is a stuffed map composed only of elementary 2-cells with the topology of a disk.

We denote $\widehat{\mathbb{M}}_{l_1,\dots,l_n}^{[g]}$ the set of stuffed maps of genus $g$ and $n$ boundaries of lengths $l_1,\ldots,l_n$.

To every stuffed map $M$ we assign a Boltzmann weight as follows: 
\begin{itemize}
\item a symmetry factor $\left|{\rm Aut}(M)\right|^{-1}$ as previously for maps,
\item a weight $t_{m_1,\ldots,m_k}^h$ per unlabeled elementary 2-cell of genus $h$ and $k$ boundaries, depending symmetrically on the lengths $\bold{m}=(m_1,\ldots,m_k)$.
\end{itemize}

Slightly extending the notation, we add a hat to the previous symbols to denote the generating series of stuffed maps of topology $(g,n)$:
$$
\widehat{W}_n^{[g]}(x_1,\ldots,x_n)= \frac{\delta_{g,0}\delta_{n,1}}{x_1} + \sum_{l_1,\ldots,l_n\geq 1}\left(\prod_{j=1}^n x_j^{-(l_j+1)}\right)\left(\sum_{M\in\widehat{\mathbb{M}}_{l_1,\ldots,l_n}^g} {\rm weight} (M)\right).
$$
We have that $\widehat{W}_n^{[g]}(x_1,\ldots,x_n)\in \mathbb{Q}\left[(x_j^{-1})_j\right][[ (t^h_{\bold{m}})_{\bold{m},h}]]$.

A slight generalization of the permutational model for maps also works for stuffed maps. Let $F$ be the number of unlabeled elementary $2$-cells, considered as the internal faces of the stuffed map. A {\it combinatorial stuffed map} $((\sigma,\alpha),\bigsqcup^F_{p=1} f_p, (h_p)_{p=1}^F)$ consists of the following data:
\begin{itemize}
\item As for maps, a pair of permutations $(\sigma,\alpha)$ on the set of half-edges $H = H^u\sqcup H^{\partial}$, where $\alpha$ is a fixed-point free involution whose cycles represent the edges of the stuffed map, and $\mathcal{C}(\sigma)$ corresponds to the set of vertices. 
The cycles of $\varphi \coloneqq (\sigma\circ\alpha)^{-1}$ are associated to the boundaries of elementary $2$-cells.
\item A partition $\bigsqcup^F_{p=1} f_p$ of $\mathcal{C}(\left.\varphi\right|_{H^u})$, where every part $f_p$ corresponds to an unlabeled elementary $2$-cell with boundaries given by the cycles in $f_p$.
\item A sequence of non-negative integers $(h_p)_{p=1}^F$, where every $h_p$ is the genus of the unlabeled elementary $2$-cell $f_p$.
\end{itemize}

To define the notion of connectedness for stuffed maps, we consider the following equivalence relation on the set of half-edges:
\begin{itemize}
\item[$(i)$] $h \sim \sigma(h)$ and $h \sim \alpha(h)$,
\item[$(ii)$] if $h,h'$ are in two cycles $c,c' \in f_{p}$ for some $p$, then $h \sim h'$.
\end{itemize}
Each equivalence class on $H$ corresponds to a connected component of the stuffed map. We say that a stuffed map  is $\partial$-{\it connected} if each equivalence class has a non-empty intersection with $H^{\partial}$. Observe that the notion of connectedness for maps relies on the equivalence class generated only by $(i)$.

Since the concepts of simplicity and fully simplicity that we introduced for maps in Section \ref{objects} only refer to properties of the boundaries, they clearly extend to stuffed maps because the elementary 2-cells corresponding to boundaries in stuffed maps are imposed to be of the topology of a disk as for maps.
Furthermore, all the results in Sections \ref{disks} and \ref{cyl} for the generating series of maps also do not concern the inner faces, they only affect the boundaries. As a consequence, all these statements are still  true in the more general setting of stuffed maps.

\subsection{Conjecture for maps carrying a loop model}

Usual maps carrying self-avoiding loop configurations are equivalent to stuffed maps for which we allow unlabeled elementary 2-cells to have the topology of a disk (usual faces) or of a cylinder (rings of faces carrying the loops). By equivalence, we mean here an equality of generating series after a suitable change of formal variables.

It is known \cite{BEOn,BEO} that the generating series of ordinary maps with loops obey the topological recursion, with initial data $\omega_{0,1}$ and $\omega_{0,2}$ again given by the generating series of disks and cylinders with loops. As of now, explicit expressions for $\omega_{0,1}$ and $\omega_{0,2}$ are only known for a restricted class of model with loops, \textit{e.g.} those in which loops cross only triangle faces \cite{GK,EKOn,BEOn} maybe taking into account bending \cite{BBG12b}.

Theorems~\ref{ordcyl} and~\ref{ordmaps} apply to maps carrying a loop model. Generalizing Conjecture~\ref{conj}, we propose:
\begin{conjecture}
\label{ConT02} After the symplectic transformation $(x,w) \to (w,x)$ in the initial data of TR for ordinary maps with loops, the TR amplitudes enumerate fully simple maps carrying a loop model.
\end{conjecture}
The argument of Section \ref{Section9} could be an ingredient to prove this conjecture, if one could first establish that the topological recursion governs the topological expansion in the formal matrix model
\thickmuskip=0mu
$$
\dd\mu(M) = \dd M \exp\bigg(N {\rm Tr}(MA) - N\,{\rm Tr}\,\frac{M^2}{2} + \sum_{h \geq 0} \sum_{k \geq 1} \sum_{d \geq 1} N\,\frac{t_{d}}{d}\,{\rm Tr}\,M^{d} + \sum_{d_1,d_2 \geq 1} \frac{t_{d_1,d_2}}{d_1d_2}\,{\rm Tr}\,M^{d_1}\,{\rm Tr}\,M^{d_2}\bigg) 
$$
which depends on the external hermitian matrix $A$.
\thickmuskip=3mu

According to our previous remark, the conjecture is true for disks and cylinders due to the validity of our Theorem~\ref{simplTh}.

\subsection{Vague conjecture for stuffed maps}

It was proved in \cite{Bstuff} that the generating series of ordinary stuffed maps satisfy the so-called blobbed topological recursion, which was axiomatized in \cite{BSblob}. In this generalized version of the topological recursion, the invariants $\omega_{n}^{[g]}$ are determined by $\omega_{1}^{[0]}$ and $\omega_{2}^{[0]}$ as before, and additionally by the so-called blobs $\varphi_n^{[g]}$ for stable topologies, \textit{i.e.}~$2g-2+n >0$.
We conjecture that, after the same symplectic change of variables, and a transformation of the blobs \emph{still to be described}, the blobbed topological recursion will enumerate fully simple stuffed maps. Again, according to our previous remark, this conjecture is true for disks and cylinders (whose expression do not involve the blobs).

\part{Matrix model interpretation}

\section{Ordinary \textit{vs} fully simple for unitarily invariant random matrix models}

We consider an arbitrary measure $\dd \mu(M)$ on the space $\mathcal{H}_N$ of $N\times N$ hermitian matrices which is invariant under conjugation by a unitary matrix. If $\mathcal{O}$ is a polynomial function of the entries of $M$, we denote $\langle \mathcal{O}(M) \rangle$ its expectation value with respect to $\dd \mu(M)$:
$$
\langle \mathcal{O}(M) \rangle = \frac{\int_{\mathcal{H}(N)} \dd \mu(M)\,\mathcal{O}(M)}{\int_{\mathcal{H}(N)} \dd\mu(M)}.
$$
And, if $\mathcal{O}_1,\ldots,\mathcal{O}_n$ are polynomial functions of the entries of $M$, we denote $\kappa_n(\mathcal{O}_1(M),\ldots,\mathcal{O}_n(M))$ their cumulant with respect to the measure $\dd \mu(M)$.

If $\gamma = (c_1 \,  c_2  \, \ldots \, c_{\ell})$ is a cycle of the symmetric group $\mathfrak{S}_{N}$, we denote
$$
\mathcal{P}_{\gamma}(M) \coloneqq M_{c_1,c_2}\, M_{c_2,c_3}\cdots M_{c_{\ell-1},c_{\ell}}\, M_{c_{\ell},c_{1}} = \prod_{m = 1}^{\ell} M_{c_m,\gamma(c_m)}.
$$
We denote $l(\gamma)$ the length of the cycle $\gamma$.

We will be interested in two types of expectation values:
\beq
\label{Obsdisc} \Big\langle \prod_{i = 1}^n {\rm Tr}\,M^{L_i} \Big\rangle \qquad {\rm and} \qquad \Big\langle \prod_{i = 1}^{n} \mathcal{P}_{\gamma_i}(M) \Big\rangle,
\eeq
where $(L_i)_{i = 1}^n$ is a sequence of nonnegative integers, and $(\gamma_i)_{i = 1}^n$ is a sequence of pairwise disjoint cycles in $\mathfrak{S}_{N}$ with $l(\gamma_i)=L_i$ -- the latter imposes $N$ to be larger than $L = \sum_{i = 1}^n l(\gamma_i)$. The first type of expectation value will be called \textit{ordinary}, and the second one \textit{fully simple}. The terms may be used for the ``disconnected'' version \eqref{Obsdisc}, or for the ``connected'' version obtained by taking the cumulants instead of the expectation values of the product. This terminology will be justified by their combinatorial interpretation in terms of ordinary and fully simple maps in Section~\ref{MMs}.

\begin{remark} The unitary invariance of $\mu$ implies its invariance under conjugation of $M$ by a permutation matrix of size $N$. As a consequence, fully simple expectation values only depend on the conjugacy class of the permutation $\gamma_1\cdots\gamma_n$, thus on the partition $\lambda$ which encodes the lengths $\ell_i$ of $\gamma_i$. We can then use the following notations without ambiguity:
\bea
& \Big\langle \prod_{i = 1}^n \mathcal{P}_{\gamma_i}(M) \Big\rangle  =  \big\langle\mathcal{P}_{\lambda}(M)\big\rangle = \Big\langle \prod_{i = 1}^n \mathcal{P}^{(\ell_i)}(M) \Big\rangle, \text{ and} \nonumber \\
& \kappa_n(\mathcal{P}_{\gamma_1}(M),\ldots,\mathcal{P}_{\gamma_n}(M)\big)  =  \kappa_n\big(\mathcal{P}^{(\ell_1)}(M),\ldots,\mathcal{P}^{(\ell_n)}(M)\big). \nonumber
\eea
If $N < L$, we convene that these quantities are zero.
\end{remark}

\subsection{Weingarten calculus}

If the (formal) measure on $M$ is invariant under conjugation by a unitary matrix of size $N$, it should be possible to express the fully simple observables in terms of the ordinary ones -- independently of the measure on $M$. This precise relation will be described in Theorem~\ref{Transi}. We first introduce the representation theory framework which proves and explains this result.

\subsubsection{Preliminaries on symmetric functions}

The character ring of ${\rm GL}_N(\mathbb{C})$ -- \textit{i.e.}~polynomial functions of the entries of $M$ which are invariant by conjugation -- is generated by $p_{l}(M) = {\rm Tr}\,M^{l}$ for $l \geq 0$. It is isomorphic to the ring of symmetric functions in $N$ variables
$$
\mathcal{B}_{N} = \mathbb{Q}[x_1,\ldots,x_N]^{\mathfrak{S}_{N}},
$$
tensored over $\mathbb{C}$.

If $\lambda$ is a partition of an integer $L\geq 0$, we denote $\mathbb{Y}_{\lambda}$ the corresponding Young diagram, $|\lambda|=L$ its number of boxes and $\ell(\lambda)$ its number of rows. By convention, we consider the empty partition $\lambda = \emptyset$ as the partition of $0$. If $\beta$ is a permutation of $L$ elements, we denote $[\beta]$ its conjugacy class, and $|\beta| = |[\beta]|$ the number of elements in this conjugacy class. Associated to the permutation $\beta$, we can form a partition $\lambda=\lambda_{[\beta]}$ -- by collecting the lengths of cycles in $\beta$ -- for which we have $|\mathcal{C}(\beta)| = \ell(\lambda_{[\beta]})$. Recall that actually the set of conjugacy classes in $\mathfrak{S}_{L}$ is in bijection with the set of partitions of $L$. We also denote $t(\beta) = t([\beta]) = L - \ell(\lambda_{[\beta]})$, which can be checked to be the minimal number of transpositions in a factorization of $\beta$.  We sometimes use the notation $C_{\lambda}$ for the conjugacy class in $\mathfrak{S}_{L}$ described by the partition $\lambda$. We denote
$$
|{\rm Aut}\,\lambda |:= \frac{L!}{|C_{\lambda}|},\qquad L = |\lambda|.
$$
The power sum functions $p_{[\beta]}(M) = p_{\lambda}(M) \coloneqq \prod_{i = 1}^{\ell(\lambda)} p_{\lambda_{i}}(M)$  with $\ell(\lambda) \leq N$ form a linear basis of the character ring of ${\rm GL}_N(\mathbb{C})$. Another linear basis is formed by the Schur functions $s_{\lambda}(M)$ with $\ell(\lambda) \leq N$, which have the following expansion in terms of power sum functions:
\beq
\label{SWform} s_{\lambda}(M) = \frac{1}{L!}\,\sum_{\mu \vdash L} |C_{\mu}|\,\chi_{\lambda}(C_{\mu})\,p_{\mu}(M) ,\qquad L = |\lambda|,
\eeq
where $\chi_{\lambda}$ are the characters of $\mathfrak{S}_{L}$. 

The $\mathcal{B}_{N}$ are graded rings, where the grading comes from the total degree of a polynomial. We will work with the graded ring of symmetric polynomials in infinitely many variables, defined as
$$
\mathcal{B} = \mathop{{\rm lim}}_{\infty \leftarrow N} \mathcal{B}_{N}.$$
This is the projective limit using the restriction morphisms $\mathcal{B}_{N + 1} \rightarrow \mathcal{B}_{N}$ sending $p(x_1,\ldots,x_{N + 1})$ to $p(x_1,\ldots,x_{N},0)$. By construction, if $r \in \mathcal{B}$, it determines for any $N \geq 0$ an element $\iota_{N}[r] \in \mathcal{B}_{N}$ by setting
$$
\iota_{N}[r](x_1,\ldots,x_{N}) = r(x_1,\ldots,x_{N},0,0,\ldots).
$$
We often abuse notation and write $r(x_1,\ldots,x_{N})$ for this restriction to $N$ variables. In fact, $\mathcal{B}$ is a free graded ring over $\mathbb{Q}$ with one generator $p_{k}$ in each degree $k \geq 1$. The power sums $p_{\lambda}$ and the Schur elements $s_{\lambda}$ are two homogeneous linear basis for $\mathcal{B}$, abstractly related \textit{via} \eqref{SWform}. A description of the various bases for $\mathcal{B}$ and their properties in relation to representation theory can be found in \cite{Fultonrep}.

Let $\mathcal{B}^{(d)}$ denote the (finite-dimensional) subspace of homogeneous elements of $\mathcal{B}$ of degree $d$. We later need to consider the tensor product of $\mathcal{B}$ with itself, defined by
$$
\mathcal{B} \hat{\otimes} \mathcal{B} := \bigoplus_{d \geq 0} \Big( \bigoplus_{d_1 + d_2 = d} \mathcal{B}^{(d_1)} \otimes \mathcal{B}^{(d_2)}\Big).
$$

\subsubsection{Moments of the Haar measure}

Unlike $\prod_{i = 1}^n {\rm Tr}\,M^{L_i}$, the expression $\prod_{i = 1}^n \mathcal{P}_{\gamma_i}(M)$ is not unitarily invariant. However, the unitary invariance of the measure implies that
\beq
\label{PMun}\Big\langle \prod_{i = 1}^n \mathcal{P}_{\gamma_i}(M) \Big\rangle  = \Big\langle \int_{\mathcal{U}(N)} \dd U\,\prod_{i = 1}^n \mathcal{P}_{\gamma_i}(UMU^{\dagger})\Big\rangle ,
\eeq
where $\dd U$ is the Haar measure on the unitary group. Moments of the entries of a random unitary matrix distributed according to the Haar measure can be computed in terms of representation theory of the symmetric group: this is Weingarten calculus \cite{CollinsWeingarten}. If $N \geq 1$ and $L \geq 0$ are two integers, the Weingarten function is defined as
$$
G_{N,L}(\beta) \coloneqq \frac{1}{L!^2} \sum_{\lambda \vdash L} \frac{\chi_{\lambda}({\rm id})^2 \chi_{\lambda}(\beta)}{s_{\lambda}(1_{N})} ,\ \ \text{ for } \beta \in \mathfrak{S}_{L}.
$$
Note that it only depends on the conjugacy class of $\beta$.

\begin{theorem} \cite{CollinsWeingarten}
\label{UNmoment}
$$
\int_{\mathcal{U}(N)} \dd U \bigg(\prod_{l = 1}^L U_{a_l,b_l} U^{\dagger}_{b_l',a_l'}\bigg) = \sum_{\beta,\tau \in \mathfrak{S}_{L}} \bigg(\prod_{l = 1}^L \delta_{a_l,a_{\beta(l)}'} \delta_{b_{l},b_{\tau(l)}'}\bigg)\,G_{N,L}(\beta\tau^{-1}).
$$
\end{theorem}

\subsubsection{From fully simple to ordinary}

We will use this formula to compute \eqref{PMun}. Let $(\gamma_i)_{i = 1}^n$ be pairwise disjoint cycles:
$$
\gamma_i = (j_{i,1} \ \   j_{i,2} \  \ldots \  j_{i,L_i}).
$$
We denote $H^{\partial} = \bigsqcup_{i = 1}^n \{i\} \times (\mathbb{Z}/L_i\mathbb{Z})$,
$$
L = |H^{\partial} | = \sum_{i = 1}^n L_i
$$
and $\varphi^{\partial} \in \mathfrak{S}_{H^{\partial} }$ the product of the cyclic permutations sending $(i,l)$ to $(i,l + 1\,\,{\rm mod}\,\,L_i)$. Our notations here are motivated by the fact that when we take a certain specialization of the measure $\dd\mu$ in Section \ref{MMs}, $H^{\partial}$ will refer to the set of half-edges belonging to boundaries of a map and $\varphi^{\partial}$ will be the permutation whose cycles correspond to the boundaries. 

\begin{proposition}\label{wein}
$$
\Big\langle \prod_{i = 1}^n \mathcal{P}_{\gamma_i}(M) \Big\rangle = \sum_{\mu \vdash L} \tilde{G}_{N,L}(C_{\mu},\varphi^{\partial})\,\big\langle p_{\mu}(M)\big\rangle = \sum_{\lambda \vdash L} \frac{ \chi_{\lambda}(\varphi^{\partial})\chi_{\lambda}({\rm id})}{L!\,s_{\lambda}(1_{N})}\,\big\langle s_{\lambda}(M) \big\rangle,  \nonumber 
$$
with 
$$
\tilde{G}_{N,L}(C,\beta) = \frac{1}{L!^2} \sum_{\lambda \vdash L} |C|\chi_{\lambda}(C)\chi_{\lambda}(\beta)\,\frac{\chi_{\lambda}({\rm id})}{s_{\lambda}(1_N)},\ \ \text{ for } \beta \in \mathfrak{S}_{L}.
$$
\end{proposition}
\noindent \textbf{Proof.}
If $M$ is a hermitian matrix, we denote $\Lambda$ its diagonal matrix of eigenvalues -- defined up to permutation. We then have
$$
\int \dd U\, \prod_{i = 1}^n \mathcal{P}_{\gamma_i}(UMU^{\dagger}) = \sum_{\substack{1 \leq a_{i,l} \leq N \\ (i,l) \in H^{\partial}}} \int_{\mathcal{U}(N)} \dd U\,\prod_{(i,l) \in H^{\partial}} \Lambda_{a_{i,l}} U_{j_{i,l},a_{i,l}} U^{\dagger}_{a_{i,l},j_{\varphi^{\partial}(i,l)}},
$$
in which we can substitute Theorem~\ref{UNmoment}. We obtain a sum over $\rho,\tau \in \mathfrak{S}_{H^{\partial}}$ of terms involving 
$$
\sum_{\substack{1 \leq a_{i,l} \leq N \\ (i,l) \in H^{\partial}}} \prod_{(i,l) \in H^{\partial}} \Lambda_{a_{i,l}} \delta_{j_{i,l},j_{\rho(\varphi^{\partial}(i,l))}} \delta_{a_{i,l},a_{\tau(i,l)}} = p_{[\tau]}(\Lambda) \prod_{(i,l) \in H^{\partial}} \delta_{j_{i,l},j_{\rho(\varphi^{\partial}(i,l))}}.
$$
As $p_{[\tau]}(\Lambda)$ is unitarily invariant, it is also equal to $p_{[\tau]}(M)$. Since we assumed the $j_{i,l}$ pairwise disjoint, this is non-zero only if $\rho = (\varphi^{\partial})^{-1}$. Therefore
\bea
\label{eq:34}\Big\langle \prod_{i = 1}^n \mathcal{P}_{\gamma_i}(M) \Big\rangle & = & \sum_{\tau \in \mathfrak{S}_{H^{\partial}}} \big\langle p_{[\tau]}(M) \big\rangle\,G_{N,L}\big((\varphi^{\partial})^{-1}\tau^{-1}\big) = \sum_{\mu \vdash L} \big\langle p_{\mu}(M) \big\rangle\,\tilde{G}_{N,L}(C_{\mu},\varphi^{\partial}),
\eea
with
$$
\tilde{G}_{N,L}(C,\beta) = \sum_{\tau \in C} G_{N,L}(\beta^{-1}\tau),
$$
as $\tau$ and $\tau^{-1}$ are in the same conjugacy class. To go further, we recall the Frobenius formula:
\begin{lemma} See e.g. \cite[Theorem 2]{Zagierapp}. If $C_1,\ldots,C_k$ are conjugacy classes of $\mathfrak{S}_{L}$, the number of permutations $\beta_i \in C_i$ such that
$\beta_1 \circ \cdots \circ \beta_{L} = {\rm id}$ is
$$
\mathcal{N}(C_1,\ldots,C_k) = \frac{1}{L!} \sum_{\lambda \vdash L} \frac{\prod_{i = 1}^k |C_i|\chi_{\lambda}(C_i)}{\chi_{\lambda}({\rm id})^{k - 2}}.
$$
\end{lemma}

Since $G_{N,L}(\beta)$ only depends on the conjugacy class of $\beta$, we compute:
\bea
\tilde{G}_{N,L}(C,\beta) & = & \frac{1}{|\beta|} \sum_{\mu \vdash L} \mathcal{N}(C,C_{\mu},[\beta])\,G_{N,L}(C_{\mu}) \nonumber \\ 
& = & \sum_{\mu,\lambda,\lambda' \vdash L} \bigg(\frac{|C|\chi_{\lambda'}(C)\,|C_{\mu}|\chi_{\lambda'}(C_{\mu}) \chi_{\lambda'}(\beta)}{L!\,\chi_{\lambda'}({\rm id})}\bigg)\,\frac{\chi_{\lambda}({\rm id})^2 \chi_{\lambda}(C_{\mu})}{s_{\lambda}(1_N)\,L!^2}. \nonumber
\eea 
The orthogonality of characters of the symmetric group gives
$$
\frac{1}{L!} \sum_{\mu \vdash L} |C_{\mu}|\,\chi_{\lambda'}(C_{\mu})\chi_{\lambda}(C_{\mu}) = \delta_{\lambda,\lambda'}.
$$
Therefore
$$
\tilde{G}_{N,L}(C,\beta) = \frac{1}{L!^2} \sum_{\lambda \vdash L} |C|\chi_{\lambda}(C) \chi_{\lambda}(\beta)\,\frac{\chi_{\lambda}({\rm id})}{s_{\lambda}(1_N)}.
$$
The claim in terms of Schur functions is found by performing the sum over conjugacy classes $C$ in \eqref{eq:34} with the help of \eqref{SWform}. \hfill $\Box$

\subsubsection{Dependence in $N$}

In Theorem~\ref{UNmoment},  the only dependence in the matrix size $N$ comes from the denominator. For a cell  $(i,j)$ in a Young diagram $\mathbb{Y}_{\lambda}$, let ${\rm hook}_{\lambda}(i,j)$ be the hook length at $(i,j)$, where $i = 1,\ldots,\ell(\lambda)$ is the row index and $j = 1,\ldots,\lambda_i$ is the column index. We have the following hook-length formulas, see \textit{e.g.} \cite{Fultonrep}
\bea
\label{dimS} \chi_{\lambda}({\rm id}) & = & \frac{L!}{\prod_{(i,j) \in \mathbb{Y}_{\lambda}} {\rm hook}_{\lambda}(i,j)}, \\
\label{dimU} s_{\lambda}(1_N) & = & \prod_{(i,j) \in \mathbb{Y}_{\lambda}} \frac{(N + j - i)}{{\rm hook}_{\lambda}(i,j)}.
\eea
Therefore
$$
\tilde{G}_{N,L}(C,\beta) = \frac{1}{L!} \sum_{\lambda \vdash L} \frac{|C|\chi_{\lambda}(C)\chi_{\lambda}(\beta)}{\prod_{(i,j) \in \mathbb{Y}_{\lambda}} (N + j - i)}.
$$

\vspace{0.2cm}

The specialization of formula \eqref{SWform} gives another expression of $s_{\lambda}(1_N)$, and thus of $\tilde{G}_{N,L}(C,\beta)$:
$$
s_{\lambda}(1_N) = \frac{N^{L}\chi_{\lambda}({\rm id})}{L!}\bigg(1 + \sum_{\substack{\mu \vdash L \\ C_{\mu} \neq [1]}} N^{-t(C_{\mu})}\,\frac{|C_{\mu}|\chi_{\lambda}(C_{\mu})}{\chi_{\lambda}({\rm id})}\bigg),
$$
where $t(C) = L - \ell(C)$. We obtain that
$$
\tilde{G}_{N,L}(C,\beta) = \frac{N^{-L}}{L!} \sum_{\lambda \vdash L} \sum_{k \geq 0} (-1)^k \sum_{\substack{\mu_1,\ldots,\mu_k \vdash L \\ C_{\mu_i} \neq [1]}}  \frac{|C|\chi_{\lambda}(C)\chi_{\lambda}(\beta)\prod_{i = 1}^k N^{-t(C_{\mu_i})}|C_{\mu_i}|\chi_{\lambda}(C_{\mu_i})}{\chi_{\lambda}({\rm id})^k}.
$$
If we introduce
$$
A^{(d)}_{L,k}(C,\beta) =\frac{1}{|\beta|} \sum_{\substack{\mu_1,\ldots,\mu_k \vdash L \\ \sum_i t(C_{\mu_i}) = d \\ t(C_{\mu_i}) > 0}} \mathcal{N}(C,[\beta],C_{\mu_1},\ldots,C_{\mu_k}),
$$
we can write $\tilde{G}_{N,L}(C,\beta)$ in a compact way
\beq
\label{exptildeG} \tilde{G}_{N,L}(C,\beta) = \sum_{d \geq 0} N^{-L - d}\bigg( \sum_{k = 0}^d (-1)^k A^{(d)}_{L,k}(C,\beta)\bigg).
\eeq
Recall that $\varphi^{\partial}$ has $n$ cycles. \textit{A priori}, $\tilde{G}_{N,L}(C,\varphi^{\partial}) \in O(N^{-L})$, but in fact there are stronger restrictions:
\begin{lemma}
We have a large $N$ expansion of the form:
$$
\tilde{G}_{N,L}(C,\varphi^{\partial}) = \sum_{g \geq 0} N^{-(L + \ell(C) - n + 2g)}\tilde{G}^{(g)}_{L}(C,\varphi^{\partial}).
$$
where $\tilde{G}^{(g)}_{L}$ does not depend on $N$.
\end{lemma}
\noindent\textbf{Proof.} The argument follows \cite{CollinsWeingarten}. $A^{(d)}_{L,k}(C,\varphi^{\partial})$ counts the number of permutations $\tau,\beta_1,\ldots,\beta_k\in \mathfrak{S}_L$ such that $\tau \in C$, $\beta_i \neq {\rm id}$, $\sum_{i = 1}^k t(\beta_i) = d$ and 
\beq
\label{prodperm} \tau \circ \varphi^{\partial}  \circ \beta_1 \circ \cdots \circ \beta_k = {\rm id}.
\eeq
Note that $|t(\sigma) - t(\sigma')| \leq t(\sigma\sigma') \leq t(\sigma) + t(\sigma')$ and thus,
$$
|\ell(C) - n| = |t(\varphi^{\partial}) - t(\tau)| \leq t(\beta_1\cdots \beta_k) \leq \sum_{i = 1}^k t(\beta_i) = d.
$$
Therefore, the coefficient of $N^{-(L + d)}$ in \eqref{exptildeG} is zero unless $d \geq |\ell(C) - n|$. \textit{A fortiori} we must have $d \geq \ell(C) - n$. Also, computing the signature of \eqref{prodperm} we must have
$$
(-1)^{(L - n) + (L - \ell(C)) + \sum_{i} t(\beta_i)} = 1,
$$
i.e. $n - \ell(C) + d$ is even. We get the claim by calling this even integer $2g$.
\hfill $\Box$

\subsection{Transition matrix via monotone Hurwitz numbers}

We dispose of a general theory relating representation theory, Hurwitz numbers and 2d Toda tau hierarchy, which was pioneered by Okounkov~\cite{Okounkov} and to which many authors contributed. For instance, it is clearly exposed in \cite{WorkOkounkov,HarnadGuay}. It relies on three isomorphic descriptions of the vector space $\mathcal{B}$: as the ring of symmetric functions in infinitely many variables, the direct sum of the centers of the group algebras of the symmetric groups, and the charge $0$ subspace of the Fock space (\textit{aka} semi-infinite wedge). After reviewing the aspects of this theory which are relevant for our purposes, we apply it in Section~\ref{SecM} to obtain a nicer form of Proposition~\ref{wein}, namely expressing the transition matrix between ordinary and fully simple observables in terms of monotone Hurwitz numbers.

\subsubsection{The center of the symmetric group algebra}

The center of the group algebra $Z(\mathbb{Q}[\mathfrak{S}_{L}])$ of the symmetric group $\mathfrak{S}_{L}$ has two interesting bases, both labelled by partitions. The most obvious one is defined by
$$
\hat{C}_{\lambda} = \sum_{\gamma \in C_{\lambda}} \gamma, \;\;\; \lambda \vdash L.
$$
The second one is the basis of orthogonal idempotents, which can be related to the first one by
\beq\label{changebasis}
\hat{\Pi}_{\lambda} = \frac{\chi_{\lambda}({\rm id})}{L!} \sum_{\mu \vdash L}\chi_{\lambda}(C_{\mu})\,\hat{C}_{\mu} \ \  \text{ and } \ \  \hat{C}_{\mu} = \frac{1}{|{\rm Aut}\,\mu |} \sum_{\lambda \vdash L} \frac{L!}{\chi_{\lambda}({\rm id})}\chi_{\lambda}(C_{\mu})\hat{\Pi}_{\lambda}.
\eeq
The orthogonality of the characters of $\mathfrak{S}_{L}$ implies that
$$
\hat{\Pi}_{\lambda}\hat{\Pi}_{\mu} = \delta_{\lambda,\mu} \hat{\Pi}_{\lambda}.
$$
The Jucys-Murphy elements of $\mathbb{Q}[\mathfrak{S}_{L}]$ are defined (see \cite{Jucys,Murphy}) by
$$
\hat{J}_{1}=0, \ \  \hat{J}_{k} = \sum_{i = 1}^{k-1} (i \ k), \ \  k=2,\ldots, L.
$$
Their key property is that the symmetric polynomials in the elements $(\hat{J}_{k})_{k = 2}^L$ span $Z(\mathbb{Q}[\mathfrak{S}_{L}])$, see \textit{e.g.} \cite{Meliot}.

\subsubsection{Action of the center on itself and Hurwitz numbers}

Let $r$ be a symmetric polynomial in infinitely many variables, \textit{i.e.} $r \in \mathcal{B}$. We define 
$$
r(\hat{J})\coloneqq r((\hat{J}_k)_{k=2}^L, 0,0,\ldots).
$$
The operator of multiplication by $r(\hat{J})$ in $\mathbb{Q}[\mathfrak{S}_{L}]$ acts diagonally on the basis $\hat{\Pi}_{\lambda}$ of idempotents, with eigenvalues equal to the evaluation on the content of the partition $\lambda$:
$$
r(\hat{J})\hat{\Pi}_{\lambda} = r({\rm cont}(\lambda),0,0,\ldots)\hat{\Pi}_{\lambda},\qquad {\rm cont}(\lambda) = (j-i)_{(i,j) \in \mathbb{Y}_{\lambda}}.
$$
A function of the form $\lambda \mapsto r({\rm cont}(\lambda))$ is called {\it content function}. We denote $r({\rm cont}(\lambda),0,0,\ldots)$ by just $r({\rm cont}(\lambda))$.

In the conjugacy class basis, the action of multiplication by $r(\hat{J})$ has a combinatorial meaning \cite{HarnadGuay}. We define the {\it double Hurwitz numbers} associated with $r$ by the formula
$$
R_{\mu,\lambda} = \frac{1}{L!}\,{\rm tr}\,r(\hat{J})\hat{C}_{\mu}\hat{C}_{\lambda}.
$$
It is obviously symmetric when we exchange the role of $\mu$ and $\lambda$. The following expression is well-known to be equivalent to our definition:
\beq\label{characters_def}
R_{\mu,\lambda} = \frac{1}{|{\rm Aut\, \lambda}||{\rm Aut\, \mu}|}\sum_{\nu \vdash L} \chi_{\nu}(C_{\mu})\, r(\text{cont}\, \nu) \,\chi_{\nu}(C_{\lambda}).
\eeq
We can also characterize double Hurwitz numbers with the following decomposition:
\beq
\label{rjC}r(\hat{J})\hat{C}_{\mu} = \sum_{\lambda \vdash L} |{\rm Aut}\,\lambda|\,R_{\mu,\lambda}\,\hat{C}_{\lambda},
\eeq
which can be checked to be equivalent to \eqref{characters_def} using the formulas \eqref{changebasis} to change between the conjugacy class basis and the idempotent basis. The definition of the Jucys-Murphy elements implies that $|{\rm Aut}\,\lambda|\,R_{\mu,\lambda}$ is a weighted number of paths in the Cayley graph of $\mathfrak{S}_{L}$ generated by transpositions, starting at an arbitrary permutation with cycle type $\mu$ and ending at an (arbitrary but) fixed permutation with cycle type $\lambda$. We can hence define several variations of the Hurwitz numbers using the standard bases of symmetric polynomials evaluated at the Jucys-Murphy elements.

\vspace{0.2cm}

\noindent \textbf{Ordinary.} $p_1(\hat{J})$ is the sum of all transpositions. Therefore
$$
p_1(\hat{J})^{k} = \sum_{\tau_1,\ldots,\tau_k} \tau_1 \cdots \tau_k
$$
and thus $|{\rm Aut}\,\lambda|\,[P_1^{k}]_{\mu,\lambda}$ is the number of sequences $(\tau_1,\ldots,\tau_k,\sigma)$ such that $\tau_i$ are transpositions, $[\sigma] = \mu$ and $\tau_1 \circ \cdots \circ \tau_k \circ \sigma$ is a given permutation with conjugacy class $\lambda$.

\vspace{0.1cm}

\noindent \textbf{Strictly monotone.} For the elementary symmetric polynomial $e_k$, we have
$$
e_k(\hat{J}) = \sum_{\substack{\tau_1,\ldots,\tau_k \\ (\max \tau_i)_{i = 1}^k\,\,\text{strictly increasing}}} \tau_1 \cdots \tau_k,
$$
where $\max \tau$ for a transposition of $a$ and $b$ is defined as $\max(a,b)$. Then, $|{\rm Aut}\,\lambda|\,[E_k]_{\mu,\lambda}$ is the number of strictly monotone $k$-step paths from $C_\mu$ to a given element in $C_\lambda$.

\vspace{0.1cm}

\noindent \textbf{Weakly monotone.} For the complete symmetric polynomials,
$$
h_k(\hat{J}) = \sum_{\substack{\tau_1,\ldots,\tau_k \\ (\max \tau_i)_{i = 1}^k\,\,{\rm weakly}\,\,{\rm increasing}}} \tau_1 \cdots \tau_k.
$$
Therefore, $|{\rm Aut}\,\lambda|\,[H_k]_{\mu,\lambda}$ is the number of weakly monotone $k$-step paths from $C_{\mu}$ to a given element in $C_\lambda$.

\vspace{0.1cm}

We refer to either of the two last cases as {\it monotone} Hurwitz numbers.

\subsubsection{Topological interpretation of Hurwitz numbers}

Hurwitz numbers enumerate branched coverings of $\mathbb{S}^2$ with various constraints.
Let $Y=\{y_1,\ldots,y_{k+2}\}$ be an ordered finite set of points in the topological sphere $\mathbb{S}^{2}$. There is a one-to-one correspondence between the set of topological branched coverings of the sphere $\mathbb{S}^{2}$ with ramification locus $Y$ and with $L$ sheets, and the set of representations of $\pi_1(\mathbb{S}^2\setminus Y)$ into $\mathfrak{S}_{L}$. Actually, loops $\gamma_j$ around every special point $y_j$ in $Y$ give a presentation of the fundamental group:
$$
\pi_1(\mathbb{S}^2 \setminus Y) = \Big\{\gamma_1,\ldots,\gamma_{k+2} \quad \Big|\quad  \prod_{j = 1}^{k+2} \gamma_j = {\rm id}\Big\}.
$$
and hence a representation of $\pi_1(\mathbb{S}^2\setminus Y)$ into $\mathfrak{S}_{L}$ is a sequence of permutations $(\beta_i)_{j = 1}^{k+2}$ such that
\beq
\label{Factor}\beta_1 \circ \cdots \circ \beta_{k+2} = {\rm id}.
\eeq
Starting from a representation of $\pi_1(\mathbb{S}^2\setminus Y)$ into $\mathfrak{S}_{L}$, we can construct a branched covering of the sphere
\beq
\label{branchedcover} \Big(\bigsqcup_{i = 1}^L \mathbb{S}^2\Big)/\sim \,\,\,\stackrel{\pi}{\longrightarrow}\,\,\, \mathbb{S}^2
\eeq
as follows.
For every $y_j \in Y$, we denote $y_j^{(i)} \in \pi^{-1}(y_j)$ its preimage in the $i$-th copy of $\mathbb{S}^2$. In the source of \eqref{branchedcover}, the equivalence relation identifies the points $y_j^{(i)}$ and $y_j^{(\beta_j(i))}$  for every $y_j \in Y$. In this way, we get $|\pi^{-1}(y_j)|=|\mathcal{C}(\beta_j)|$. Conversely, if we are given a branched covering, the monodromy representation indeed gives a representation of $\pi_1(\mathbb{S}^2\setminus Y)$ into $\mathfrak{S}_{L}$. The total space of the covering is connected if and only if $\beta_1,\ldots,\beta_{k+2}$ act transitively on $\llbracket 1,L \rrbracket$.

Let $\lambda_j$ be the partition of $L$ describing the conjugacy class of $\beta_j$, that is also the ramification profile over $y_j$. The Euler characteristic of the total space is given by the Riemann-Hurwitz formula:
$$
\chi = 2L - \sum_{j = 1}^{k+2} t(\lambda_j).
$$
In particular, if there are $k$ transpositions and two conjugacy classes described by partitions $\lambda,\mu$ of $L$, we have
$$
\chi = \ell(\lambda) + \ell(\mu) - k.
$$
This special case corresponds to the one described previously: double Hurwitz numbers $[R_k]_{\lambda,\mu}$, which count the number of ways ${\rm id} \in\mathfrak{S}_L$ may be factorized into a product of $k$ transpositions and two permutations of cycle types given by $\lambda$ and $\mu$, and equivalently the number of $L$-sheeted branched coverings of the sphere with ramification profile given by $\lambda$ and $\mu$ over two points and $k$ other simple ramifications. Various constraints can be put on the factorization \eqref{Factor}, which can in turn be interpreted as properties of the branched coverings. The symmetry factor $|{\rm Aut}\,\lambda|^{-1}$ is reflected in the fact that we count branched covers up to automorphism -- in particular up to relabelling of the sheets.

\subsubsection{Hypergeometric tau-functions}

We consider Frobenius' characteristic map
$$
\Phi\,:\,\bigoplus_{L \geq 0} Z(\mathbb{Q}[\mathfrak{S}_{L}]) \longrightarrow \mathcal{B}
$$
defined by
$$
\Phi(\hat{C}_{\lambda}) = \frac{p_{\lambda}}{|{\rm Aut}\,\lambda|} = \frac{|C_{\lambda}|\,p_{\lambda}}{L!},\qquad |\lambda| = L.
$$
This map is linear and it is a graded isomorphism -- namely it sends $Z(\mathbb{Q}[\mathfrak{S}_{L}])$ to $\mathcal{B}^{(L)}$. This definition together with the formula \eqref{SWform} and the formula for change of basis \eqref{changebasis} imply that
$$
\Phi(\hat{\Pi}_{\lambda}) = \frac{\chi_{\lambda}({\rm id})}{L!}\,s_{\lambda}.
$$
The action of $Z(\mathbb{Q}[\mathfrak{S}_{L}])$ on itself by multiplication can then be assembled into an action of $\mathcal{B}$ on itself. Concretely, if $r \in \mathcal{B}$ this action is given by
$$
r(\hat{J}) := \Phi \circ\bigg(\bigoplus_{L \geq 0} r\big((\hat{J}_{k})_{k = 2}^L,0,0,\ldots\big)\bigg)\circ \Phi^{-1}.
$$

\begin{definition} 
A {\it hypergeometric tau-function} is an element of $\mathcal{B} \hat{\otimes} \mathcal{B}$ of the form $\sum_{\lambda} A_{\lambda}\,s_{\lambda} \otimes s_{\lambda}$ for some scalar-valued $\lambda \mapsto A_{\lambda}$ function which is a content function.
\end{definition}

\begin{remark} A \emph{2d Toda tau-function} is an element of $\mathcal{B} \hat{\otimes} \mathcal{B}$ which satisfies the Hirota bilinear equations -- these are the analog of Pl\"ucker relations in the Sato Grassmannian. It is known that, if $A$ is a content function, $\sum_{\lambda}  A_{\lambda} s_{\lambda} \otimes s_{\lambda}$ is a 2d Toda tau-function \cite{Carrell,OrlovSch}. We adopt here the name ``hypergeometric'' coined by Orlov and Harnad for those particular 2d Toda tau-functions. Let us mention there exist 2d Toda tau-functions which are diagonal in the Schur basis but with coefficients which are not content functions. 
\end{remark}

We can identify $\mathcal{B}\hat{\otimes} \mathcal{B}$ with the ring of symmetric functions in two infinite sets of variables $\mathbf{z}=(z_1,z_2,\ldots)$ and $\widetilde{\mathbf{z}}=(\widetilde{z}_1,\widetilde{z}_2,\ldots)$.
There is a trivial hypergeometric tau-function:
$$
\mathcal{T}_{\emptyset} := \exp\Big(\sum_{k \geq 1} \frac{1}{k}\,p_{k}(\mathbf{z})\,p_{k}(\widetilde{\mathbf{z}})\Big) = \prod_{i,j = 1}^{\infty} \frac{1}{1 - z_{i}\widetilde{z}_{j}} = \sum_{\lambda} s_{\lambda}(\mathbf{z})\,s_{\lambda}(\widetilde{\mathbf{z}}) = \sum_{\mu} \frac{1}{|{\rm Aut}\,\mu|}\,p_{\mu}(\mathbf{z})\,p_{\mu}(\widetilde{\mathbf{z}}),
$$
where the two last sums are over all partitions and for the equality in the middle we have used Cauchy-Littlewood formula \cite[Chapter 1]{Macdonald}. 

An element $r \in \mathcal{B}$ acts on the set of hypergeometric tau-functions by action on the first factor via $r(\hat{J}) \otimes {\rm Id}$.  More concretely, the action on $\mathcal{T}_{\emptyset}$ reads
\beq
\label{TauQ2} \mathcal{T}_{r} = \sum_{\lambda} r({\rm cont}\,\lambda)\,s_{\lambda} \otimes s_{\lambda} = \sum_{L \geq 0} \sum_{|\lambda| = |\mu| = L} R_{\lambda,\mu}\,p_{\lambda}\otimes p_{\mu}\,.
\eeq

\subsubsection{Main result}
 \label{SecM}
We prove that the transition matrix from ordinary to fully simple expectation values is given by double, weakly monotone Hurwitz numbers (with signs), while the transition matrix from fully simple to ordinary is given by the double, strictly monotone Hurwitz numbers.

\begin{theorem}
\label{Transi}With respect to any $\mathcal{U}_{N}$-invariant measure on the space $\mathcal{H}_N$ of $N \times N$ hermitian matrices, we obtain
\bea
\frac{\big\langle \mathcal{P}_{\lambda}(M)\big\rangle}{|{\rm Aut}\,\lambda|} & = & \sum_{\mu \vdash |\lambda|} N^{-|\mu|} \Big(\sum_{k \geq 0} (-N)^{-k} [H_{k}]_{\lambda,\mu}\Big)\big\langle p_{\mu}(M) \big\rangle, \nonumber \\
\frac{\big\langle p_{\mu}(M) \big\rangle}{|{\rm Aut}\,\mu|} & = & \sum_{\lambda \vdash |\mu|} N^{|\lambda|} \Big(\sum_{k \geq 0} N^{-k}[E_{k}]_{\mu,\lambda}\Big) \big\langle \mathcal{P}_{\lambda}(M) \big\rangle, \nonumber 
\eea
where $[E_k]_{\mu,\lambda}$ (resp. $[H_k]_{\lambda,\mu}$) are the double Hurwitz numbers related to the elementary symmetric (resp. complete symmetric) polynomials.
\end{theorem}

\noindent \textbf{Proof.} We introduce an auxiliary diagonal matrix $\widetilde{M}$ and deduce from Proposition \ref{wein} and Equation~\eqref{SWform} that
\beq
\label{idgen} \frac{1}{L!} \sum_{\lambda \vdash L} |C_{\lambda}|\, p_{\lambda}(\widetilde{M}) \big\langle \mathcal{P}_{\lambda}(M) \big\rangle = \sum_{\mu \vdash L} \frac{\chi_{\mu}({\rm id})}{L!\,s_{\mu}(1_N)}\,s_{\mu}(\widetilde{M})\,\big\langle s_{\mu}(M)\big\rangle.
\eeq
The formulas \eqref{dimS}-\eqref{dimU} show that $\frac{\chi_{\mu}({\rm id})}{L!\,s_{\mu}(1_N)}$ is a content function coming from the complete symmetric polynomials\footnote{We would like to remark that we could have used a result of Novak \cite[Theorem 1.1]{NovakJMelements} phrasing our \eqref{contentfct} in an interesting form, but we keep using our formula since it is more elementary.}
\beq\label{contentfct}
\frac{\chi_{\mu}({\rm id})}{L!\,s_{\mu}(1_N)} = \prod_{(i,j) \in \mathbb{Y}_{\mu}} \frac{1}{N + {\rm cont}(i,j)} = N^{-|\mu|} \sum_{k \geq 0} (-N)^{-k}\,h_k({\rm cont}\,\mu).
\eeq
We denote $r_N$ the corresponding element of $\mathcal{B}[[N^{-1}]]$. The identity \eqref{idgen} then translates into:
\beq\label{twistedtau}
 \frac{1}{L!} \sum_{\lambda \vdash L}|C_{\lambda}|\, p_{\lambda}(\widetilde{M})\,\big\langle \mathcal{P}_{\lambda}(M)\big\rangle
=  \sum_{\mu \vdash L}  r_N({\rm cont}\,\,\mu)\,s_{\mu}(\widetilde{M})  \big\langle s_{\mu}(M) \big\rangle.
\eeq
To interpret these expressions as $r_N$ acting on $\mathcal{T}_{\emptyset}$ as in \eqref{TauQ2}, we remind that $\mathcal{T}$ in $\mathcal{B} \hat{\otimes} \mathcal{B}$ can be seen as a function of two sets of infinitely many variables $\Lambda$ and $\widetilde{\Lambda}$. Moreover, we consider $\mathcal{T}$ evaluated at two matrices $M$ and $\widetilde{M}$ of size $N$, by substituting $\Lambda$ (resp. $\widetilde{\Lambda}$) by the set of $N$ eigenvalues of $M$ (resp. $\widetilde{M}$) completed by infinitely many zeros. We then write $\mathcal{T}(M,\widetilde{M})$ to stress that we have a function of two matrices. In this way, we identify the summation of \eqref{twistedtau} over $L \geq 0$ with $\langle \mathcal{T}_{r_N}(M,\widetilde{M})\rangle$, where the expectation value is taken with respect to any unitarily-invariant measure on $M$ -- while $\widetilde{M}$ is a matrix-valued parameter. 

Now comparing with \eqref{TauQ2}, we find that
\beq
\label{2ndequation} \frac{\big\langle \mathcal{P}_{\lambda}(M) \big\rangle}{|{\rm Aut}\,\lambda|} = \sum_{\mu \vdash |\lambda|} (R_N)_{\lambda,\mu}\, \big\langle  p_{\mu}(M) \big\rangle,
\eeq
which yields the first formula we wanted to prove. To obtain the second formula, we observe that
$$
s_N({\rm cont}\,\,\lambda) = \prod_{(i,j) \in \mathbb{Y}_{\lambda}} (N + {\rm cont}(i,j)) = N^{|\lambda|} \sum_{k \geq 0} N^{-k}\,e_k({\rm cont}\,\,\lambda)
$$
defines an element $s_N \in \bigoplus_{d \geq 0} \mathcal{B}^{(d)} \otimes (N^{d}\cdot\mathbb{C}[[N^{-1}]])$, which is inverse to $r_N$ in $\mathcal{B}[N,N^{-1}]]$. We denote $[S_{N}]_{\lambda,\mu}$ the Hurwitz numbers it determines via \eqref{rjC}. 
If we act by $s_N(\hat{J})$ on $\mathcal{T}_{r_N}$, we recover the trivial tau-function:
\beq
\label{T222}\langle \mathcal{T}_{\emptyset}(\widetilde{M},M) \rangle = \sum_{\mu} \frac{p_{\mu}(\widetilde{M})}{|{\rm Aut}\,\mu|}\,\langle p_{\mu}(M) \rangle.
\eeq
On the other hand, representing this action in the power sum basis using \eqref{TauQ2} and \eqref{twistedtau} yields
\beq
\label{T222a}\langle \mathcal{T}_{\emptyset}(\widetilde{M},M) \rangle =\sum_{L \geq 0} \sum_{|\lambda| = |\mu| = L} [S_{N}]_{\mu,\lambda}\,  p_{\mu}(\widetilde{M}) \langle \mathcal{P}_{\lambda}(M) \rangle.
\eeq
Finally, we can identify the coefficients of \eqref{T222} and \eqref{T222a} to obtain the desired formula. \hfill $\Box$

\vspace{0.15cm}

With our proof, we obtain some intermediate formulas that will be useful later. However, for the derivation of Theorem \ref{Transi} it is not crucial to write our generating series in the form of $\tau$-functions. Using Proposition \ref{wein}, \eqref{contentfct} and the expression \eqref{characters_def} for Hurwitz numbers, Theorem \ref{Transi} is straightforward. Apart from the reason already mentioned, we also consider it is interesting to illustrate the relation with the world of $\tau$-functions.

\subsection{Relation with the matrix model with external field}

The Itzykson-Zuber integral \cite{IZ} is a function of an integer $N$ and two matrices $A$ and $B$ of size $N$ defined by
\beq
\label{IZint} \mathcal{I}_{N}(A,B) := \int_{U_N} \dd U\,\exp\big[N\,{\rm Tr}(AUBU^{\dagger})\big],
\eeq
where $\dd U$ is the Haar measure on $U_N$ normalized to have mass $1$. It admits a well-known expansion in terms of characters of the unitary group, \textit{i.e.} Schur functions:
\begin{theorem} \cite[Eq. (4.6)]{Balentekin}\label{balen}
$$
\mathcal{I}_{N}(A,B) = \sum_{\lambda} \frac{N^{|\lambda|}\chi_{\lambda}({\rm id})}{L!\,s_{\lambda}(1_{N})}\,s_{\lambda}(A)\,s_{\lambda}(B).
$$
\end{theorem}
For any unitarily invariant measure $\mu$ on $\mathcal{H}_{N}$, we define
$$
\check{Z}(A) = \int_{\mathcal{H}_{N}} \dd\mu(M)e^{N\,{\rm Tr}(AM)}.
$$

\begin{corollary}
\label{gudn}We denote $\langle \cdot \rangle$ the expectation value and $\kappa_n(\cdot)$ the n-th order cumulant with respect to any unitarily invariant measure $\mu$ on $M\in\mathcal{H}_{N}$.
We have the formulas
\bea  
\frac{\check{Z}(A)}{\check{Z}(0)} & = & 1+ \sum_{\lambda\neq \emptyset} \frac{|C_{\lambda}|}{|\lambda|!}\,N^{|\lambda|} p_{\lambda}(A) \big\langle\mathcal{P}_{\lambda}(M)\big\rangle = \big\langle \mathcal{I}_{N}(A,M) \big\rangle, \nonumber \\
\ln\Big(\frac{\check{Z}(A)}{\check{Z}(0)}\Big) & = & \sum_{n \geq 1}\frac{1}{n!} \sum_{\ell_1,\ldots,\ell_n \geq 1} N^{L}\, \kappa_n
\big(\mathcal{P}^{(\ell_1)}(M),\ldots,\mathcal{P}^{(\ell_n)}(M)\big)\,\prod_{i = 1}^n \frac{p_{\ell_i}(A)}{\ell_i}, \nonumber
\eea  
where we recall that the corresponding expectation value is zero whenever $|\lambda|$ or $L\coloneqq\sum_{i} \ell_i$ exceeds $N$.
\end{corollary}
\noindent \textbf{Proof.} Comparing Theorem \ref{balen} with \eqref{twistedtau} gives the first line. We introduce the factor which allows us to go from (ordered) tuples $(\ell_1,\ldots,\ell_n)$ with $L\coloneqq\sum_{i=1}^n \ell_i$ to (unordered) partitions of $L$:
\beq\label{g}
g_{\lambda}\coloneqq \frac{\ell(\lambda)!}{\prod_{i=1}^L m_i(\lambda) !} = \frac{\ell(\lambda)!\, |C_{\lambda}|\, \prod_{i=1}^{\ell(\lambda)} \lambda_i}{|\lambda| !},
\eeq
where $L = \sum_{i=1}^L i\, m_i(\lambda) = \sum_{i=1}^n \lambda_i$. If we replace now the sum over partitions by the sum over tuples of positive integers multiplying by $\frac{1}{g_{\lambda}}$, we find
\bea
\frac{\check{Z}(A)}{\check{Z}(0)} & = & 1 + \sum_{n \geq 1} \frac{1}{n!} \sum_{\ell_1,\ldots,\ell_n \geq 1}  N^{L}\,\big\langle \mathcal{P}^{(\ell_1)}(M)\cdots \mathcal{P}^{(\ell_n)}(M)\big\rangle\,\prod_{i = 1}^n\frac{p_{\ell_i}(A)}{\ell_i} \nonumber \\ 
& = & \left\langle \exp\left(\sum_{i\geq 1} \frac{N^{\ell_i} p_{\ell_i}(A)}{\ell_i} \mathcal{P}^{(\ell_i)}(M)\right) \right\rangle. \nonumber
\eea
Taking the logarithm gives precisely the cumulant generating series as in the second formula.
\hfill  $\Box$

\vspace{0.2cm}

In other words, the fully simple observables for the matrix model $\mu$ are naturally encoded in the corresponding matrix model with an external source $A$.  Compared to Theorem~\ref{Transi}, this result is in agreement with the combinatorial interpretation of the Itzykson-Zuber integral in terms of double monotone Hurwitz numbers \cite{HCIZ}.

\section{Combinatorial interpretation in terms of ordinary and fully simple maps}\label{MMs}
\label{Matcomb}
The relation between ordinary and fully simple observables through monotone Hurwitz numbers is universal in the sense that it does not depend on the unitarily invariant measure considered. This section is devoted to the relation between matrix models and the enumeration of maps for a specific unitarily invariant measure. This relation is well-known for ordinary maps, but here we give a detailed derivation gluing polygons, that is working directly with maps, in contrast to the classical derivation in physics which is in terms of gluing stars, that is working on the dual. We include this calculation to give a different (but completely equivalent) detailed derivation which will also make clearer the refinement of the argument that we need to provide a matrix model for fully simple maps. This specialization motivated our study of the general ordinary and fully simple observables.

We introduce the Gaussian probability measure on the space $\mathcal{H}_N$ of $N \times N$ hermitian matrices:
$$
\dd\mu_0(M) = \frac{\dd M}{Z_0}\,e^{-N\mathrm{Tr}\,\frac{M^2}{2}},\qquad Z_0 = \int_{\mathcal{H}_{N}} \dd M\,e^{-N\mathrm{Tr}\,\frac{M^2}{2}},
$$
and the generating series:
\bea
\tilde{T}_{h,k}(w_1,\ldots,w_k) &=& \sum_{m_1,\ldots,m_k \geq 1} \frac{t^h_{m_1,\ldots,m_k}}{m_1\cdots m_k}\,\prod_{i = 1}^k w_i^{m_i},\nonumber \\
T_{h,k}(w_1,\ldots,w_k) &= &\tilde{T}_{h,k}(w_1,\ldots,w_k) - \delta_{h,0}\delta_{k,1}\frac{w_1^2}{2}. \nonumber
\eea
We consider the formal measure
\beq
\label{eq:mu0}\dd\mu(M) = \frac{\dd M}{Z_0}\,\exp\left(\sum_{h\geq 0,k\geq 1} \frac{N^{2 - 2h - k}}{k!} \mathrm{Tr}\,T_{h,k}\left(M^{(1)}_k,\ldots,M^{(k)}_k\right)\right),
\eeq
where $M^{(i)}_k\coloneqq\bigotimes_{j=1}^{i-1} I_N \otimes M \otimes \bigotimes_{j=i+1}^{k} I_N$, in the sense that the expectation value of any polynomial function of $M$ with respect to this measure is defined as a formal series in the $t$'s.

For a combinatorial map $(\sigma, \alpha)$, we consider here a special structure for the set of half-edges $H= H^u \sqcup H^{\partial}$ which will be convenient for our derivation:
$$
H^{\partial} = \bigsqcup_{i = 1}^n \{i\} \times (\mathbb{Z}/L_i\mathbb{Z}),\qquad H^u = \bigsqcup_{m = 1}^r \{m\} \times (\mathbb{Z}/k_m\mathbb{Z}).
$$
The permutation $\varphi \coloneqq (\sigma\circ\alpha)^{-1}$ acting on $H$, whose cycles correspond to faces of the map, is hence given by $\varphi((i, l))=(i,l+1)$. With this special structure of the set of half-edges, counting the number of relabelings of $H^u$ amounts to choosing an order of the unmarked faces and a root for each of them. Therefore
$$
{\rm Rel}(\sigma,\alpha) = r!\,\prod_{m= 1}^r k_m.
$$

\subsection{Ordinary usual maps}
\label{Maps}
Consider first the case where $T_{h,k} = 0$ for $(h,k) \neq (0,1)$, \textit{i.e.}
\beq
\label{eq:mes}\dd\mu(M) = \frac{\dd M}{Z_{\rm GUE}}\,\exp\left\{N\,{\rm Tr}\Big(-\frac{M^2}{2} + \sum_{k \geq 1} \frac{t_k\,M^k}{k}\Big)\right\},\qquad Z = \int_{\mathcal{H}(N)} \dd\mu(M).
\eeq
We denote $\langle \mathcal{\cdot} \rangle_{{\rm GUE}}$ the expectation value with respect to the Gaussian measure $\dd\mu_0$. The matrix elements have covariance:
\beq
\label{contrac} \langle M_{a,b}M_{c,d} \rangle_{{\rm GUE}} = \frac{1}{N}\delta_{a,d}\delta_{b,c}.
\eeq

Let $(L_i)_{i = 1}^n$ be a sequence of nonnegative integers, and $L = \sum_{i = 1}^n L_i$. The expectation values with respect to $\dd\mu$ are computed, as formal series in $(t_k)_k$:
\beq\label{eq:ordev}
\Big\langle \prod_{i = 1}^n {\rm Tr}\,M^{L_i} \Big\rangle =  \frac{1}{Z} \sum_{r \geq 0} \sum_{k_1,\ldots,k_r \geq 1} \frac{N^r}{r!} \prod_{m = 1}^r \frac{t_{k_m}}{k_m} \sum_{\substack{1 \leq j_{h} \leq N, \\ h \in H}} \Big\langle \prod_{h\in H} M_{j_h,j_{\varphi(h)}}\Big\rangle_{{\rm GUE}}. 
\eeq

With the help of Wick's theorem for the Gaussian measure and (\ref{contrac}), we obtain:
$$
\Big\langle \prod_{h\in H} M_{j_h,j_{\varphi(h)}}\Big\rangle_{{\rm GUE}} =  N^{-\frac{|H|}{2}} \sum_{\alpha \in \mathfrak{I}_H} \prod_{h\in H} \delta_{j_{h},j_{\alpha(\varphi(h))}}.
$$
where $\mathfrak{I}_H\subset \mathfrak{G}_H$ is the set of all fixed-point free involutions, \textit{i.e.} all pairwise matchings, on $H$.

We observe that the product on the right hand side is $1$ if $h\mapsto j_h$ is constant over the cycles of $\alpha\circ\varphi$, otherwise it is $0$. Therefore,
$$
\sum_{\substack{1 \leq j_{h} \leq N, \\ h \in H}} \Big\langle \prod_{h\in H} M_{j_h,j_{\varphi(h)}}\Big\rangle_{{\rm GUE}} =  N^{-\frac{|H|}{2}} \sum_{\alpha \in \mathfrak{I}_H} N^{|\mathcal{C}(\alpha\circ\varphi)|}.
$$
To recognize \eqref{eq:ordev} as a sum over combinatorial maps, we let $\alpha\in\mathfrak{I}_H$ correspond to the edges of maps whose faces are given by cycles of $\varphi$, and $\sigma\coloneqq (\alpha\circ\varphi)^{-1}$, whose cycles will correspond to the vertices. Observe that $2|\mathcal{C}(\alpha)|=|H|$.

$$
\Big\langle \prod_{i = 1}^n {\rm Tr}\,M^{L_i} \Big\rangle = \frac{1}{Z} \sum_{(\sigma,\alpha)} \frac{N^{|\mathcal{C}(\sigma)| - |\mathcal{C}(\alpha)| + |\mathcal{C}(\varphi)|-n}}{{\rm Rel}\,(\sigma,\alpha)}\,\prod_{f \in \mathcal{C}(\left.\varphi\right|_{H^u})} t_{\ell(f)},
$$
where the sum is taken over non-connected combinatorial maps $(\sigma,\alpha)$ with $n$ boundaries of lengths $L_1,\ldots,L_n$.

To transform this sum over combinatorial maps into a generating series for (unlabeled) maps, we have to multiply by the number of combinatorial maps which give rise to the same unlabeled combinatorial map $[(\sigma,\alpha)]$, which is ${\rm Gl}(\sigma,\alpha)= \frac{{\rm Rel}(\sigma,\alpha)}{|{\rm Aut}(\sigma,\alpha)|}$, as we explained in section \ref{aut}.

A similar computation can be done separately for $Z$, and we find it is the generating series of maps with empty boundary. The contribution of connected components without boundaries factorizes in the numerator and, consequently,
$$
\Big\langle \prod_{i = 1}^n {\rm Tr}\,M^{L_i} \Big\rangle = \sum_{\substack{\partial \text{-connected}\  \mathcal{M} = [(\sigma,\alpha)] \\ \text{with } \partial \text{ lengths}\,\,(L_i)_{i = 1}^n}} \frac{N^{\chi(\sigma,\alpha)}}{|{\rm Aut}(\sigma,\alpha)|}\,\prod_{f \in \mathcal{C}(\left.\varphi\right|_{H^u})}t_{\ell(f)}.
$$
We remark that the power of $N$ sorts maps by their Euler characteristic. 
Finally, a standard argument shows that taking the logarithm for closed maps or the cumulant expectation values for maps with boundaries, we obtain the generating series of connected maps:

\begin{proposition}\cite{EynardBook}
$$
\ln Z = \sum_{g \geq 0} N^{2-2g} F^{[g]}, \quad \quad
\kappa_n({\rm Tr}\,M^{L_1},\ldots,{\rm Tr}\,M^{L_n}) = \sum_{g \geq 0} N^{2-2g-n} F^{[g]}_{L_1,\ldots,L_n}.
$$
\end{proposition}
This kind of results appeared first for planar maps in \cite{BIPZ}, but for a modern and general exposition see also \cite{EynardBook}.

\subsection{Ordinary stuffed maps}

For the general formal measure
\bea
\label{eq:mesS}\dd\mu(M) & = & \frac{\dd M \,e^{-N\,{\rm Tr} \frac{M^2}{2}}}{Z_{\rm GUE}}\,\exp\left(\sum_{h\geq 0,k\geq 1} \frac{N^{2 - 2h - k}}{k!} \mathrm{Tr}\,\tilde{T}_{h,k}\left(M^{(1)}_k,\ldots,M^{(k)}_k\right)\right) \nonumber \\
& = & \frac{\dd M \,e^{-N\,{\rm Tr} \frac{M^2}{2}}}{Z_{\rm GUE}}\,\exp\left(\sum_{h\geq 0,k\geq 1} \frac{N^{2 - 2h - k}}{k!} N^{2 - 2h - k}\sum_{m_1,\ldots,m_k \geq 1} t^h_{m_1,\ldots,m_k}\,\prod_{i = 1}^k \frac{\mathrm{Tr}\, M^{m_i}}{m_i}\right),
\eea
the expectation values are generating series of stuffed maps. We denote here $Z_S$, $\langle \cdot \rangle_S$ and $\kappa_n(\cdot)_S$ the partition function, the expectation value and the n-th order cumulant expectation values with respect to this general measure to distinguish these more general expressions from the previous ones.

A generalization of the technique reviewed in \S~\ref{Maps} shows that
$$
\big\langle {\rm Tr}\,M^{L_1} \cdots {\rm Tr}\,M^{L_n} \big\rangle_S = \sum_{\substack{\partial\text{-connected stuffed map } \mathcal{M} \\ {\rm with}\, \partial \,  {\rm lengths} \,(L_i)_{i = 1}^n}} \frac{N^{\chi(\mathcal{M})}}{|{\rm Aut}(\mathcal{M})|}\,\prod_{p = 1}^F t^{h_p}_{(\ell(c))_{c\in f_p}}.
$$
As in the previous subsection, taking the logarithm or the cumulant expectation values in absence or presence of boundaries respectively, we obtain the generating series of connected stuffed maps:
\begin{proposition}\cite{Bstuff}\label{propStuffed}
$$
\ln Z_S = \sum_{g \geq 0} N^{2-2g} \widehat{F}^{[g]}, \quad \quad
\kappa_n({\rm Tr}\,M^{L_1},\ldots,{\rm Tr}\,M^{L_n})_S = \sum_{g \geq 0} N^{2-2g-n} \widehat{F}^{[g]}_{L_1,\ldots,L_n}.
$$
\end{proposition}

\subsection{Fully simple maps}

Let $(\gamma_i)_{i = 1}^n$ be pairwise disjoint cycles:
\beq
\label{gammapre}\gamma_i = (j_{i,1} \rightarrow j_{i,2} \rightarrow \cdots \rightarrow j_{i,L_i})
\eeq
and $L = \sum_{i = 1}^n L_i$.
We want to compute $\big\langle \prod_{i = 1}^n \mathcal{P}_{\gamma_i}(M) \big\rangle$ and the idea is that only fully simple maps will make a non-zero contribution, so we will be able to express it as a generating series for fully simple maps. Let us describe this expression for the measure \eqref{eq:mes} in terms of maps. Repeating the steps of \S~\ref{Maps}, we obtain 
\bea
\Big\langle \prod_{i = 1}^n \mathcal{P}_{\gamma_i}(M)  \Big\rangle & = &  \frac{1}{Z} \sum_{r \geq 0} \sum_{k_1,\ldots,k_r \geq 1} \frac{N^r}{r!} \prod_{m = 1}^r \frac{t_{k_m}}{k_m} \sum_{\substack{1 \leq j_{h} \leq N, \\ h \in H^u}} \Big\langle \prod_{h\in H} M_{j_h,j_{\varphi(h)}}\Big\rangle_{0} \nonumber\\
& = &  \frac{1}{Z} \sum_{r \geq 0} \sum_{k_1,\ldots,k_r \geq 1} \frac{N^r}{r!} \prod_{m = 1}^r \frac{t_{k_m}}{k_m} N^{\frac{-|H|}{2}} \sum_{\substack{1 \leq j_{h} \leq N, \\ h \in H^u}}  \sum_{\alpha \in \mathfrak{I}_H} \prod_{h\in H} \delta_{j_{h},j_{\alpha(\varphi(h))}}. \nonumber 
\eea
We consider as before the permutation $\sigma \coloneqq (\alpha\circ\varphi)^{-1}$ which will correspond to the vertices of the maps. The difference with \eqref{eq:ordev} lies in the summation over indices $j_{h}$ between $1$ and $N$ only for $h \in H^u$, while $j_{h}$ for $h = (i,l) \in H^{\partial}$ is prescribed by \eqref{gammapre}. As $j_{i,l}$ are pairwise distinct, the only non-zero contributions to the sum will come from maps for which $(i,l) \in H^{\partial}$ belong to pairwise distinct cycles of $\sigma$ and this is the characterization for fully simple maps in the permutational model setting. The function $h \mapsto j_h$ must be constant along the cycles of $\sigma$, and its value for every $h \in H^{\partial}$ is prescribed by \eqref{gammapre}. So, the number of independent indices of summation among $(j_{h})_{h \in H^u}$ is
$$
|\mathcal{C}(\sigma)| - L.
$$

Thus,
\bea
\Big\langle \prod_{i = 1}^n \mathcal{P}_{\gamma_i}(M)  \Big\rangle & = & \frac{1}{Z} \sum_{r \geq 0} \sum_{k_1,\ldots,k_r \geq 1} \frac{N^r}{r!} \prod_{m = 1}^r \frac{t_{k_m}}{k_m} N^{-\frac{|H|}{2}} \sum_{\alpha \in \mathfrak{I}_H} N^{|\mathcal{C}(\alpha\circ\varphi)|-L}. \nonumber \\
&= & \sum_{\substack{\partial \text{-connected fully}\\ \text{simple } \mathcal{M} = [(\sigma,\alpha)] \\ \text{with } \partial \text{ lengths}\,\,(L_i)_{i = 1}^n}} \frac{N^{\chi(\sigma,\alpha)- L}}{|{\rm Aut}(\sigma,\alpha)|}\,\prod_{f \in \mathcal{C}(\left.\varphi\right|_{H^u})}t_{\ell(f)}. \nonumber
\eea

The generalization to the measure \eqref{eq:mesS} is straightforward and gives rise to generating series for fully simple stuffed maps:
$$
\Big\langle \prod_{i = 1}^n \mathcal{P}_{\gamma_i}(M)  \Big\rangle_S = \sum_{\substack{\mathcal{M} \\ {\rm fully}\,\,{\rm simple}\,\,{\rm stuffed}\,\,{\rm map} \\ \partial\,{\rm perimeters}\,\,(L_i)_{i = 1}^n \\ \partial {\rm -connected}}} \frac{N^{\chi(\mathcal{M}) - L}}{|{\rm Aut}(\mathcal{M})|}\,\prod_{p = 1}^F t^{h_p}_{(\ell(c))_{c\in f_p}}.
$$

As before, the cumulant expectation values give the generating series of connected fully simple maps and stuffed maps for the more general measure:
\begin{proposition}\label{PPPPP}
\bea
\kappa_n(\mathcal{P}_{\gamma_1}(M) ,\ldots,\mathcal{P}_{\gamma_n}(M) ) & = & \sum_{g \geq 0} N^{2-2g-n-L}  H^{[g]}_{L_1,\ldots,L_n}, \nonumber \\ 
\kappa_n(\mathcal{P}_{\gamma_1}(M) ,\ldots,\mathcal{P}_{\gamma_n}(M) )_S & = & \sum_{g \geq 0} N^{2-2g-n-L} \widehat{H}^{[g]}_{L_1,\ldots,L_n}. \nonumber
\eea
\end{proposition}

\section{Towards a proof of the conjecture for usual maps}
\label{PRof}
\label{Section9}

In this section we give a sketch on the ideas towards a proof of the Conjecture~\ref{conj} for usual maps, indicating all the technicalities we skip. We manage to reduce the problem to a technical condition concerning a weaker version of symplectic invariance for the exchange transformation of the spectral curve of the hermitian matrix model with external field. We confirmed experimentally that this condition is satisfied for some particular cases. However, we do not have a justification for this condition to be satisfied in general for the moment.

We believe the further analysis of our problem could help shed some clarity on this fundamental question of TR.

The starting point of our argument\footnote{The idea of this argument first appeared in the derivation of Bouchard-Mari\~{n}o conjecture proposed in \cite{BEMS}. In that article, generating series of simple Hurwitz numbers were represented in terms of a matrix model with external field for a complicated $V$, albeit it was later pointed out by D.~Zvonkine that this representation was ill-defined even in the realm of formal series. This issue is not relevant here as we start with a well-defined matrix model in formal series, and are careful to justify all steps by legal operations within formal series.} is the representation of the generating series of connected fully simple maps as the free energies $\ln \frac{\check{Z}(A)}{\check{Z}(0)}$ of the $1$-hermitian matrix model with external field \cite{Semenoff}. This model is considered here to be valued in formal series. The topological expansion of its correlators satisfies Eynard-Orantin topological recursion, for a well-characterized spectral curve $(\mathcal{S}_{A},x,y)$ \cite{EPf}. On the other hand, the generating series $X_{n}^{[g]}$ we are after are encoded into the $n$-th order Taylor expansion of $\ln \frac{\check{Z}(A)}{\check{Z}(0)}$ around $A = 0$. Using a milder version of symplectic invariance and the properties of topological recursion under deformations of the spectral curve \cite{EORev}, we relate these $n$-th order Taylor coefficients to TR amplitudes of the topological recursion applied to the curve $(\mathcal{S}_{0},y,x)$. As the matrix model $\check{Z}(0)$ generates usual maps, the spectral curve $\mathcal{S}_{0}$ must be the initial data mentioned in Theorem~\ref{TRRRR}.  Unfortunately, our idea of proof is not combinatorial and relies on the symplectic invariance itself.

Prior to applying the result of \cite{EPf}, we give the definition of the topological expansion of $\frac{\check{Z}(A)}{\check{Z}(0)}$ and sketch the computation of the spectral curve from Schwinger-Dyson equations. These aspects are well-known to physicists.

\subsection{The topological expansion}

Corollary~\ref{gudn} together with Proposition~\ref{PPPPP} to access the genus $g$ part yields 
\beq\label{gsHMMEF}
\hat{F}^{[g]}_A = \sum_{n \geq 1}\frac{1}{n!} \sum_{\ell_1,\ldots,\ell_n \geq 1} \frac{p_{\ell_1}(A)}{N\ell_1}\cdots \frac{p_{\ell_n}(A)}{N\ell_n} \kappa_n^{[g]}(\mathcal{P}^{(\ell_1)}(M),\ldots,\mathcal{P}^{(\ell_n)}(M)\big).
\eeq
Recall that we had identified $\kappa_n^{[g]}(\mathcal{P}^{(\ell_1)}(M),\ldots,\mathcal{P}^{(\ell_n)}(M)\big)$ with the generating series of fully simple maps $H^{[g]}_{\ell_1,\ldots,\ell_n}$. We will actually think of $\hat{F}^{[g]}_A$ in general as an element 
$$
\hat{F}^{[g]}_A(q_1,q_2,q_3,\ldots)\in \mathcal{R} :=\mathcal{R}_0[[q_1,q_2,\ldots]], \text{ where } \mathcal{R}_0\coloneqq\mathbb{Q}[[t_3,t_4,\ldots]].
$$
In our concrete case, $\hat{F}^{[g]}_A=\hat{F}^{[g]}_A\left(\frac{{\rm Tr}\,A}{N},\frac{{\rm Tr}\,A^2}{N},\frac{{\rm Tr}\,A^3}{N},\ldots\right)$.

The same procedure gives the topological expansion of the correlators
\beq\label{HMMEFcorrelators}
\hat{W}_{n;A}(x_1,\ldots,x_n) = \kappa_{n}\Big({\rm Tr}\,\frac{1}{x_1 - M},\ldots,{\rm Tr}\,\frac{1}{x_n - M}\Big)_A = \sum_{g \geq 0} N^{2 - 2g - n}\,\hat{W}_{n;A}^{[g]}(x_1,\ldots,x_n),
\eeq
with $\hat{W}_{n;A}^{[g]} \in \mathcal{R}[[x_1^{-1},\ldots,x_n^{-1}]]$, where we think of $q_i$ as replacing ${\rm Tr}\,A^i/N$ as above.

\subsection{The spectral curve}

To deduce the spectral curve for the hermitian matrix model with external field, one would need to generalize the notion of topological expansion here in order to extract the first term in the topological expansion of the first Schwinger-Dyson equation of this model, which is proved in \cite{EPf} in the more general context of the chain of matrices.

As a matter of fact, one should handle more general observables, which involve expectation values of products of ${\rm Tr}\,[M^{k}]_{i,i}$ and of ${\rm Tr}\,M^{k'}$. We are skipping the details, but their topological expansion can also be defined, and the coefficients of their expansion will now belong to
$$
\tilde{\mathcal{R}} = \lim_{\infty \leftarrow \nu} \mathbb{Q}[[t_3,t_4,\ldots]][[a_1,\ldots,a_{\nu}]][[q_1,q_2,\ldots]],
$$
where $A$ is specialized to the matrix ${\rm diag}(a_1,\ldots,a_{\nu},0,\ldots,0)$. The restriction morphisms to define the projective limit consist in specializing some $a$'s to $0$. Taking $q_i = \frac{1}{N}\big(\sum_{j} a_j\big)^i$, we can often work in the ring
$$
\mathcal{R}_{-} := \mathbb{Q}[[t_3,t_4,\ldots,]][[a_1,\ldots,a_{\nu}]][[N^{-1}]].
$$

\begin{lemma}
We have the following identity in $\tilde{\mathcal{R}}[N,N^{-1}]][[x^{-1},y^{-1}]]$:
\bea
 0 & = & \kappa_{2}\Big({\rm Tr}\,\frac{1}{x - M},{\rm Tr}\,\frac{1}{x - M}\frac{1}{y - A}\Big)_A + (\hat{W}_{1;A}(x) - NV'(x) + Ny)\Big\langle {\rm Tr}\,\frac{1}{x - M}\,\frac{1}{y - A}\Big\rangle_{A} \nonumber \\
\label{SD1} && - N\hat{W}_{1;A}(x) + \Big\langle {\rm Tr}\,\frac{V'(x) - V'(M)}{x - M}\,\frac{1}{y - A} \Big\rangle_{A}\,. \nonumber
\eea
\end{lemma}
\noindent \textbf{Proof.} This is obtained from the relation
$$
0 = \sum_{i,j = 1}^N \sum\int \dd M\,\partial_{M_{i,j}}\bigg(\Big(\frac{1}{x - M}\,\frac{1}{y - A}\Big)_{j,i} \exp\big[N\,{\rm Tr}(MA - V(M))\big]\bigg)\,.
$$
\hfill $\Box$

Let us introduce simplified notations
$$
\hat{W}_{A}^{(i)}(x) = \bigg\langle \Big[\frac{1}{x - M}\Big]_{i,i}\Big\rangle_{A}^{[0]},\qquad \hat{W}_{A}(x) = \hat{W}_{1;A}^{[0]}(x) = \sum_{i = 1}^N \hat{W}_{A}^{(i)}(x)\,.
$$
We can write the planar limit of the first Schwinger-Dyson equation
\begin{proposition}
We have
\beq
\label{Sdd2}\hat{W}_{A}(x)^2 - [V'(x)\hat{W}_{A}(x)]_{-} + \sum_{i = 1}^N \tfrac{a_i}{N} \hat{W}_{A}^{(i)}(x) = 0\,,
\eeq
and for any $i \in \{1,\ldots,N\}$, we have
\beq
\label{Sdd3}\hat{W}_{A}^{(i)}(x)\big(\hat{W}_{A}(x) + a_i) - \tfrac{1}{N} [V'(x)\hat{W}_{A}^{(i)}(x)]_{-} = 0\,,
\eeq
where $[\ \cdot\ ]_{-}$ takes the negative part of the Laurent expansion when $x \rightarrow \infty$.
\end{proposition}
\noindent \textbf{Proof.} In the planar limit of \eqref{SD1}, $U_2$ disappears
$$
0 = (\hat{W}_{A}(x) - V'(x)  + y)\Big\langle {\rm Tr}\,\frac{1}{x - M}\frac{1}{y - A}\Big\rangle_{A} - \hat{W}_{A}(x) + \frac{1}{N}\,\Big\langle {\rm Tr}\,\frac{V'(x) - V'(M)}{x - M}\,\frac{1}{y - A}\Big\rangle_{A}^{[0]}\,.
$$
The right-hand side rational function of $y$, with simple poles at $y \rightarrow a_i$ and $y \rightarrow \infty$. Identifying the coefficient of these poles gives an equivalent set of equations. At $y \rightarrow \infty$ we get a trivial relation. At $y \rightarrow a_i$ we get
\beq
(\hat{W}_{A}(x) - V'(x) + \tfrac{a_i}{N})\hat{W}_{A}^{(i)}(x) + \Big\langle \Big[\frac{V'(x) - V'(M)}{x - M}\Big]_{i,i}\Big\rangle_{A}^{[0]} = 0\,,
\eeq
in which we recognize \eqref{Sdd3}. Summing this relation over $i \in \{1,\ldots,N\}$, we obtain \eqref{Sdd2}. \hfill $\Box$

For $A = 0$, we would obtain the planar limit of the first Schwinger-Dyson equation of the hermitian matrix model
\beq
\label{SD0} \hat{W}_{A=0}(x)^2 - [V'(x)\hat{W}_{A=0}(x)]_{-} = 0\,,\eeq
which is equivalent to Tutte's equation for the generating series of disks. Its solution is well-known, see \textit{e.g.} \cite{EynardBook}. Observe that for $A=0$, $\hat{W}_{A=0}(x)=W(x)$.

\begin{lemma}
\label{Tutte000}The equation \eqref{SD0} determines $\hat{W}_{A=0}(x)$ completely. Let $\mathcal{R}^0 = \mathbb{Q}[[t_3,\ldots,t_d]]$. There exist unique $\alpha,\gamma \in \mathcal{R}^0$ such that
$$
x(\zeta;0) = \alpha + \gamma(\zeta + \zeta^{-1}),\qquad w(\zeta;0) = [V'(x(\zeta))]_{+}
$$
satisfy $\hat{W}_{1;A=0}^{[0]}(x(\zeta;0)) = V'(x(\zeta;0)) - w(\zeta;0)$. We have that $x(\zeta;0)=x(\zeta)$. Here, $[\ \cdot\ ]_+$ takes the polynomial part in $\zeta$. More precisely, $\alpha$ and $\gamma$ are determined by the conditions
$$
[\zeta^0]\,V'(x(\zeta)) = 0,\qquad [\zeta^{-1}] \,V'(x(\zeta)) = \gamma^{-1}\,.
$$
and we have $\gamma = 1 + O(\mathbf{t})$.
\hfill $\Box$
\end{lemma}

\begin{lemma}
\label{UniqueW}The equations \eqref{Sdd2}-\eqref{Sdd3} determine $\hat{W}_A^{(i)}(x)$ uniquely for all $i \in \{1,\ldots,p\}$.
\end{lemma}
\noindent \textbf{Proof.} By specializing $A$ to diagonal matrices of arbitrary size $\nu$ and $q_i$ to ${\rm Tr}\,A_i/N$ it is enough to work here in the ring $\mathcal{R}_- := \mathbb{C}[[t_3,\ldots,t_d,a_1,\ldots,a_{\nu},N^{-1}]]$. We introduce two gradings on $\mathcal{R}_{-}$: the first one denoted $\deg_{A}$ assigns a degree $1$ to the variables $a_i$, and $0$ to the other generators; the second one $\deg_{t}$ assigns a degree $1$ to the variables $t_{j}$ and $0$ to the other generators.  We denote $\hat{W}^{(i)}_{\alpha;A}$  and $\hat{W}_{\alpha;A}$ the homogeneous part of $\hat{W}_A^{(i)}$ and $\hat{W}_A$ with ${\rm deg}_{A} = \alpha$. We remark that $\hat{W}_{0;A}^{(i)}(x) = \hat{W}_{A=0}^{(i)}(x)$ is independent of $A$ and $i$ and thus
$$
\hat{W}_{0;A}^{(i)}(x) = \frac{\hat{W}_{A=0}(x)}{N}.
$$
Besides, we observe that
\beq 
\label{secondr} \forall \alpha \geq 1,\quad \forall i \in \{1,\ldots,N\},\qquad \hat{W}^{(i)}_{\alpha;A}(x) = O(x^{-2}).
\eeq
We proceed by induction on ${\rm deg}_{A}$. We already know that the $\deg_{A} = 0$ part of \eqref{Sdd2}-\eqref{Sdd3} has a unique solution given by Lemma~\ref{Tutte000}. Let $\alpha \geq 1$, and assume \eqref{Sdd2}-\eqref{Sdd3} determine uniquely $\hat{W}_{\alpha'}^{(i)}$ for $\alpha' < \alpha$. Decomposing \eqref{Sdd2} in homogeneous degree $\alpha \geq 1$, we find
\beq
\label{kalpha}\mathcal{K}[\hat{W}_{\alpha;A}(\cdot)](x) + \sum_{\substack{0 < \alpha_1,\alpha_2 < \alpha \\ \alpha_1 + \alpha_2 = \alpha}} \hat{W}_{\alpha_1;A}(x)\hat{W}_{\alpha_2;A}(x) + \sum_{i = 1}^N \frac{a_i}{N} \hat{W}^{(i)}_{\alpha - 1;A}(x) = 0,
\eeq
where $\mathcal{K}$ is the linear operator
$$
\mathcal{K}[f](x) =  2\hat{W}_{A=0}(x)f(x) - [V'(x)f(x)]_{-}.
$$
Let us write
$$
V(x) = \frac{x^2}{2} + \delta V(x),\qquad {\rm deg}_{t}( \delta V) = 1.
$$
When $f(x) = O(x^{-2})$, we have
$$
\mathcal{K}[f](x) = (2\hat{W}_{A=0}(x) - x)f(x) - [(\delta V)'(x)f(x)]_{-}.
$$
Under this assumption, the equation
\beq
\label{eqk}\mathcal{K}[f](x) = g(x)
\eeq
in $\mathcal{R}_{-}[x^{-1}]$ has a unique solution, which we denote $f(x) = \mathcal{K}^{-1}[g](x)$. Indeed, $(2\hat{W}_{A=0}(x) - x)$ is invertible and \eqref{eqk} determine recursively the $\deg_{t}$ homogeneous components of $f$ recursively in terms of those of $g$. Taking into account \eqref{secondr}, we can apply this remark to \eqref{kalpha} and find that
$$
\hat{W}_{\alpha;A}(x) = -\mathcal{K}^{-1}\bigg[\sum_{\substack{0 < \alpha_1,\alpha_2 < \alpha \\ \alpha_1 + \alpha_2 = \alpha}} \hat{W}_{\alpha_1;A}(\cdot)\hat{W}_{\alpha_2;A}(\cdot) + \sum_{i = 1}^N \frac{a_i}{N} \hat{W}^{(i)}_{\alpha - 1;A}(\cdot)\bigg](x)
$$
is determined. We turn to the $\deg_{A} = \alpha$ part of \eqref{Sdd3}
$$
\mathcal{K}[\hat{W}^{(i)}_{\alpha;A}(\cdot)](x) + \tfrac{1}{N}\big(\hat{W}_{A=0}(x)\hat{W}_{\alpha;A}(x) + \frac{a_i}{N} \hat{W}^{(i)}_{\alpha - 1;A}(x)\big) + \sum_{\substack{0 < \alpha_1,\alpha_2 < \alpha \\ \alpha_1 + \alpha_2 = \alpha}} \hat{W}^{(i)}_{\alpha_1;A}(x)\hat{W}^{(i)}_{\alpha_2;A}(x) = 0.
$$
Hence
$$
\hat{W}_A^{(i)}(x) = -\mathcal{K}^{-1}\bigg[\tfrac{1}{N}\big(\hat{W}_{A=0}(-)\hat{W}_{\alpha;A}(\cdot) + \frac{a_i}{N} \hat{W}^{(i)}_{\alpha - 1;A}(\cdot)\big) + \sum_{\substack{0 < \alpha_1,\alpha_2 < \alpha \\ \alpha_1 + \alpha_2 = \alpha}} \hat{W}^{(i)}_{\alpha_1;A}(\cdot)\hat{W}^{(i)}_{\alpha_2;A}(\cdot) \bigg](x)
$$
is determined as well. We conclude by induction.
\hfill $\Box$

\begin{lemma}
There exists a unique polynomial $w(z;A) \in \mathcal{R}_{-}[z]$ and $b_{k} = b_{k}(A) \in \mathcal{R}_{-}$ for $k \in \{1,\ldots,N\}$, such that
\beq
\label{xzA} x(z;A) = z + \frac{1}{N} \sum_{i = 1}^N \frac{1}{w'(b_i;A)(z - b_i)}\,,
\eeq
together with
\beq
\label{wVW} w(z;A) = V'(x(z)) + O(z^{-1}),\qquad w(b_k;A) = a_k\,.
\eeq
Besides, this unique data is such that
\beq
\label{911} w(z;A) = V'(x(z)) + z^{-1} + O(z^{-2}).
\eeq
\end{lemma} 
\noindent \textbf{Proof.} Let $\beta_i(A),b_i(A)$ be elements of $\mathcal{R}_{-}$, so far undetermined. We assume $\beta_i(A)$ invertible, and $\beta := \beta_i(0)$ and $b := b_i(0)$ independent of $i$. We define
\beq
\label{POlu}x(z;A) = z + \frac{1}{N}\,\sum_{i = 1}^N \frac{1}{\beta_i(A)(z - b_i(A))}
\eeq
and introduce the polynomial $w(z;A) \in \mathcal{R}_{-}[z]$ such that
\beq
\label{POlu2}V'(x(z;A)) = w(z;A) + O(z^{-2}),\qquad z \rightarrow \infty
\eeq
Equivalently
$$
w(z;A) = \oint \frac{\dd \tilde{z}}{2{\rm i}\pi}\,\frac{V'(x(\tilde{z};A))}{\tilde{z} - z},
$$
where the contour is close enough to $\tilde{z} = \infty$. We are going to prove that the system of equations
\beq
\label{sys}\forall i \in \{1,\ldots,N\},\qquad  \left\{\begin{array}{lll} w(b_{i}(A);A) & = & a_i \\ w'(b_i(A);A) & = & \beta_{i}(A)  \end{array}\right.
\eeq
has unique solutions $b_i(A)$ and $\beta_{i}(A)$ in $\mathcal{R}_{-}$, which we will adopt to define \eqref{POlu}-\eqref{POlu2}.

For $A = 0$, our definition gives
$$
x(z;0) = z + \frac{1}{\beta(z - b)}.
$$
Let $\alpha,\gamma \in \mathcal{R}^0$ be as in Lemma~\ref{Tutte000}. We recall that $\gamma = 1 + O(t)$, hence $\gamma$ is invertible in $\mathcal{R}^0$.  By making the change of variable $z = \gamma\zeta + \alpha$ and choosing
$$
\beta_{i}(A) = \gamma^{-2} + O(A),\qquad b_{i}(A) = \alpha + O(A),
$$
we find by comparing with Lemma~\ref{Tutte000} that
$$
\hat{W}_{A=0}(x(z)) = V'(x(z)) - w(z;0),\qquad w(b;0) = 0,\qquad w'(b;0) = \beta,
$$
in terms of the functions introduced in \eqref{POlu}-\eqref{POlu2}. 

Next, we introduce a grading in $\mathcal{R}_{-}$ by assigning degree $1$ to each $a_i$, and $0$ to all other generators. We write $x_{d}(z;A)$, $w_{d}(z;A)$, $\beta_{i,d}(A)$ and $b_{i,d}(A)$ for the degree $d$ component of the corresponding quantities. Fix $d \geq 1$, and assume we have already determined all these quantities in degree $d' < d$. Let us examine the degree $d$ part of the system \eqref{sys}, and isolate the pieces involving $\beta_{i,d}(A)$ and $b_{i,d}(A)$. We find for all $i \in \{1,\ldots,N\}$
\beq
\label{sys2} \left\{\begin{array}{lll} w'(\gamma^{-2};0) b_{i,d}(A) + \frac{c_2}{N}\Big( \sum_{i = 1}^N \frac{\beta_{i,d}(A)}{\gamma^{-4}}\Big)+ \frac{c_3}{N}\Big(\sum_{i = 1}^N \frac{b_{i,d}(A)}{\gamma^{-2}}\Big)  & = & Y_{i,d}(A), \\
w''(\gamma^{-2};0) b_{i,d}(A) + \frac{c_3}{N}\Big(\sum_{i = 1}^N \frac{\beta_{i,d}(A)}{\gamma^{-4}}\Big) + \frac{c_4}{N}\Big(\sum_{i = 1}^N \frac{b_{i,d}(A)}{\gamma^{-2}}\Big) & = & \tilde{Y}_{i,d}(A), \end{array}\right. 
\eeq
where
$$
c_{k} = \oint \frac{\dd\tilde{z}}{2{\rm i}\pi}\,\frac{V''(x(\tilde{z};0))}{(\tilde{z} - \alpha)^{k}}
$$
and $Y_{i,d}$ and $\tilde{Y}_{i,d}$ are polynomials in the $t_k$s, $a_i$s, and $\beta_{j,d'}(A)$ and $b_{j,d'}(A)$ with $d' < d$. In this formula we  used that $\beta_{j,0} = \gamma^{-2}$ and $b_{j,0} = \alpha$ for all $j$. For instance, in degree $1$
$$
Y_{i,1}(A) = a_i,\qquad \tilde{Y}_{i,1}(A) = 0.
$$
As $V''(x) = 1 + O(t)$, we deduce by moving the contour to surround $\tilde{z} = \alpha$ that
$$
\forall k \geq 2,\qquad c_{k} = O(t).
$$
Therefore, the system \eqref{sys2} takes the matrix form
$$
\Bigg[\left(\begin{array}{cc} \gamma^{-2}\,{\rm Id}_{N} & 0 \\ 0 & -{\rm Id}_{N}\end{array}\right) + N^{-1}O(t)\Bigg]\left(\begin{array}{c} b_{\bullet,d} \\ \beta_{\bullet,d} \end{array}\right) = \left(\begin{array}{l} Y_{\bullet,d}(A) \\ \tilde{Y}_{\bullet,d}(A) \end{array}\right).
$$
The matrix in the left-hand side is invertible in $\mathcal{R}_{-}$, hence $\beta_{i,d}(A)$ and $b_{i,d}(A)$ are uniquely determined. By induction, we conclude to the existence of unique $\beta_{i}(A)$ and $b_{i}(A)$ in $\mathcal{R}_{-}$ satisfying \eqref{sys}.  

Let $c_{\infty} \in \mathcal{R}_{-}$ be such that 
\beq
\label{Vexp} V'(x(z;A)) = w(z;A) + c_{\infty}\,z^{-1} + O(z^{-2}).
\eeq
We claim that $c_{\infty} = 1$. Indeed, the second set of equations in \eqref{sys} imply
$$
\sum_{i = 1}^N \Res_{z \rightarrow b_i} x(z;A)\dd w(z;A) = 1,
$$
while as $x(z;A) = O(z)$ when $z \rightarrow \infty$, we obtain by moving the contour to $\infty$ and using \eqref{Vexp}
$$
\sum_{i = 1}^N \Res_{z \rightarrow b_i} x(z;A)\dd w(z;A) = -\Res_{z \rightarrow \infty} x(z;A)\dd w(z;A) = c_{\infty} - \Res_{x \rightarrow \infty} x\,\dd V' = c_{\infty}.
$$
The last equality holds because $V'$ is a polynomial in $x$. \hfill $\Box$

\begin{lemma}
There exists a unique polynomial $P_i(\xi;A) \in \mathcal{R}_{-}[\xi]$ of degree $d - 2$ with leading coefficient $\tfrac{t_{d}}{N}$, such that
\beq
\label{WPs} V'(x(z;A)) - w(z;A) = \sum_{i = 1}^N \frac{P_i(x(z);A)}{w(z;A) - a_i}.
\eeq
\end{lemma}
\noindent \textbf{Proof.} We have
$$
x(z;A) = \frac{S_{N + 1}(z)}{T_{N}(z)},\qquad w(z;A) = U_{d - 1}(z),
$$
where $S,T,U$ are polynomials in $z$ with coefficients in $\mathcal{R}_{-}$, of degree indicated by the subscript, and $T_N$ is monic. Therefore, the resultant
$$
Q(X,Y) = {\rm res}_{z}\big[XT_N(z) - S_{N + 1}(z),T_N(z)(Y - U_{d - 1}(z))\big]
$$
is a polynomial with coefficients in $\mathcal{R}_{-}$ and of degree $N + d - 1$ in $X$, and $N + 1$ in $Y$, which gives a polynomial relation
$$ 
Q(x(z;A),w(z;A)) = 0.
$$ 

\begin{center}
\begin{figure}[h!]
\begin{center}
\includegraphics[width=0.65\textwidth]{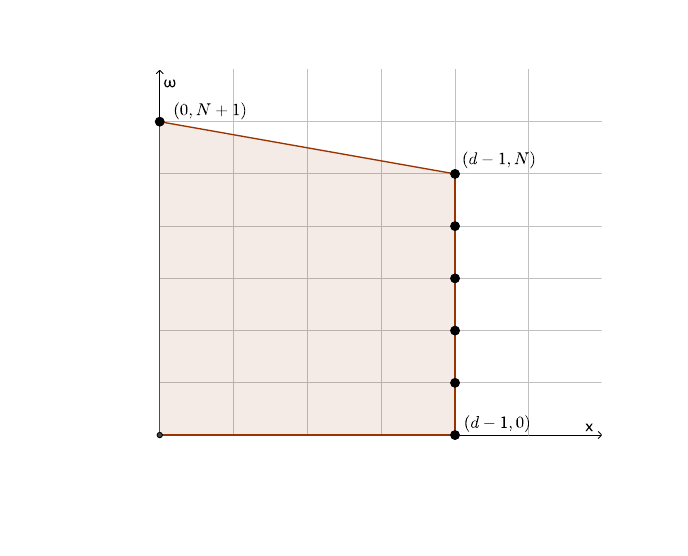}
\vspace{-1.3cm}
\caption{\label{Poly} The Newton polygon of $Q(x,w)$. The coefficients identified in \eqref{QXY} correspond to the nodes in the picture.}
\end{center}
\end{figure}
\end{center}

In fact, we can argue that the degree in $X$ is smaller, as follows. We study the slopes of the Newton polygon of $Q$ associated to $x \rightarrow \infty$ or $w \rightarrow \infty$. We have $w \rightarrow \infty$ if and only if $z \rightarrow \infty$, and in this case $x \rightarrow \infty$ and we have
\beq
\label{wtd} w \sim -t_dx^{d - 1}.
\eeq
Besides, the only other situation where $x \rightarrow \infty$ is when $w \rightarrow a_i$ for some $i \in \{1,\ldots,N\}$, and in this case we read from \eqref{xzA} that
\beq
\label{wai} (w - a_i)x \  \sim \  1.
\eeq
This determines slopes which must be in the Newton polygon of $Q$.  A closer look to the determinant defining the resultant shows, using $T_N(0) = (-1)^{N} \prod_{i = 1}^N a_i$ and that $XT_N(z) - S_{N + 1}(z) = -z^{N + 1} + O(z^{N})$, that the top degree term of $Q(X,Y)$ in the variable $Y$ is $Y^{N + 1} \prod_{i = 1} a_i^{N}$. In particular, the coefficient of $X^iY^{N + 1}$, which could a priori exist, vanishes for $i \in \{1,\ldots,d - 1 + N\}$. The existence of the previous slopes then forces the Newton polygon to be included in the shaded region of Figure~\ref{Poly}. In particular, $Q$ must be irreducible. And the precise behaviors \eqref{wtd}-\eqref{wai} leads to a decomposition
\beq
\label{QXY}Q(X,Y) = c \Big(Y^{N + 1} + t_{d}X^{d - 1} \prod_{i = 1}^N (Y - a_i)\Big) + \tilde{Q}(X,Y),\qquad c := \prod_{i = 1}^N a_i^{N},\eeq
for some polynomial $\tilde{Q}(X,Y)$ with coefficients in $\mathcal{R}_{-}$ such that
\beq
\label{degP}\deg_{X} \tilde{Q} \leq d - 2,\qquad \deg_{Y} \leq N.
\eeq
Now, let us examine
$$
L(z) = \big(V'(x(z;A)) - w(z;A)\big) \prod_{i = 1}^N (w(z;A) - a_i).
$$
It can also be written
$$
L(z) = -w(z;A)^{N + 1} - t_{d}x(z;A)^{d - 1} \prod_{i = 1}^N (w(z;A) - a_i) + \tilde{L}(x(z;A),w(z;A)),
$$
where $\tilde{L}(X,Y)$ is a polynomial satisfying the same degree bound as \eqref{degP}. Using the relation \eqref{QXY}, we can eliminate the first terms and get
$$
L(z) = \hat{L}(x(z;A),w(z;A)),
$$
for some polynomial $\hat{L}(X,Y)$ with coefficients in the localization $\mathcal{R}_{-}^{(c)}$ of $\mathcal{R}_{-}$ at $c$, and with $\deg_{X} \hat{L} \leq d - 2$ and $\deg_{Y} \hat{L} \leq N$. We deduce the existence of polynomials $(P_i(X))_{i = 1}^{N}$ and $P_{\infty}(X)$ of degree $\leq d - 2$ and with coefficients in $\mathcal{R}_{-}^{(c)}$, such that
$$ 
V'(x(z;A)) - w(z;A) = P_{\infty}(x(z;A)) + \sum_{i = 1}^N \frac{P_i(x(z;A))}{w(z;A) - a_i},
$$
after partial fraction decomposition. According to \eqref{wVW}, the left-hand side behaves is $O(z^{-1})$ when $z \rightarrow \infty$. As $w(z;A) = -t_{d}z^{d - 1}$ and $P_i(x(z;A)) = O(z^{d - 2})$, this is also true for the sum over $i$. Any non-zero monomial in $P_{\infty}$ would disagree with this behavior at $z \rightarrow \infty$. Hence $P_{\infty} = 0$, and we have prove the existence result for polynomials $P_i(x(z;A))$ with coefficients in $\mathcal{R}_{-}^{(c)}$.

Now, we prove uniqueness. By construction, $w(z;A) - a_i$ is a polynomial of degree $d - 1$ in $z$ with coefficients in the local ring $\mathcal{R}_{-}$, and $b_i(A)$ is a root. We remark that
$$
w(z;A) = -t_{d}z^{d - 1} + z + \tilde{w}(z;A),
$$
where $\tilde{w}(z;A) = O(A,t)$ is a polynomial of degree $\leq (d - 2)$. Therefore, $w(z;A) - a_i$ has $(d - 2)$-other roots, counted with multiplicity, which belong to the local ring $\widehat{\mathcal{R}}_{-}$ obtained from $\mathcal{R}_{-}$ by adjunction of a finite set $\mathfrak{b}$ consisting of $t_{d}^{-1/(d - 2)}$ and all its Galois conjugates. Besides, in $\widehat{\mathcal{R}}_{-}$ those $(d - 2)$ roots are pairwise distinct. 

By construction, $V'(x(z);A) - w(z;A)$ is a rational function of $z$, with poles at $z = b_i$ and $z = \infty$. As the denominator of the $i$-th term has $(d - 2)$ roots $\mathfrak{b}_{i}$ which are not poles of $\mathcal{W}(z;A)$ as set of root, we deduce that $P_i(x(z);A)$ must have roots at $\mathfrak{b}_i$, hence
$$
P_i(\xi;A) =  \tfrac{t_d}{N}\,\prod_{\rho(A) \in \mathfrak{b}_{i}(A)} (\xi - x(\rho;A)).
$$
Therefore, \eqref{WPs} determines uniquely the polynomial $P_i(\xi;A)$. Note that the coefficients of this polynomial belong to the localization of $\mathcal{R}_{-}$ at $t_{d}$, and not only to the localization of $\widehat{\mathcal{R}}_{-}$ at $t_{d}$, as the product runs over Galois orbits. By comparison, the $P_i$ constructed in the existence part had coefficients in $\mathcal{R}_{-}^{(c)}$. We deduce that $P_i$ has coefficients in $\mathcal{R}_{-}^{(c)} \cap \mathcal{R}_{-}{(t_{d})} = \mathcal{R}_{-}$.
\hfill $\Box$

\begin{corollary}
We have
$$
\hat{W}_A(x(z)) = V'(x(z);A) - w(z;A),\qquad \hat{W}_A^{(i)}(x(z)) = \frac{P_i(x(z);A)}{w(z;A) - a_i}.
$$
\end{corollary}
\noindent \textbf{Proof.} As $x = z + O(z^{-1})$, we can perform a Lagrange inversion and define unique elements $\hat{\mathcal{W}}(x;A)$ and $\hat{\mathcal{W}}_A^{(i)}(x)$ in $ \mathcal{R}_{-}[[x^{-1}]]$ such that
$$
\hat{\mathcal{W}}_A(x(z;A)) = V'(x(z;A)) - w(z;A),\qquad \hat{\mathcal{W}}_A^{(i)}(x(z;A)) = \frac{P_i(x(z;A);A)}{w(z;A) - a_i}.
$$
By construction, we have
\beq
\label{WSum}\sum_{i = 1}^N \hat{\mathcal{W}}^{(i)}_A(x) = \hat{\mathcal{W}}_A(x),
\eeq
and
$$
\forall i \in \{1,\ldots,N\},\qquad \hat{\mathcal{W}}_A^{(i)}(x)(\hat{\mathcal{W}}_A(x) - V'(x) + a_i) + P_i(x) = 0.
$$
Therefore, $\hat{\mathcal{W}}_A^{(i)}$ and $\hat{\mathcal{W}}_A$ satisfy \eqref{Sdd2}-\eqref{Sdd3}. Note that \eqref{911} ensures that
$$
\hat{\mathcal{W}}_A(x) = \tfrac{1}{x} + O(\tfrac{1}{x^2}),\qquad x \rightarrow \infty,
$$
and that $V'(x) = -t_{d}x^{d - 1}$ also implies
$$
\hat{\mathcal{W}}_A^{(i)}(x) = \tfrac{1}{Nx} + O(\tfrac{1}{x^{2}}),\qquad x \rightarrow \infty.
$$
Since the solution of these equations is unique according to Lemma~\ref{UniqueW}, we conclude that $\hat{W}_A = \hat{\mathcal{W}}_A$ and $\hat{W}^{(i)}_A = \hat{\mathcal{W}}_A^{(i)}$. 
\hfill $\Box$

\subsection{Topological recursion}

Eynard and Prats-Ferrer analyzed in \cite{EPf} the (topological expansion) of tower of Schwinger-Dyson equations which results from variation of the potential $V$, and involve the $n$-point correlators. Their final result reads:

\begin{theorem}
Let $\omega_{g,n}^{A}$ be the TR amplitudes for the initial data
\beq
\label{rare}\left\{\begin{array}{l} \mathcal{C} = \mathbb{P}^1 \nonumber \\ p(z) = x(z;A) \\ \lambda(z) = -w(z;A) \\ B(z_1,z_2) = \frac{\dd z_1\dd z_2}{(z_1 - z_2)^2}\,.  \end{array}\right. 
\eeq
For any $g,n \geq 0$, the equality
$$
\hat{W}_{n;A}^{[g]}(p(z_1),\ldots,p(z_n))\prod_{i = 1}^n \dd p(z_i) = \omega_{g,n}^{A}(z_1,\ldots,z_n) - \delta_{g,0}\delta_{n,2}\frac{\dd p(z_1)\dd p(z_2)}{(p(z_1) - p(z_2))^2} + \delta_{g,0}\delta_{n,1}\,\dd V(p(z_1)) 
$$
holds in Laurent expansion near $z_i \rightarrow \infty$.
\end{theorem}

\subsubsection{Deformations of the spectral curve}

\begin{lemma}
We have, for any $i \in \{1,\ldots,N\}$,
\beq
\label{omedef}\Omega_{i}(z;A) = \partial_{a_i} x(z;A) w'(z;A) - \partial_{a_i} w(z;A) x'(z;A) = \frac{1}{w(b_{i};A)}\,\frac{1}{(z - b_i)^2},\qquad
\eeq
where the derivative with respect to $a_i$ is taken at $z$ fixed, and $\,\,'$ denotes the derivative with respect to the variable $z$.
\end{lemma} 
\noindent \textbf{Proof.} From the form of $x(z;A)$ and $w(z;A)$, we know that $\Omega_{i}(z,A)$ is a rational function of $z$, with at most double poles at $z \rightarrow b_{k}$ for $k \in \{1,\ldots,N\}$, and maybe a pole at $\infty$. We are going to identify $\Omega_{i}(z,A)$ from its singular behavior at these poles. It is easier to start by computing
$$
\tilde{\Omega}_{j}(z;A) := \partial_{b_j} x(z;A) w'(z;A) - \partial_{b_j} w(z;A) x'(z;A)
$$
and then use the relation
\beq
\label{Ometil} \Omega_{i}(z;A) = \sum_{j = 1}^N \frac{\partial b_j}{\partial a_i}\,\tilde{\Omega}_{j}(z;A)\,.
\eeq
We start by examining $z \rightarrow \infty$. From the equation
$$
w(z;A) = V'(x(z;A)) + \tfrac{1}{z} + O(\tfrac{1}{z^2})\,,
$$
we deduce
\bea
w'(z;A) & = & x'(z;A)V''(x(z;A)) - \tfrac{1}{z^2} + O(\tfrac{1}{z^3})\,, \nonumber \\
\partial_{b_j} w(z;A) & = & \partial_{b_{j}}x(z;A) V''(x(z;A)) + O(\tfrac{1}{z^2})\,, \nonumber
\eea 
and from the form of $x$
$$
\partial_{b_j} x(z;A) = O(1),\qquad x'(z;A) = O(1)\,.
$$
This implies
\bea
\tilde{\Omega}_{i}(z;A) & = & \partial_{b_j} x(z;A)\big(x'(z;A)V''(x(z;A)) + O(\tfrac{1}{z^2})\big) \nonumber \\
&-& \big(\partial_{b_{j}} x(z;A) V''(x(z;A) + O(\tfrac{1}{z^2})\big)x'(z;A) \nonumber \\
& = & O(\tfrac{1}{z^2})\,, \nonumber
\eea
and therefore $\tilde{\Omega}_{i}(z;A)$ has no pole at $\infty$.

Next, we examine $z \rightarrow b_{k}$. We have
\bea
\partial_{b_{j}} x(z;A) & = & \frac{\delta_{j,k}}{N}\bigg(-\frac{w''(b_{k};A)}{(w'(b_{k};A))^2}\,\frac{1}{z - b_{k}} + \frac{1}{w'(b_{k};A)}\,\frac{1}{(z - b_{k})^2}\bigg) \nonumber \\
&- &\frac{1}{N}\,\frac{\partial_{b_{j}}w'(b_{k};A)}{(w'(b_{k};A))^2}\,\frac{1}{z - b_{k}} + O(1)\,, \nonumber \\
w'(z;A) & = & w'(b_{k};A) + w''(b_{k};A)(z - b_{k}) + O(z - b_{k})^2\,, \nonumber \\ 
\partial_{b_{j}} w(z;A) & = & \partial_{b_{j}} w(b_{k};A) + \partial_{b_{j}} w'(b_{k};A)(z - b_{k}) + O(z - b_{k})^2\,, \nonumber \\
x'(z) & = & -\frac{1}{w'(b_{k};A)}\,\frac{1}{(z - b_{k})^2} + O(1)\,. \nonumber
\eea
Hence, we obtain after simplification
$$
\tilde{\Omega}_{j}(z;A) = \frac{1}{N}\bigg(\delta_{j,k} + \frac{\partial_{b_{j}} w'(b_{k};A)}{w'(b_{k};A)}\bigg)\,\frac{1}{(z - b_{k})^2} + O(1) \,.
$$
So, the only singularities of $\tilde{\Omega}_{j}(z;A)$ are double pole without residues at $b_{k}$, and we get
$$
\tilde{\Omega}_{j}(z;A) = \frac{1}{N} \sum_{k = 1}^N \bigg(\delta_{j,k} + \frac{\partial_{b_{j}} w'(b_{k};A)}{w'(b_{k};A)}\bigg)\,.
$$

We finally return to $\Omega_j(z;A)$. Differentiating the relation $w(b_k;A) = a_k$ with respect to $a_i$ we get
$$
\partial_{a_i} w(b_{k};A) + \partial_{a_i}b_{k} w'(b_{k};A) = \delta_{i,k}\,.
$$
We insert it in \eqref{Ometil} and find a simplification
$$
\Omega_i(z;A) = \frac{1}{N} \frac{1}{w'(b_{i};A)}\,\frac{1}{(z - b_i)^2}\,.
$$

\begin{flushright}
$\Box$
\end{flushright}

A general property of the TR amplitudes, which we only state here for spectral curves of genus $0$, is the following:
\begin{theorem}\label{deformationsTR}\cite{EOFg}
Let $(\mathbb{P}^1,p_t,\lambda_t,B = \frac{\dd z_1\dd z_2}{(z_1 - z_2)^2})$ be a holomorphic family of initial data for TR, depending on a parameter $t \in \mathbb{C}$, and $\omega_{g,n}^{t}$ its TR amplitudes. Assume there exists a generalized cycle $\gamma$ in $\mathbb{P}^1$ whose support does not contain the zeroes of $\dd p$, such that 
$$
\partial_{t} \lambda_t(z) \dd p_t(z) - \partial_{t}p_t(z) \dd \lambda_t(z) = \int_{\gamma} B(z,\cdot).
$$
where the derivatives are taken with $z$ fixed.
Then, for any $g+n > 1$, $n,m > 0$,
$$
\partial_{t}^m \omega_{g,n}(z_1,\ldots,z_n) = -\int_{\gamma^m} \omega_{g,n + m}^{t}(z_1,\ldots,z_n,\zeta_1,\ldots, \zeta_m),
$$
where the derivatives are taken at $z_i$ fixed and the $\zeta_i$ are integrated over the cycles $\gamma$. 

For $n = 0$ and $g\geq 1$, or $g =0$ and $m\geq 3$, we have:
$$
\partial_{t}^m \mathcal{F}_g= \int_{\gamma^m} \omega_{g, m}^{t}(\zeta_1,\ldots, \zeta_m).
$$ 
\end{theorem}

We use the notation $\mathcal{F}_g[x,y]$ for the topological recursion $n=0$ invariants, emphasizing the spectral curve they come from. For instance, in Theorem~\ref{deformationsTR}, the last equation concerns $\mathcal{F}_{g}[p_{t},\lambda_{t}]$.
\begin{remark}
We remark that the Theorem~\ref{deformationsTR} cannot be applied to our spectral curve 
$$
(x(z;A),-w(z;A)), \text{ when } A\rightarrow 0.
$$ 
It may seem there is no problem from a naive perspective, but the behavior of the branchpoints is pathological in the limit, since they coalesce to $\infty$. This gives a setting in which the usual topological recursion does not apply.

On the other hand, the exchanged spectral curve $(w(z;A),x(z;A))$ behaves like a regular spectral curve to which we can apply the usual topological recursion, and hence Theorem~\ref{deformationsTR}. We denote by $\check{\omega}_{g,n}^A(z_1,\ldots,z_n)$ its corresponding TR amplitudes.
\end{remark}

If we specialize to $n = 0$ and the deformation \eqref{omedef}, we find
\begin{corollary}
\label{coniu}
We have for any $n \geq 1$
$$
\partial_{a_1}\cdots\partial_{a_n} \mathcal{F}_{g}^A[w,x] = \frac{\check{\omega}_{g,n}^{A}(z_1,\ldots,z_n)}{\dd w(z_1;A)\cdots \dd w(z_n;A)},
$$
where the $z_i$ are points in $\mathcal{C}$ defined by $w(z_i;A) = a_i$.
\end{corollary}
\noindent \textbf{Proof.} $\Omega_i$ represents the infinitesimal deformation of the TR initial data $(\lambda = w(z;A),p = x(z;A))$, and an equivalent form of \eqref{omedef} is
$$
\Omega_{i}(z;A)\dd z = \Res_{\zeta \rightarrow b_i} \frac{B(z,\zeta)}{w(\zeta;A) - w(b_i;A)}.
$$
Therefore, from Theorem \ref{deformationsTR}, we obtain
\bea
\partial_{a_1}\cdots\partial_{a_n} \mathcal{F}_{g}^A[w,x] & = & \Res_{\zeta_1 \rightarrow b_{1}} \cdots \Res_{\zeta_n \rightarrow b_n} \frac{\check{\omega}_{g,n}^{A}(\zeta_1,\ldots,\zeta_n)}{\prod_{i = 1}^n (w(\zeta_i;A) - w(b_i;A))} \nonumber \\
\label{reg} & = & \frac{\check{\omega}_{g,n}^{A}(b_1,\ldots,b_n)}{\dd w(b_1;A) \cdots \dd w(b_n;A)},
\eea 
where the differential acts on the first variable of $w$.
\hfill $\Box$

\subsection{Conclusion}

Now we give the technical condition under which our conjecture would be true for usual maps. For $n>0$:
\beq\label{technicalCond}
\partial_{a_1}\cdots\partial_{a_n}\hat{F}^{[g]}_A =\partial_{a_1}\cdots\partial_{a_n}\mathcal{F}_{g}^A[x,-w]=\partial_{a_1}\cdots\partial_{a_n}\mathcal{F}_{g}^A[w,x].
\eeq
Observe that we only need this equality for $n>0$, \textit{i.e.} for the non-constant Taylor coefficients.
We believe the first equality is true since the free energy coming from the matrix model and the TR invariants should differ by a constant not depending on $A$.
We also think the second equality is true and can be proved by checking that the correction terms of symplectic invariance, which are still being analyzed in general, do not depend on $A$.

In any case, the computations we did for quadrangulations to support our conjecture indicate that this difference is indeed zero for topology $(0,3)$ and that if there existed a non-zero difference in other cases, it would also have a combinatorial interpretation, which may help understand the nature of the still mysterious property of symplectic invariance.

We sketch the final argument which would lead to a proof of the conjecture if we suppose \eqref{technicalCond} is true.

We view $\mathcal{R}$ as a graded ring by assigning degree $1$ to each generator $q_i$.
Recall from \eqref{gsHMMEF} that, as an element of $\mathcal{R}$, the free energy decomposes as
$$
\hat{F}^{[g]}_A = \sum_{n \geq 1}\frac{1}{n!} \sum_{\ell_1,\ldots,\ell_n} H_{\ell_1,\ldots,\ell_n}^{[g]}\,\prod_{i = 1}^n \frac{p_{\ell_i}(A)}{N\ell_i},\qquad H_{\ell_1,\ldots,\ell_n}^{[g]} \in \mathcal{R}^0.
$$

\begin{corollary}
If we assume \eqref{technicalCond} is true, we have in $\tilde{\mathcal{R}}$, for $2g - 2 + n > 0$,
$$
\frac{\check{\omega}_{g,n}^{A = 0}(b_1,\ldots,b_n)}{\prod_{i = 1}^n\dd w(b_i;0)} = \sum_{\ell_1,\ldots,\ell_n \geq 1} H_{\ell_1,\ldots,\ell_n}^{[g]}\prod_{i = 1}^n a_i^{\ell_i - 1},\qquad \text{with}\quad w(b_i;0) = a_i.
$$
\end{corollary}
\noindent\textbf{Proof.} 
Let $n \geq 1$ be an integer. We denote $\pi_{\mathcal{R},n}$ the projection from $\mathcal{R}$ to its degree $n$ subspace. We introduce the ring $\mathcal{Q}^{(n)} = \mathcal{R}^0[[a_1,\ldots,a_n]][[q_1,q_2,\ldots]]$, where $q_i\equiv \tfrac{p_i(A)}{N}= \tfrac{{\rm Tr}\, A^i}{N}$. We make it a graded ring by assigning degree $1$ to each generator $q_{\ell}$.  We denote $\pi_{\mathcal{Q},0}$ the projection from $\mathcal{Q}^{(n)}$ to its degree~$0$ subspace. We can define a linear map $\Upsilon^{(n)}\,:\,\mathcal{R}^{\mathfrak{S}_{N}} \rightarrow \mathcal{Q}^{(n)}$ by
$$
\Upsilon^{(n)}(f) = a_1\partial_{a_1}\cdots a_n\partial_{a_n} f.
$$
The map $\Upsilon^{(n)}$ is homogeneous of degree $-n$, in particular it sends $p_{\lambda}$ with $\ell(\lambda) < n$ to zero. Besides, from the degree $n$ part to the degree $0$ part it induces an isomorphism, which is just the change of basis from power sums $\tfrac{p_i(A)}{N}$ to (unnormalized) symmetric monomials in the $a_i$'s, and we have
\beq
\label{comutd}\Upsilon^{(n)} \circ \pi_{\mathcal{R},n} = \pi_{\mathcal{Q},0}\circ \Upsilon^{(n)}.
\eeq

We would like to access
\beq
\label{dgfsg222}\Upsilon^{(n)} \circ \pi_{\mathcal{R},n}(\hat{F}^{[g]}_A) = \sum_{\ell_1,\ldots,\ell_n \geq 1} H_{\ell_1,\ldots,\ell_n}^{[g]}\,\prod_{i  = 1}^n a_i^{\ell_i}.
\eeq
Using the technical condition \eqref{technicalCond} and that $w(b_i;A) = a_i$, we obtain from Corollary~\ref{coniu} that
$$
\Upsilon^{(n)}(\hat{F}^{[g]}_A) = \check{\omega}_{g,n}^A(b_1,\ldots,b_n) \prod_{i = 1}^n \frac{ w(b_i;A)}{\dd w(b_i;A)}.
$$
According to \eqref{comutd} the degree $0$ part of the right-hand side computes the quantity in \eqref{dgfsg222}. The spectral curve $(x(z;A), w(z;A))$ is symmetric in $a_1,\ldots,a_{N}$. Therefore when we compute the right-hand side with TR, we obtain a formal series in $\tfrac{p_{\ell}(A)}{N}$, whose term of degree $0$ is
$$
\check{\omega}_{g,n}^{A = 0}(b_1,\ldots,b_n)\,\prod_{i = 1}^n \frac{w(b_i;0)}{\dd w(b_i;0)},\qquad {\rm with}\quad  w(b_i;0) = a_i.
$$ 
\hfill $\Box$

\part{Applications}

\section{Relation with free probability}
\label{SFree}
\label{Section10}

We explain how the results of Part 1 fit in the context of free probability, and give a possible application of our Conjecture~\ref{conj} and more general conjectures of Section \ref{Section6}. More concretely, we give a combinatorial interpretation of higher order free cumulants in terms of fully simple maps.

\subsection{Review of higher order free probability}

\label{Highfree}
Voiculescu \cite{Voiculescu2,Voiculescu3} introduced the notion of freeness (or free independence) to capture simultaneously a property of algebraic and probabilistic independence of two subsets $\mathcal{A}_{1},\mathcal{A}_{2}$ of a non-commutative probability space $(\mathcal{A},\varphi)$. Whereas Voiculescu's original approach is quite analytical and operator algebraic in nature, we will focus on the combinatorial aspects of free probability. These were first studied by Speicher \cite{Speicher}, who introduced the concept of free cumulants using the lattice of non-crossing partitions, in an analogous way as in classical probability moments can be expressed in terms of classical cumulants using partitions. 

As for classical independence, if $\mathcal{A}_{1},\mathcal{A}_{2}$ are free, moments of any element in the subalgebra generated by $\mathcal{A}_{1}$ and $\mathcal{A}_{2}$ can be computed solely in terms of moments of elements in $\mathcal{A}_{1}$, and moments of elements in $\mathcal{A}_{2}$, but the precise relation in terms of moments is a bit more intricate for the notion of free independence. Moreover, for $a, b \in \mathcal{A}$, the free cumulants $(k_{\ell}(a))_{\ell \geq 0}$ are determined by the moments $(\varphi(a^{\ell}))_{\ell \geq 0}$ and  tailored such that
$$
\forall \ell \geq 0,\qquad k_{\ell}(a + b) = k_{\ell}(a) + k_{\ell}(b), \;\; \text{ if } a \text{ and } b \text{ are free.}
$$

The precise definitions can be consulted in \cite{NicaSpeicher}, where the combinatorial approach to (first order) free probability is nicely and exhaustively introduced. 

The combinatorics behind the definition of free cumulants is compactly handled at the level of generating series by the $R$-transform. Departing from the usual normalizations in free probability, we introduce the \emph{Cauchy transform} and the $R$\emph{-transform}:
$$
G(x) = \sum_{\ell \geq 0} \frac{\varphi(a^{\ell})}{x^{\ell + 1}},\qquad \mathcal{R}(w) = \sum_{\ell \geq 1} k_\ell(a)\,w^{\ell - 1}.
$$
\begin{theorem} \cite{Voiculescu2} Free cumulants are computed in terms of moments via the following functional relation:
\label{thX1}
\beq\label{voiculescu}
\frac{1}{G(x)}+\mathcal{R}(G(x)) = x.
\eeq
\end{theorem}
The first formula \eqref{voiculescu} is well-known in free probability, was given by Voiculescu in \cite{Voiculescu2} and is sometimes referred to as the $R$-transform machinery.

This theory has been generalized to second order cumulants by Mingo and Speicher \cite{MingoSpeicher}, and Collins, Mingo, Speicher and \'{S}niady for higher order cumulants \cite{Secondorderfreeness}. They introduce a notion of higher order non-commutative probability space $\mathcal{A}$. Beyond a linear trace $\varphi = \varphi_{1}$, it is equipped with multilinear traces $\varphi_{n}\,:\,\mathcal{A}^{n} \rightarrow \mathbb{C}$ which are often called $n$-th order correlation moments. While first order free cumulants are defined using non-crossing partitions, second order free cumulants use annular non-crossing permutations and higher order free cumulants are defined in terms of the correlation moments through complicated combinatorial objects called partitioned permutations.

Roughly speaking, a family of pairwise disjoint subsets is free of all orders if all mixed free cumulants vanish.

An explicit characterization of freeness in terms of moments exists also for second order, but was not found for higher order.
The relation between second order moments and free cumulants is elucidated in \cite[Theorem 2.12]{Secondorderfreeness}:

\begin{theorem}
\label{thX2} Let
\bea
G(x_1,x_2) & = & \sum_{\ell_1,\ell_2 \geq 1} \frac{\varphi_{2}(a^{\ell_1},a^{\ell_2})}{x_1^{\ell_1 + 1}x_2^{\ell_2 + 1}}, \nonumber \\
\mathcal{R}(w_1,w_2) & = & \sum_{\ell_1,\ell_2 \geq 1} k_{\ell_1,\ell_2}(a,a)\,w_1^{\ell_1 - 1}w_2^{\ell_2 - 1}. \nonumber
\eea
The moment-cumulant relations for second order are equivalent to the functional relation:
\beq\label{speicher}
G(x_1,x_2)=G^{\prime}(x_1)G^{\prime}(x_2)\left(\mathcal{R}(G(x_1),G(x_2))+\frac{1}{(G(x_1)-G(x_2))^2}\right)-\frac{1}{(x_1-x_2)^2}.
\eeq
\end{theorem}
For $n \geq 3$, the computation of $n$-th order free cumulants via generating series has not been devised yet. Even if the conceptual framework is the same for any order, the complexity of the combinatorial objects involved makes the computations in higher orders too complicated.

This theory is partly driven by its application to asymptotics of unitarily invariant random matrices. Indeed, if $M$ and $\widetilde{M}$ are two independent hermitian random matrices of size $N$, and the distribution of $M$ is unitarily invariant, $M$ and $\widetilde{M}$ determine in the large $N$ limit (higher order) free elements of a (higher order) non-commutative probability space when this limit exists. A precise statement can be found in \cite{Voiculescu3} at first order, \cite{MMS07} at second order, and \cite{Secondorderfreeness} in what regards higher order.

\subsection{Enumerative interpretation of higher order free cumulants}\label{HOFCsFS}

Let $M=(M_N)_{N\in\mathbb{N}}$ be a unitarily invariant hermitian random matrix ensemble. The scaled large $N$ limits of classical cumulants of $n$ traces of powers of our matrices -- in case they exist for $n\geq 1$ --
\beq\label{correlationMom}
\varphi_{\ell_1,\ldots,\ell_n}^M\coloneqq \lim_{N\to \infty} N^{n-2}\kappa_n({\rm Tr}\, M_N^{\ell_1},\ldots,{\rm Tr}\, M_N^{\ell_n})
\eeq
define a higher order non-commutative probability space generated by a single element $M$. 
They are called $n$-th order correlation moments and constitute the limiting distribution of all orders of $M$, turning the space of hermitian unitarily invariant random matrix ensembles into a higher order probability space.

In the important setting of random matrices, Theorem~4.4 in \cite{Secondorderfreeness} expressed the $n$th order free cumulants also as scaled limits of classical cumulants, but this time of entries of the matrices $M_N=(M^{(N)}_{r,s})_{r,s=1}^{N}$:
\beq
k_{\ell_1,\ldots,\ell_n}^M = \lim_{N \rightarrow \infty} N^{n - 2 + \sum_{i} \ell_i} \kappa_{
\sum_{i = 1}^n \ell_i}\big((M^{(N)}_{i_{j,k},i_{j,k + 1\,\,{\rm mod}\,\,\ell_j}})\big),
\eeq
where $\gamma_j := (i_{j,1},\ldots,i_{j,\ell_j})_{j = 1}^n$ are pairwise disjoint cycles of respective lengths $\ell_j$. Note also that we can express the free cumulants in a more compact way in terms of the fully simple observables we introduced:

\begin{lemma}
\label{bgsfgK}We have
$$ 
\kappa_{\sum_{i = 1}^n \ell_i}\big(M_{i_{j,k},i_{j,k + 1\,\,{\rm mod}\,\,\ell_j}}\big) = \kappa_n\big(\mathcal{P}_{\gamma_1}(M),\ldots,\mathcal{P}_{\gamma_n}(M)\big).
$$
\end{lemma} 
\noindent \textbf{Proof.} Both sides can be expressed as a linear combination of terms of the form 
$$
\prod_{\alpha}\Big\langle \prod_{(j,k) \in I_{\alpha}} M_{i_{j,k},i_{j,k + 1\,\,{\rm mod}\,\,\ell_j}}\Big\rangle,
$$
where $(I_{\alpha})_{\alpha}$ is a partition of $\mathbf{I} = \big\{(j,k)\,\,\big|\,\,1 \leq j \leq n,\,\,1 \leq k \leq \ell_j\big\}$. The term corresponding to the partition consisting of a single set $I = \mathbf{I}$ appears on both sides with a coefficient $1$. We have in general
\bea
\label{Iaprof}&& \Big\langle \prod_{(j,k) \in I_{\alpha}} M_{i_{j,k},i_{j,k + 1\,\,{\rm mod}\,\,\ell_j}}\Big\rangle \\
& = & \sum_{b\,:\,I_{\alpha} \rightarrow \llbracket 1,N \rrbracket} \Big\langle \prod_{(j,k) \in I_{\alpha}} \lambda_{b_{j,k}} \Big\rangle  \int_{U_N} \dd U\,\bigg[\prod_{(j,k) \in I_{\alpha}} U_{i_{j,k},b_{j,k}} U^{\dagger}_{b_{j,k},i_{j,k + 1\,\,{\rm mod}\,\,\ell_j}} \bigg], \nonumber 
\eea
where we have used the $U_N$-invariance of the distribution of $M$, and $(\lambda_{a})_{a = 1}^N$ are the (unordered) eigenvalues of $M$. The integral over $U_N$ is computed by Weingarten calculus, Theorem~\ref{UNmoment}. To match the notations of Theorem~\ref{UNmoment} we have
\bea
A_{\alpha} & := & \{a_{l}\,\,|\,\,1 \leq l \leq N\} = \{i_{j,k}\,\,|\,\,(j,k) \in I_{\alpha}\} \nonumber \\
A'_{\alpha} & := & \{a'_{l}\,\,|\,\,1 \leq l \leq N\} = \{i_{j,k + 1\,\,{\rm mod}\,\,\ell_j}\,\,|\,\,(j,k) \in I_{\alpha}\} \nonumber
\eea
and the product of Kronecker deltas in Theorem~\ref{UNmoment} tells us that if $A_{\alpha} \neq A'_{\alpha}$ as subsets of $\llbracket 1,N \rrbracket$, then \eqref{Iaprof} will be zero. This is indeed the case when $I_{\alpha}$ is strictly included in $\mathbf{I}$,  for the $i_{j,k}$ are pairwise disjoint. This claimed equality follows.
\hfill $\Box$

\vspace{0.1cm}

We have hence expressed the $n$th order correlation functions and the $n$th order free cumulants for a higher order probability space given by a hermitian unitarily invariant random matrix ensemble in terms of the connected ordinary and fully simple correlators that we defined (in their disconnected versions) in \eqref{Obsdisc}.

More concretely, for a general measure \eqref{eq:mesS}, the propositions ~\ref{propStuffed} (for ordinary stuffed) and \ref{PPPPP} (for fully simple stuffed) represent the $n$th order correlation functions and the $n$th order free cumulants as generating series of planar ordinary and fully simple stuffed maps, respectively:
\bea\label{FP-Maps1}
\varphi_{\ell_1,\ldots,\ell_n}^M & = & \widehat{F}_{\ell_1,\ldots,\ell_n},\\
k_{\ell_1,\ldots,\ell_n}^M & = & \widehat{H}_{\ell_1,\ldots,\ell_n}.\label{FP-Maps2}
\eea

For the particular case of matrix ensembles given by the measure \eqref{eq:mes}, we obtain generating series of planar ordinary and fully simple usual maps instead.

\subsection{$R$-transform machinery in terms of maps}

With the identifications \eqref{FP-Maps1}-\eqref{FP-Maps2} we just exposed in mind, we see that the relations from Propositions~\ref{0,1} and \ref{0,2} between fully simple and ordinary generating series in the case of disks and cylinders recover the important $R$-transform formulas from Theorems~\ref{voiculescu} and ~\ref{speicher}, which related the generating series of first and second order moments and free cumulants. The relation between the generating series of fully simple disks $X$ and the $R$-transform (generating series of free cumulants) $\mathcal{R}$ is as follows: $\mathcal{R}(w)=X(w)-w^{-1}$. We have proved the formulas relating $X_{1}$ with $W_1$  and $X_{2}$ with $W_2$ via combinatorics of maps -- instead of non-crossing partitions -- independently of \cite{Secondorderfreeness}. We remark that in the context of topological recursion it is clear that the second order formula \eqref{speicher} can be re-written in a symmetric way, which was not obvious from the free probability point of view. We recall here our formula for cylinders in this symmetric form:
\beq
W_2(x_1,x_2)\dd x_1 \dd x_2 + \frac{\dd x_1 \dd x_2}{(x_1-x_2)^2} = X_2(w_1,w_2)\dd w_1 \dd w_2 + \frac{\dd w_1 \dd w_2}{(w_1-w_2)^2}.
\eeq
Recovering these formulas was one of the first motivations for considering fully simple maps.

Although Weingarten calculus and the HCIZ integral is also used in \cite{Secondorderfreeness} to relate higher order free cumulants to moments, we have explained in Section~\ref{SecM} that the relation is naturally expressed in terms of monotone Hurwitz numbers. As Hurwitz theory develops rapidly, this fact may give insight into the structure of higher order cumulants generating series. 

Our Conjecture~\ref{conj} applies to matrix ensembles in which $\ln\big(\frac{\dd\mu(M)}{\dd M}\big)$ is linear in the trace of powers of $M$, \textit{i.e.} with $T_{h,k} = 0$ for $(h,k) \neq (0,1)$, and it turns into Conjecture~\ref{ConT02} if we turn on $T_{0,2}$ as well. These are the models governed by the topological recursion, and whose combinatorics is captured by usual maps, or maps carrying a loop model. These conjectures would therefore provide a computational tool for the free cumulants in these models, via the topological recursion restricted to genus $0$: going free amounts to performing the exchange transformation $(x,w) \mapsto (w,x)$. 

More general unitarily invariant ensembles are rather governed by the blobbed topological recursion of \cite{Bstuff} and related to stuffed maps. Concretely, the initial data for the blobbed topological recursion is a spectral curve as in \eqref{inidataordmap}, supplemented with blobs $(\phi_{g,n})_{2g - 2 + n > 0}$ which play the role of extra initial data intervening in topology $(g,n)$ and beyond. For matrix ensembles of the form \eqref{eq:mesS}, there exist specific values for the blobs, such that the blobbed topological recursion computes the large $N$ expansion of the correlators. It would be interesting to know if the $x \leftrightarrow y$ can be supplemented with a transformation of the blobs, in such a way that the blobbed topological recursion for the transformed initial data computes the free cumulants. Restricting to genus $0$, it would give a computational scheme to handle free cumulants of any order -- in full generality -- via the blobbed topological recursion. 

The higher genus theory should capture finite size corrections to freeness, and the universality of the topological recursion suggests that it may be possible to formulate a universal theory of approximate freeness, for which unitarily invariant matrix ensembles would provide examples.

Generalizing the $R$-transform machinery to higher order free cumulants proved to be a very complicated problem and in the most general article \cite{Secondorderfreeness} they only managed to do it for second order, and in a quite intricate way. If our conjecture for usual maps~\ref{conj} is true, topological recursion for the pair of pants, \textit{i.e.} for the topology $(0,3)$, gives directly the $R$-transform formula of third order \ref{03formula} for the specific measure \eqref{eq:mes}, as illustrated in Section~\ref{RtransPairPants}. This is already interesting to the free probability community and gives a hint on how to generalize those complicated relations.

\section{An ELSV-like formula for monotone Hurwitz numbers}

\label{Section11}

ELSV-type formulas relate connected Hurwitz numbers to the intersection theory of the moduli space of stable curves $\overline{\mathcal{M}}_{g,n}$, enabling the tranfer of results from one world to the other. In this section, we provide a new ELSV-like formula for $2$-orbifold monotone Hurwitz numbers.

\subsection{Connected $2$-orbifold monotone Hurwitz numbers}

We are going to consider double strictly monotone Hurwitz numbers, with ramification profiles $\mu$ arbitrary and $\lambda = (2,\ldots,2)$: $[E_{k}]_{\mu,(2,\ldots,2)}$, which are called {\it $2$-orbifold} strictly monotone Hurwitz numbers. Notice that
$$
|{\rm Aut}\,(2,\ldots,2)| = 2^{L/2}\,(L/2)!,\text{ with } L = |\mu|, \text{ since } \ell((2,\ldots,2))=\frac{|\mu|}{2},
$$ 
and remember that $2^{L/2}\,(L/2)!\,[E_k]_{\mu,(2,\ldots,2)}$ is the number of strictly monotone $k$-step paths from $C_{\mu}$ to some (arbitrary but) fixed permutation $\sigma \in C_{(2,\ldots,2)}$ in the Cayley graph of $\mathfrak{S}_{L}$. We say that such a path is connected when the group generated by all permutations met along the path acts transitively on $\llbracket 1,L \rrbracket$.  This matches the usual definition of connectedness in the language of branched coverings. We define $2^{L/2}(L/2)!\,[E^{\circ}_{k}]_{\mu,(2,\ldots,2)}$ to solve the same weighted enumeration but restricted to connected paths. We can express these disconnected Hurwitz numbers in terms of the connected ones as follows. A disconnected path from $C_{\mu}$ to $\sigma$ can be broken in connected components, i.e in paths in $\prod_{i = 1}^{s} \mathfrak{S}_{J_i}$ where $(J_i)_{i = 1}^{s}$ is an unordered partition of $\llbracket 1,L \rrbracket$ into subsets such that each $2$-cycle in $\sigma$ is contained in some $J_i$. In particular these sets all have even cardinalities. Each connected component starts from a conjugacy class $C_{\mu^{(i)}}$, where $(\mu^{(i)})_{i = 1}^s$ are partitions whose concatenation is $\mu$, and ends at $\sigma|_{J_i}$ -- which has again type $(2,\ldots,2)$ with $\ell((2,\ldots,2))=\frac{|\mu^{(i)}|}{2}$. Therefore, we get
$$ 
2^{\frac{L}{2}}(L/2)!\,[E_k]_{\mu,(2,\ldots,2)} =\sum_{s \geq 1} \frac{1}{s!} \sum_{\substack{\mu^{(1)} \cup \cdots \cup \mu^{(s)} = \mu \\ k_1 + \cdots + k_s = k}} \frac{\frac{L}{2}!}{\frac{|\mu^{(1)}|}{2}!\cdots \frac{|\mu^{(s)}|}{2}!}\,\prod_{i = 1}^{s} 2^{|\mu^{(i)}|/2}\,\big(|\mu^{(i)}|/2\big)!\,[E^{\circ}_{k_i}]_{\mu^{(i)},(2,\ldots,2)}.
$$
Hence, the symmetry factors disappear and we get
$$ 
[E_k]_{\mu,(2,\ldots,2)} = \sum_{s \geq 1} \frac{1}{s!} \sum_{\substack{\mu^{(1)} \cup \cdots \cup \mu^{(s)} = \mu \\ k_1 + \cdots + k_s = k}} \prod_{i = 1}^s [E^{\circ}_{k_i}]_{\mu^{(i)},(2,\ldots,2)}.
$$
It is convenient to rename $[E^{\circ,g}]_{\mu,(2,\ldots,2)}$ the $2$-orbifold connected Hurwitz numbers of genus $g$, which is related to the above $k$ by Riemann-Hurwitz formula in the branched covering interpretation:
$$
k = 2g - 2 + \ell(\mu) + \frac{|\mu|}{2}.
$$

\subsection{GUE and monotone Hurwitz numbers}
The Gaussian Unitary Ensemble is the probability measure on the space of hermitian matrices of size $N$
\bea
\dd\mu(M) & = &  2^{-\frac{N}{2}}\pi^{-\frac{N^2}{2}}\dd M\,\exp(- N\,{\rm Tr}\,\tfrac{M^2}{2}) \nonumber \\
& = & 2^{-\frac{N}{2}}\pi^{-\frac{N^2}{2}}\prod_{i = 1}^N \dd M_{i,i}\,e^{-NM_{i,i}^2/2} \prod_{i < j} \dd {\rm Re}\,M_{i,j}\,e^{-N({\rm Re} M_{i,j})^2}  \dd {\rm Im}\,M_{i,j}\,e^{- N({\rm Im}\,M_{i,j})^2}. \nonumber 
\eea
In this section we consider expectation values and cumulants with respect to the GUE measure.

We recall that the cumulants of the GUE have a topological expansion
$$
\kappa_{n}({\rm Tr}\,M^{\mu_1},\ldots,{\rm Tr}\,M^{\mu_n}) = \sum_{g \geq 0} N^{2 - 2g - n}\,\kappa_{n}^{[g]}({\rm Tr}\,M^{\mu_1},\ldots,{\rm Tr}\,M^{\mu_n}).
$$
For any fixed positive integers $\mu_i$, this sum is finite. The coefficients $\kappa_{n}^{[g]}$ count genus $g$ maps whose only faces are the $n$ marked faces, and are sometimes called generalized Catalan numbers. They have been extensively studied by various methods \cite{LehmanWalsh,HarerZagier,GNica,ChapuyZ,AndersenGaussian}. We are able to relate them to $2$-orbifold strictly monotone Hurwitz numbers, observing that the only fully simple maps without internal faces are the degenerate ones:

\begin{proposition}
\label{propdh} For any $g \geq 0$, $n \geq 1$ and partition $\mu$,
$$
\frac{\kappa_{n}^{[g]}({\rm Tr}\,M^{\mu_1},\ldots,{\rm Tr}\,M^{\mu_n})}{|{\rm Aut}\,\mu|} = [E^{\circ,g}]_{\mu,(2,\ldots,2)},
$$
where $[E_{g}^{\circ}]_{\mu,\lambda}$ are the connected double weakly monotone Hurwitz numbers of genus $g$. 
\end{proposition}
\noindent \textbf{Proof.} As the entries of $M$ are independent and gaussian, we can easily evaluate
$$
\langle \mathcal{P}_{\lambda}(M) \rangle = \prod_{i = 1}^{\ell(\lambda)} \frac{\delta_{\lambda_i,2}}{N}.
$$
From Theorem~\ref{Transi} we deduce
$$
|{\rm Aut}\,\mu|^{-1}\Big\langle \prod_{i = 1}^n {\rm Tr}\,M^{\mu_i}\Big \rangle =N^{\frac{|\mu|}{2}}\sum_{k \geq 0} N^{-k}[E_{k}]_{\mu,(2,\ldots,2)}.
$$
For the purpose of this proof, we name $\tilde{p}_{\mu}$ the power sum basis of $\mathcal{B}$. We have
\bea
\sum_{\mu} \frac{\langle p_{\mu}(M) \rangle \tilde{p}_{\mu}}{|{\rm Aut}\,\mu|} & = &  \Big\langle \exp\Big(\sum_{m \geq 1}  \frac{{\rm Tr}\,M^m\,\tilde{p}_{m}}{m}\Big)\Big\rangle \\ 
& = & \exp\bigg(\sum_{n \geq 1} \frac{1}{n!} \sum_{m_1,\ldots,m_n \geq 1} \kappa_{n}({\rm Tr}\,M^{m_1},\cdots,{\rm Tr}\,M^{m_n}) \prod_{i = 1}^n \frac{\tilde{p}_{m_i}}{m_i}\bigg) \nonumber \\
& = & \exp\bigg(\sum_{\mu} \frac{\kappa(p_{\mu_1}(M),\ldots,p_{\mu_{\ell(\mu)}}(M))\,\tilde{p}_{\mu}}{|{\rm Aut}\,\mu|}\bigg) \nonumber \\
& = & \exp\bigg(\sum_{\mu} \sum_{g \geq 0} N^{2 - 2g - \ell(\mu)}\,\kappa_{n}^{[g]}({\rm Tr}\,M^{\mu_1},\ldots,{\rm Tr}\,M^{\mu_{\ell(\mu)}})\,\frac{\tilde{p}_{\mu}}{|{\rm Aut}\,\mu|} \bigg).
\eea 
But on the other hand we have
\bea
\sum_{\mu} \frac{\langle p_{\mu}(M) \rangle\,\tilde{p}_{\mu}}{|{\rm Aut}\,\mu|} & = & \sum_{\mu} N^{\frac{|\mu|}{2}} \Big(\sum_{k \geq 0} N^{-k}[E_{k}]_{\mu,(2,\ldots,2)}\Big)\tilde{p}_{\mu} \nonumber \\
& = & \sum_{s \geq 1} \frac{1}{s!} \sum_{\substack{\mu^{(1)},\ldots,\mu^{(s)} \\ k_1,\ldots,k_s \geq 0}} \prod_{i = 1}^s N^{\frac{|\mu^{(i)}|}{2} -k_i}\,[E_{k_i}^{\circ}]_{\mu^{(i)},(2,\ldots,2)}\tilde{p}_{\mu^{(i)}} \nonumber \\
& = & \exp\bigg(\sum_{\mu} \sum_{k \geq 0} N^{\frac{|\mu|}{2} - k} [E_{k}^{\circ}]_{\mu,(2,\ldots,2)} \tilde{p}_{\mu} \bigg) \nonumber \\
& = & \exp\bigg(\sum_{\mu}  \sum_{g \geq 0} N^{2 - 2g - \ell(\mu)}\,[E^{\circ,g}]_{\mu,(2,\ldots,2)} \tilde{p}_{\mu} \bigg). \nonumber
\eea
Comparing the two formulas yields the claim. \hfill $\Box$
\vspace{0.2cm}

This specialization of our result recovers a particular case of \cite[Prop. 4.8]{ALS} which says that the enumeration of hypermaps is equivalent to the strictly monotone orbiforld Hurwitz problem. This suggests it is natural to investigate if our results can be extended to the more general setting of hypermaps.

It is well-known that the GUE correlation functions are computed by the topological recursion for the spectral curve,
$$
\mathcal{C} = \mathbb{P}^1,\qquad p(z) = z + \frac{1}{z},\qquad \lambda(z) = \frac{1}{z},\qquad B(z_1,z_2) = \frac{\dd z_1\dd z_2}{(z_1 - z_2)^2}\,.
$$ 
Therefore, Proposition~\ref{propdh} gives a new proof that the $2$-orbifold strictly monotone Hurwitz numbers are computed by the topological recursion, a fact already known as a special case of more general results, see \textit{e.g.} \cite{DOPS14,EynardHarnad}.

\subsection{GUE and Hodge integrals}

Dubrovin, Liu, Yang and Zhang \cite{DiYang} recently discovered a relation between Hodge integrals and the even GUE moments. For $2g - 2 + n > 0$, the Hodge bundle $\mathbb{E}$ is the holomorphic vector bundle over Deligne-Mumford compactification of the moduli space of curves $\overline{\mathcal{M}}_{g,n}$ whose fiber above a curve with punctures $(\mathcal{C},p_1,\ldots,p_n)$ is the $g$-dimensional space of holomorphic $1$-forms on $\mathcal{C}$. We denote
$$
\Lambda(t) = \sum_{j = 0}^{g} {\rm c}_{j}(\mathbb{E})\,t^{j}
$$
its Chern polynomial. Let $\psi_i$ be the first Chern class of the line bundle $T^*_{p_i}\mathcal{C}$, and introduce the formal series
$$ 
Z_{{\rm Hodge}}(t;\hbar) = \exp\bigg(\sum_{2g - 2 + n > 0} \frac{\hbar^{2g - 2 + n}}{n!} \sum_{i_1,\ldots,i_n \geq 0} \int_{\overline{\mathcal{M}}_{g,n}} \Lambda(-1)\Lambda(-1)\Lambda(\tfrac{1}{2}) \prod_{i = 1}^n \psi_{i}^{d_i} t_{d_i}\bigg).
$$ 
On the GUE side, the cumulants have a topological expansion
$$
\kappa_{n}({\rm Tr}\,M^{\ell_1},\ldots,{\rm Tr}\,M^{\ell_n}) = \sum_{g \geq 0} N^{2 - 2g - n}\,\kappa_{n}^{[g]}({\rm Tr}\,M^{\ell_1},\ldots,{\rm Tr}\,M^{\ell_n}),
$$
where $\kappa_{n}^{[g]}$ are independent of $N$, and the sum is always finite. We introduce the formal series 
\beq 
\label{Zeven} Z_{{\rm even}}(s;N) = \frac{e^{-A(s;N)}\,\prod_{j = 1}^{N - 1}j!}{2^{N}\pi^{\frac{N(N + 1)}{2}}}\,\int_{\mathcal{H}_{N}} \dd M\,\exp\bigg[N\,{\rm Tr}\,\Big(-\tfrac{M^2}{2} + \sum_{j \geq 1} s_{j}\,{\rm Tr}\,M^{2j}\Big)\bigg],
\eeq
where we choose
\bea  
A(s;N) & =  & \frac{\ln N}{12} - \zeta'(-1) \nonumber \\
&& + N^2\bigg[-\frac{3}{4} + \sum_{j \geq 1} \frac{1}{j + 1}\,{2j \choose j}s_j + \frac{1}{2} \sum_{j_1,j_2 \geq 1} \frac{j_1j_2}{j_1 + j_2}{2j_1 \choose j_1}{2j_2 \choose j_2}s_{j_1}s_{j_2}\bigg]. \nonumber
\eea
The normalization factor in \eqref{Zeven} is related to the volume of $U_N$ and the factor $e^{-A(s;N)}$ cancels the non-decaying terms in its large $N$ asymptotics, as well as the contributions of $\kappa_{1}^{[0]}$ and $\kappa_{2}^{[0]}$. The large $N$ asymptotics of the outcome  reads
\bea
Z_{{\rm even}}(s;N) & = &  \exp\bigg( \sum_{g \geq 2} \frac{N^{2 - 2g}B_{2g}}{4g(g - 1)} \nonumber \\
&& + \sum_{2g - 2 + n > 0} \frac{N^{2g - 2 + n}}{n!}\,\sum_{\ell_1,\ldots,\ell_n \geq 0}  \kappa_n^{[g]}({\rm Tr}\,M^{2\ell_1},\ldots,{\rm Tr}\,M^{2\ell_n}) \prod_{i = 1}^n s_{\ell_i} \bigg),  \nonumber 
\eea 
and we consider it as an element of $N^2\mathbb{Q}[N^{-1}][[s_1,s_2,\ldots]]$.

\begin{theorem} \cite{DiYang}
\label{thDi}With the change of variable
$$
T_{i,\pm}(s;N) = \sum_{k \geq 1} k^{i + 1}{2k \choose k} s_{k} - \delta_{i \geq 2} \pm \frac{\delta_{i,0}}{2N},
$$
we have the identity of formal series
$$
Z_{{\rm even}}(s;N) = Z_{{\rm Hodge}}(T_+(s;N),\sqrt{2}N^{-1}) Z_{{\rm Hodge}}(T_{-}(s;N),\sqrt{2}N^{-1}).
$$
\end{theorem}

We can extract from this result an explicit formula for the even GUE moments, which gives an ELSV-like formula for the monotone Hurwitz numbers with even ramification above $\infty$, and $(2,\ldots,2)$ ramification above $0$.
\begin{corollary}
For $g \geq 0$ and $n \geq 1$ such that $2g - 2 + n > 0$ and $m_1,\ldots,m_n \geq 0$, we have
\bea
&& |{\rm Aut}\,(2m_1,\ldots,2m_n)|\,[E^{\circ}_{g}]_{(2m_1,\ldots,2m_n),(2,\ldots,2)} \nonumber \\
& = &\!\!\! \kappa_n^{[g]}({\rm Tr}\,M^{2m_1},\ldots,{\rm Tr}\,M^{2m_n}) \nonumber \\
& = &\!\!\! 2^{g} \int_{\overline{\mathcal{M}}_{g,n}} [\Delta]\,\cap\,\Lambda(-1)\Lambda(-1)\Lambda(\tfrac{1}{2})\exp\Big(-\sum_{d \geq 1} \frac{\kappa_{d}}{d}\Big) \prod_{i = 1}^n \frac{m_i {2m_i \choose m_i}}{1 - m_i\psi_i}, \nonumber
\eea
where $\kappa_{d}$ are the pushforwards of $\psi_{n + 1}^{d + 1}$ via the morphism forgetting the last puncture. Denoting $[\Delta_h]$ the class of the boundary strata $\overline{\mathcal{M}}_{g - h,n + 2h} \subset \overline{\mathcal{M}}_{g,n}$ which comes from the pairwise gluing of the last $2h$ punctures, we have introduced
$$ 
[\Delta] = \sum_{h \geq 0} \frac{[\Delta_h]}{2^{3h}(2h)!}.
$$
\end{corollary}
\noindent \textbf{Proof.} Identifying the coefficient of $\frac{N^{2 - 2g - n}}{n!}\,s_{m_1}\cdots s_{m_n}$ in Theorem~\ref{thDi} yields
\bea
\label{kappaaaa}&& \kappa_{n}^{[g]}({\rm Tr}\,M^{2m_1},\ldots,{\rm Tr}\,M^{2m_n}) \nonumber \\
& = & \sum_{h = 0}^{\lfloor \frac{g}{2} \rfloor} \sum_{\ell \geq 0} \frac{2^{g - 3h}}{(2h)!\ell!} \int_{\overline{\mathcal{M}}_{g - h,n + \ell + 2h}} \Lambda(-1)\Lambda(-1)\Lambda(\tfrac{1}{2}) \prod_{i = 1}^n \frac{m_i {2m_i \choose m_i}}{1 - m_i\psi_i} \prod_{i = n + 1}^{n + \ell} \frac{-\psi_{i}^2}{1 - \psi_i}.
\eea
This sum is actually finite as the degree of the class to integrate goes beyond the dimension of the moduli space. We can get rid of the $\ell$ factors of $\psi$-classes by using the pushforward relation
\beq 
\label{relka} (\pi_{\ell})_*\Big(X \prod_{i = 1}^{\ell} \psi_i^{d_i + 1}\Big) = X\,\bigg(\sum_{\sigma \in \mathfrak{S}_{\ell}} \prod_{\gamma \in \mathcal{C}(\sigma)} \kappa_{\sum_{i \in \gamma} d_i}\bigg),
\eeq
where $\pi_{\ell}\,:\,\overline{\mathcal{M}}_{g',k + \ell} \rightarrow \overline{\mathcal{M}}_{g',k}$ is morphism forgetting the last $\ell$ punctures, and $X$ is the pullback via $\pi_{\ell}$ of an arbitrary class on $\overline{\mathcal{M}}_{g',k}$. In general, if we introduce formal variables $\hat{a}_{1},\hat{a}_{2},\ldots$, we deduce from \eqref{relka} the relation
$$
\sum_{\ell \geq 0} \frac{1}{\ell!} \sum_{d_1,\ldots,d_{\ell} \geq 1} (\pi_{\ell})_{*}\bigg(X\prod_{i = k + 1}^{k + \ell} \psi_{i}^{d_i + 1}\bigg) \prod_{i = 1}^{\ell} \hat{a}_{d_j} = X \exp\Big(\sum_{d \geq 1} a_{d}\kappa_{d}\Big),
$$
where
$$
\sum_{d \geq 1} a_{d}v^{d} = -\ln\Big(1 - \sum_{d \geq 1} \hat{a}_{d}v^{d}\Big).
$$
To simplify \eqref{kappaaaa} we should apply this relation with $\hat{a}_{d} = -1$ for all $d \geq 1$. Therefore $a_{d} = -\frac{1}{d}$. Consequently
\bea
&&  \kappa_{n}^{[g]}({\rm Tr}\,M^{2m_1},\cdots,{\rm Tr}\,M^{2m_n}) \nonumber \\
& = & \sum_{h = 0}^{\lfloor \frac{g}{2} \rfloor} \frac{2^{g - 3h}}{(2h)!} \int_{\overline{\mathcal{M}}_{g - h,n + 2h}} \Lambda(-1)\Lambda(-1)\Lambda(\tfrac{1}{2})\exp\Big(-\sum_{d \geq 1} \frac{\kappa_{d}}{d}\Big) \prod_{i = 1}^n \frac{m_i {2m_i \choose m_i}}{1 - m_i\psi_i}. \nonumber
\eea
The claim is a compact rewriting of this formula, using the pushforward via the inclusions $\iota_{h}\,:\,\overline{\mathcal{M}}_{g - h,n + 2h} \rightarrow \overline{\mathcal{M}}_{g,n}$. \hfill $\Box$

\bibliographystyle{plain}
\bibliography{BibliSimple}

\begin{thebibliography}{10}

\bibitem{EynardHarnad}
A.~Alexandrov, G.~Chapuy, B.~Eynard, and J.~Harnad.
\newblock Fermionic approach to weighted {H}urwitz numbers and topological
  recursion.
\newblock 2017.
\newblock math-ph/1706.00958.

\bibitem{ALS}
A.~Alexandrov, D.~Lewa\'nski, and S.~Shadrin.
\newblock Ramifications of {H}urwitz theory, {K}{P} integrability and quantum
  curves.
\newblock {\em The Journal of High Energy Physics}, 5, 2016.
\newblock math-ph/1512.07026.

\bibitem{ACM}
J.~Ambj{\o}rn, L.O. Chekhov, and Yu. Makeenko.
\newblock Higher genus correlators from the hermitian {$1$}-matrix model.
\newblock {\em Phys. Lett. B}, 282:341--348, 1992.
\newblock hep-th/9203009.

\bibitem{AndersenGaussian}
J.E. Andersen, L.O. Chekhov, P.~Norbury, and R.C. Penner.
\newblock Topological recursion for {G}aussian means and cohomological field
  theories.
\newblock {\em Theor. Math. Phys.}, 185(3):1685--1717, 2015.
\newblock math-ph/1512.09309.

\bibitem{Balentekin}
A.B. Balentekin.
\newblock Character expansions, {I}tzykson-{Z}uber integrals, and the {QCD}
  partition function.
\newblock {\em Phys. Rev. D}, 62(085017), 2000.
\newblock hep-th/0007161.

\bibitem{BernardiFusy}
O.~Bernardi and \'E. Fusy.
\newblock Bijections for planar maps with boundaries.
\newblock {\em J. Combin. Theory Ser. A}, 158:176--227, 2018.

\bibitem{Bstuff}
G.~Borot.
\newblock Formal multidimensional integrals, stuffed maps, and topological
  recursion.
\newblock {\em Annales Institut Poincar\'e - D}, 1(2):225--264, 2014.
\newblock math-ph/1307.4957.

\bibitem{BBG3}
G.~Borot, J.~Bouttier, and E.~Guitter.
\newblock Loop models on random maps via nested loops: case of domain symmetry
  breaking and application to the {P}otts model.
\newblock {\em J. Phys. A: Math. Theor}, 2012.
\newblock Special issue: Lattice models and integrability: in honour of F.Y.
  Wu, math-ph/1207.4878.

\bibitem{BBG12b}
G.~Borot, J.~Bouttier, and E.~Guitter.
\newblock More on the {$O(n)$} model on random maps via nested loops: loops
  with bending energy.
\newblock {\em J. Phys. A: Math. Theor.}, 45(275206), 2012.
\newblock math-ph/1202.5521.

\bibitem{BEOn}
G.~Borot and B.~Eynard.
\newblock Enumeration of maps with self avoiding loops and the {$O(n)$} model
  on random lattices of all topologies.
\newblock {\em J. Stat. Mech.}, (P01010), 2011.
\newblock math-ph/0910.5896.

\bibitem{BEMS}
G.~Borot, B.~Eynard, M.~Mulase, and B.~Safnuk.
\newblock A matrix model for simple {H}urwitz numbers, and topological
  recursion.
\newblock {\em J. Geom. Phys.}, 61(2):522--540, 2011.
\newblock math-ph/0906.1206.

\bibitem{BEO}
G.~Borot, B.~Eynard, and N.~Orantin.
\newblock Abstract loop equations, topological recursion, and applications.
\newblock {\em Commun. Number Theory and Phys.}, 9(1):51--187, 2015.
\newblock math-ph/1303.5808.

\bibitem{BSblob}
G.~Borot and S.~Shadrin.
\newblock Blobbed topological recursion : properties and applications.
\newblock {\em Math. Proc. Cam. Phil. Soc.}, 162(1):39--87, 2017.
\newblock math-ph/1502.00981.

\bibitem{BKMP}
V.~Bouchard, A.~Klemm, M.~Mari{\~{n}}o, and S.~Pasquetti.
\newblock Remodeling the {B}-model.
\newblock {\em Commun. Math. Phys.}, 287(1):117--178, 2009.
\newblock hep-th/0709.1453.

\bibitem{BIPZ}
\'{E}. Br{\'{e}}zin, C.~Itzykson, G.~Parisi, and J.-B. Zuber.
\newblock Planar diagrams.
\newblock {\em Commun. Math. Phys.}, 59:35--51, 1978.

\bibitem{Nonsep}
W.G. Brown.
\newblock On the enumeration of non-planar maps.
\newblock {\em Mem. Amer. Math. Soc.}, 65:42, 1966.

\bibitem{Nonsep2}
J.~Cai and Y.~Liu.
\newblock Enumeration of nonseparable planar maps.
\newblock {\em Europ. J. Combinatorics}, 23:881--889, 2002.

\bibitem{Carrell}
S.R. Carrell.
\newblock Diagonal solutions to the {$2$}-{T}oda hierarchy.
\newblock {\em Math. Res. Lett.}, 22:439--465, 2015.
\newblock math.CO/1109.1451.

\bibitem{ChapuyZ}
G.~Chapuy.
\newblock A new combinatorial identity for unicellular maps, via a direct
  bijective approach.
\newblock {\em Adv. Appl. Math.}, 47(4):874--893, 2011.
\newblock math.CO/1006.5053.

\bibitem{C05}
L.O. Chekhov.
\newblock Hermitean matrix model free energy: {F}eynman graph technique for all
  genera.
\newblock {\em JHEP}, (0603:014), 2006.
\newblock hep-th/0504116.

\bibitem{CollinsWeingarten}
B.~Collins.
\newblock Moments and cumulants of polynomial random variables on unitary
  groups, the {I}tzykson-{Z}uber integral and free probability.
\newblock {\em Int. Math. Res. Not.}, 17:953--982, 2003.
\newblock math-ph/0205010.

\bibitem{Secondorderfreeness}
B.~Collins, J.~Mingo, P.~{\'S}niady, and R.~Speicher.
\newblock Second order freeness and fluctuations of random matrices {III}.
  {H}igher order freeness and free cumulants.
\newblock {\em Documenta Math.}, 2:1--70, 2007.
\newblock math.OA/0606431.

\bibitem{dFZJ}
P.~di~Francesco, P.~Ginsparg, and J.~Zinn-Justin.
\newblock {$2d$} gravity and random matrices.
\newblock {\em Phys. Rep.}, 254(1), 1994.
\newblock hep-th/9306153.

\bibitem{DiYang}
B.~Dubrovin, S.-Q. Liu, D.~Yang, and Y.~Zhang.
\newblock Hodge-{GUE} correspondence and the discrete {K}d{V} equation.
\newblock 2016.
\newblock math-ph/1612.02333.

\bibitem{DOPS14}
P.~Dunin-Barkowski, N.~Orantin, A.~Popolitov, and S.~Shadrin.
\newblock Combinatorics of loop equations for branched covers.
\newblock {\em Int. Math. Res. Not.}, 2017.
\newblock math-ph/1412.1698.

\bibitem{E1MM}
B.~Eynard.
\newblock Topological expansion for the {$1$}-hermitian matrix model
  correlation functions.
\newblock {\em JHEP}, 0411:031, 2004.
\newblock hep-th/0407261.

\bibitem{Eformal}
B.~Eynard.
\newblock Formal matrix integrals and combinatorics of maps.
\newblock In J.~Harnad, editor, {\em Random matrices, random processes and
  integrable systems}, CRM Series in Mathematical Physics, pages 415--442, New
  York, 2011. Springer.
\newblock math-ph/0611087.

\bibitem{EynardBook}
B.~Eynard.
\newblock {\em Counting surfaces}, volume~70 of {\em Progress in Mathematics}.
\newblock Birkh\"auser, Basel, 2016.

\bibitem{EKOn}
B.~Eynard and C.~Kristjansen.
\newblock Exact solution of the {$O(n)$} model on a random lattice.
\newblock {\em Nucl. Phys. B}, 455:577--618, 1995.
\newblock hep-th/9506193.

\bibitem{EMS}
B.~Eynard, M.~Mulase, and B.~Safnuk.
\newblock The {L}aplace transform of the cut-and-join equation and the
  {B}ouchard-{M}ari\~no conjecture on {H}urwitz numbers.
\newblock {\em Publ. Res. Inst. Math. Sci.}, 47(2):629--670, 2011.
\newblock math.AG/0907.5224.

\bibitem{EOFg}
B.~Eynard and N.~Orantin.
\newblock Invariants of algebraic curves and topological expansion.
\newblock {\em Commun. Number Theory and Physics}, 1(2), 2007.
\newblock math-ph/0702045.

\bibitem{EO2MM}
B.~Eynard and N.~Orantin.
\newblock Topological expansion of mixed correlations in the hermitian {$2$}
  matrix model and {$x-y$} symmetry of the {$F_g$} invariants.
\newblock {\em J. Phys. A: Math. Theor.}, 41, 2008.
\newblock math-ph/0705.0958.

\bibitem{EORev}
B.~Eynard and N.~Orantin.
\newblock Topological recursion in random matrices and enumerative geometry.
\newblock {\em J. Phys. A: Mathematical and Theoretical}, 42(29), 2009.
\newblock math-ph/0811.3531.

\bibitem{EOxy}
B.~Eynard and N.~Orantin.
\newblock About the {$x$}-{$y$} symmetry of the {$F_g$} algebraic invariants.
\newblock 2013.
\newblock math-ph/1311.4993.

\bibitem{EPf}
B.~Eynard and A.~Prats-Ferrer.
\newblock Topological expansion of the chain of matrices.
\newblock {\em JHEP}, (096), 2009.
\newblock math-ph/0805.1368.

\bibitem{WorkOkounkov}
G.~Felder.
\newblock The work of {A}ndrei {O}kounkov.
\newblock In {\em International {C}ongress of {M}athematicians. {V}ol. {I}},
  pages 55--64. Eur. Math. Soc., Z\"urich, 2007.
\newblock math.GM/0609847.

\bibitem{Fultonrep}
W.~Fulton and J.~Harris.
\newblock {\em Representation theory}.
\newblock Graduate texts in Mathematics. Springer, 2004.

\bibitem{GK}
M.~Gaudin and I.K. Kostov.
\newblock {$O(n)$} on a fluctuating lattice. {S}ome exact results.
\newblock {\em Phys. Lett. B}, 220(1-2):200--206, 1989.

\bibitem{HCIZ}
I.P. Goulden, M.~Guay-Paquet, and J.~Novak.
\newblock Monotone {H}urwitz numbers and the {HCIZ} integral.
\newblock {\em Ann. Math. Blaise Pascal}, 21(1):71--89, 2014.
\newblock math.CO/1107.1015.

\bibitem{GNica}
I.P. Goulden and A.~Nica.
\newblock A direct bijection for the {H}arer-{Z}agier formula.
\newblock {\em J. Combin. Theory Ser. A}, 111(2):224--238, 2005.

\bibitem{HarnadGuay}
M.~Guay-Paquet and J.~Harnad.
\newblock {$2d$} {T}oda tau-functions as combinatorial generating functions.
\newblock {\em Lett. Math. Phys.}, 105(6):827--852, 2015.
\newblock math-ph/1405.6303.

\bibitem{HarerZagier}
J.~Harer and D.~Zagier.
\newblock The {E}uler characteristics of the moduli space of curves.
\newblock {\em Invent. Math.}, 85:457--485, 1986.

\bibitem{IZ}
C.~Itzykson and J.-B. Zuber.
\newblock The planar approximation {II}.
\newblock {\em J. Math. Phys.}, 21:411--421, 1980.

\bibitem{Nonsep1}
B.~Jacquard and G.~Schaeffer.
\newblock A bijective census of nonseparable planar maps.
\newblock {\em J. Comb. Theory Ser. A}, 83:1--20, 1998.

\bibitem{Jucys}
A.A. Jucys.
\newblock Symmetric polynomials and the center of the symmetric group ring.
\newblock {\em Reports on Mathematical Physics}, 5(1):107--112, 1974.

\bibitem{Kazakov86}
V.A. Kazakov.
\newblock {I}sing model on a dynamical planar random lattice: {E}xact solution.
\newblock {\em Phys. Lett. A}, 119(3):140--144, 1986.

\bibitem{Kontsevich}
M.~Kontsevich.
\newblock Intersection theory on the moduli space of curves and the matrix
  {A}iry function.
\newblock {\em Commun. Math. Phys.}, 147:1--23, 1992.

\bibitem{Krikun}
M.~Krikun.
\newblock Explicit enumeration of triangulations with multiple boundaries.
\newblock {\em Electron. J. Combin.}, 14(1):Research Paper 61, 14, 2007.
\newblock math.CO/0706.0681.

\bibitem{LandoZvonkin}
S.K. Lando and A.K. Zvonkin.
\newblock {\em Graphs on surfaces and their applications}, volume 141 of {\em
  Encyclopedia of Mathematical Sciences}.
\newblock Springer, 2004.
\newblock With an appendix by Don B. Zagier.

\bibitem{Macdonald}
I.G. Macdonald.
\newblock {\em Symmetric functions and {H}all polynomials}.
\newblock Oxford Mathematical Monographs. Oxford University Press, 1995.

\bibitem{Semenoff}
Yu. Makeenko and G.W. Semenoff.
\newblock Properties of hermitian matrix models in an external field.
\newblock {\em Mod. Phys. Lett. A}, 6(37), 1991.

\bibitem{Meliot}
P.-L. M{\'e}liot.
\newblock {\em Representation theory of symmetric groups}.
\newblock Discrete Mathematics and its applications. CRC Press, 2017.

\bibitem{MMS07}
J.~Mingo, P.~{\'S}niady, and R.~Speicher.
\newblock Second order freeness and fluctuations of random matrices: {II}.
  {U}nitary random matrices.
\newblock {\em Adv. Math.}, 209:212--240, 2007.
\newblock math.OA/0405258.

\bibitem{MingoSpeicher}
J.~Mingo and R.~Speicher.
\newblock Second order freeness and fluctuations of random matrices: {I}.
  {G}aussian and {W}ishart matrices and cyclic {F}ock spaces.
\newblock {\em J. Funct. Anal.}, 235:226--270, 2006.
\newblock math.OA/0405191.

\bibitem{Mirzakhani}
M.~Mirzakhani.
\newblock Simple geodesics and {W}eil-{P}etersson volumes of moduli spaces of
  bordered {R}iemann surfaces.
\newblock {\em Invent. Math.}, 167:179--222, 2007.

\bibitem{Murphy}
G.E. Murphy.
\newblock A new construction of {Y}oung's seminormal representation of the
  symmetric groups.
\newblock {\em Journal of Algebra}, 69:287--297, 1981.

\bibitem{NicaSpeicher}
A.~Nica and R.~Speicher.
\newblock {\em Lectures on the combinatorics of free probability}, volume 335
  of {\em London Mathematical Society Lecture Note Series}.
\newblock Cambridge University Press, Cambridge, 2006.

\bibitem{NovakJ}
J.~Novak.
\newblock Complete symmetric polynomials in {J}ucys-{M}urphy elements and the
  {W}eingarten function.
\newblock math.CO/0811.3595.

\bibitem{NovakJMelements}
J.~Novak.
\newblock Jucys-{M}urphy elements and the unitary {W}eingarten function.
\newblock In {\em Noncommutative harmonic analysis with applications to
  probability {II}}, volume~89 of {\em Banach Center Publ.}, pages 231--235.
  Polish Acad. Sci. Inst. Math., Warsaw, 2010.

\bibitem{Okounkov}
A.~Okounkov.
\newblock Toda equations for {H}urwitz numbers.
\newblock {\em Math. Res. Lett.}, 7(4):447--453, 2000.
\newblock math.AG/0004128.

\bibitem{OrlovSch}
A.Yu. Orlov and D.M. Shcherbin.
\newblock Hypergeometric solutions of soliton equations.
\newblock {\em Theoret. Math. Phys.}, 128:906--926, 2001.

\bibitem{Speicher}
R.~Speicher.
\newblock Multiplicative functions on the lattice of noncrossing partitions and
  free convolution.
\newblock {\em Math. Ann.}, 298(4):611--628, 1994.

\bibitem{tHooft}
G.~t'Hooft.
\newblock A planar diagram theory for strong interactions.
\newblock {\em Nucl. Phys. B}, 72:461--473, 1974.

\bibitem{Tutte2}
W.T. Tutte.
\newblock A census of {H}amiltonian polygons.
\newblock {\em Canad. J. Math.}, 14:402--417, 1962.

\bibitem{Tutte4}
W.T. Tutte.
\newblock A census of planar maps.
\newblock {\em Canad. J. Math.}, 15:249--271, 1963.

\bibitem{Voiculescu2}
D.-V. Voiculescu.
\newblock Addition of certain non-commuting random variables.
\newblock {\em J. Funct. Anal.}, 66:323--346, 1986.

\bibitem{Voiculescu3}
D.-V. Voiculescu.
\newblock Limit laws for random matrices and free products.
\newblock {\em Invent. Math.}, 104(1):201--220, 1991.

\bibitem{LehmanWalsh}
R.S. Walsh and A.B. Lehman.
\newblock Counting rooted maps by genus. {I}.
\newblock {\em J. Combin. Theory Ser. B}, 13:192--218, 1972.

\bibitem{Nonsep0}
R.S. Walsh and A.B. Lehman.
\newblock Counting rooted maps by genus {III}: {N}onseparable maps.
\newblock {\em J. Comb. Theory Series B}, 18(3):222--259, 1975.

\bibitem{Witten}
E.~Witten.
\newblock Two dimensional gravity and intersection theory on moduli space.
\newblock {\em Surveys in Diff. Geom.}, 1:243--310, 1991.

\bibitem{Zagierapp}
D.~Zagier.
\newblock Applications of the representation theory of finite groups.
\newblock {\em Encycl. of Math. Sciences}, 141:399--427, 2004.
\newblock Appendix to "Graphs on Surfaces and Their applications",
  Springer-Verlag, Berlin Heidelberg.

\end{thebibliography}

\end{document}